\documentclass[aps,prx,twocolumn,superscriptaddress]{revtex4-2}

\usepackage{graphicx}
\usepackage{commath}
\usepackage{tikz}
\usepackage{mdframed, color}
\usepackage{titlesec}
\usepackage{multirow}
\usepackage[hypertexnames=false]{hyperref}
\usepackage{bookmark}

\usepackage{bm}

\begin{document}

\title{The statistical geometry of material loops in turbulence}

\newcommand{\mpids}{Max Planck Institute for Dynamics and Self-Organization, Am Fa\ss berg 17,
37077 G\"{o}ttingen, Germany}
\newcommand{\bayreuth}{Theoretical Physics I, University of Bayreuth, Universit\"atsstra\ss e 30, 95447 Bayreuth, Germany}

\author{Lukas Bentkamp}
    \affiliation{\mpids}
    \affiliation{\bayreuth}
    
\author{Theodore D. Drivas}
    \affiliation{Mathematics Department, Stony Brook University, 100 Nicolls Rd., Stony Brook, NY 11794, USA}
    \affiliation{School of Mathematics, Institute for Advanced Study, 1 Einstein Dr., Princeton, NJ 08540, USA}
    
\author{Cristian C. Lalescu}
    \affiliation{\mpids}
    \affiliation{Max Planck Computing and Data Facility, Gie\ss enbachstra\ss e 2, 85748 Garching b.\ M\"unchen, Germany}

\author{Michael Wilczek}
    \email{michael.wilczek@uni-bayreuth.de}
    \affiliation{\mpids}
    \affiliation{\bayreuth}
\date{\today}

\begin{abstract}
\textbf{Abstract}\quad Material elements -- which are lines, surfaces, or volumes behaving as passive, non-diffusive markers -- provide an inherently geometric window into the intricate dynamics of chaotic flows. Their stretching and folding dynamics has immediate implications for mixing in the oceans or the atmosphere, as well as the emergence of self-sustained dynamos in astrophysical settings.
Here, we uncover robust statistical properties of an ensemble of material loops in a turbulent environment.  Our approach combines high-resolution direct numerical simulations of Navier-Stokes turbulence, stochastic models, and dynamical systems techniques to reveal predictable, universal features of these complex objects. We show that the loop curvature statistics become stationary through a dynamical formation process of high-curvature folds, leading to distributions with power-law tails whose exponents are determined by the large-deviations statistics of finite-time Lyapunov exponents of the flow. This prediction applies to advected material lines in a broad range of chaotic flows. To complement this dynamical picture, we confirm our theory in the analytically tractable Kraichnan model with an exact Fokker-Planck approach.
\end{abstract}

\maketitle

\section*{\pdfbookmark[1]{Introduction}{Introduction}Introduction}

\begin{figure}[hb!]
    \centering
    \setlength{\fboxsep}{0pt}
    \definecolor{lgray}{gray}{0.95}
    \begin{minipage}{0.5\textwidth}
      \includegraphics[width=\textwidth]{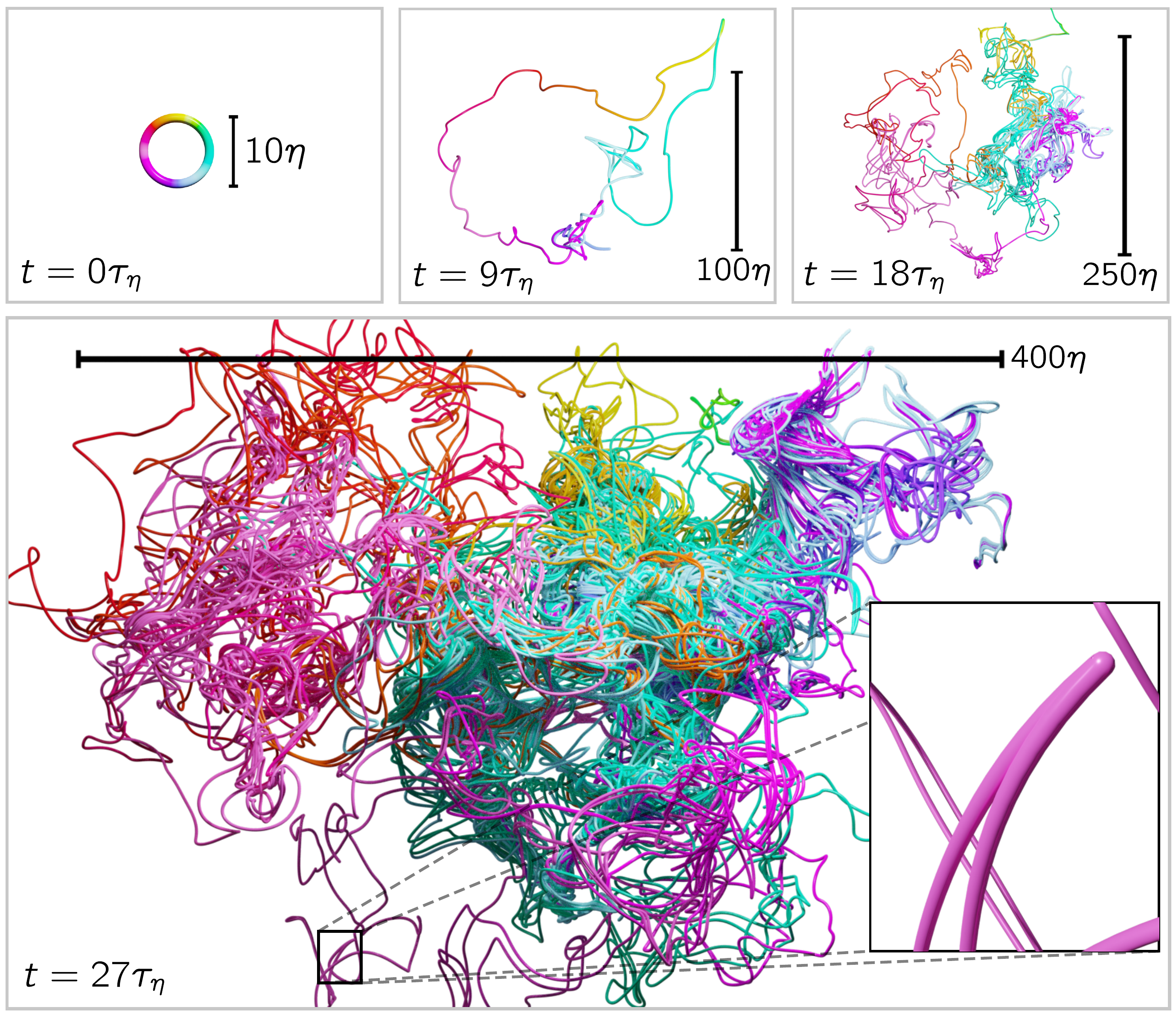}
    \end{minipage}
    \caption{\textbf{Visualization of material loop evolution.} The initially circular loop (color corresponds to initial angle) is advected by a turbulent flow field for $27\tau_\eta$, where $\tau_\eta$ is the Kolmogorov time. The twisting and folding action of the turbulent flow creates a complex loop geometry while the length of the loop increases exponentially on average (cf.\ Fig.~\ref{fig:peak_number}). The loop shown is a comparably extreme case; loops in less turbulent regions develop an extended and complex structure after a longer time. Inset: material fold causing a peak of curvature. (See also \href{https://youtu.be/1FK2nfswz1c}{supplementary movie}) \label{fig:loop_visualization}}
\end{figure}

Chaotic flows tend to fold, writhe, and wrinkle material elements into a state of seemingly infinite complexity over time (see Fig.~\ref{fig:loop_visualization} and \href{https://youtu.be/1FK2nfswz1c}{supplementary movie}). A fundamental question is whether this tumultuous process has any predictable features which persist over long periods of time. Answering this question provides insights into the process of mixing which occurs in a whole range of systems, from the diffusion of dye into water, the dispersion of plankton colonies on the ocean surface, to the blast propagation in supernovae thermonuclear explosions~\cite{dimotakis_annrev_2005}. Material lines and interfaces, in particular, provide idealized descriptions of nutrient, temperature and salinity fronts in the oceans~\cite{prants_dsr_2014}, and potential vorticity fronts in the atmosphere~\cite{haynes_jas_1987}. They are also closely related to the dynamics of vorticity filaments in fully developed turbulence~\cite{ohkitani_pre_2002, guala_jfm_2005}, the conformation of polymer chains~\cite{balkovsky_prl_2000,liberzon2005pof,bagheri2012statistics}, the dynamics of flexible phytoplankton chains \cite{musielak2009nutrient}, as well as the motion of magnetic field lines at high conductivity (or high magnetic Reynolds numbers)~\cite{davidsonTurbulenceRotatingStratified2013}. The latter is related to the dynamo problem, in which chaotic stretching, folding, and twisting processes are essential for sustaining the growth of a magnetic field. The progress we make in understanding how material elements react to turbulent flows stands to advance our understanding of these fundamental problems.

\begin{figure*}[ht]
    \begin{minipage}{1\textwidth}
      \includegraphics[width=1.\textwidth]{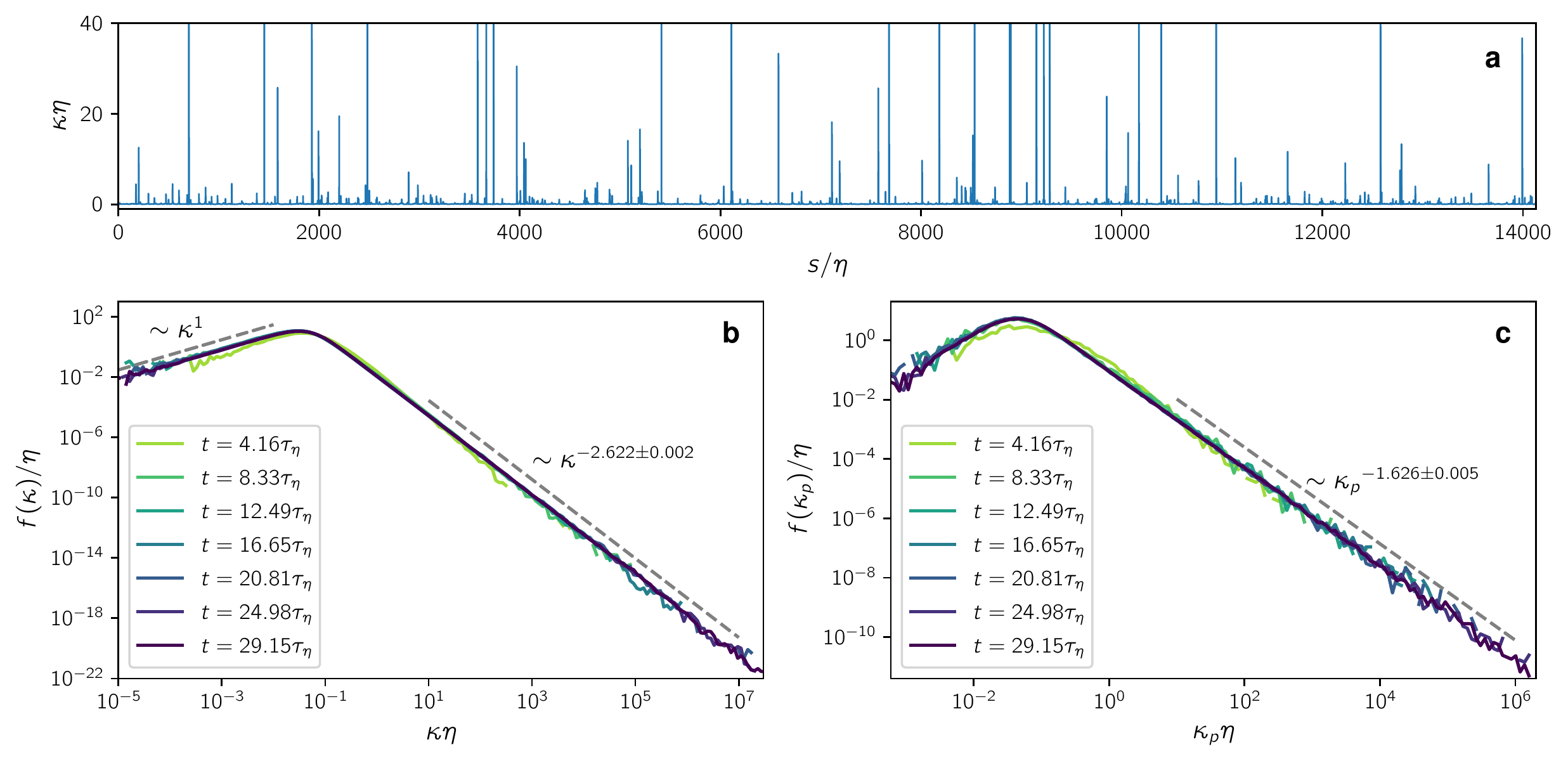}
    \end{minipage}
    \caption{\textbf{Localized peaks of curvature along the loop cause heavy-tailed curvature distributions.} \textbf{a} Curvature along a material loop at $t=29.15\tau_\eta$ as a function of arc length $s$. The function is highly spiked, indicating that high curvature only occurs in isolated narrow regions. These isolated peaks contribute to the high-curvature tails of the curvature PDF. \textbf{b} Curvature PDF of material loops at times $t=4.16\tau_\eta$ (light green) up to $t=29.15\tau_\eta$ (violet). \textbf{c} PDF of curvature peaks of material loops at the same times. The high-curvature regime is fitted by power laws in the regions indicated by the dashed lines by means of a linear fit to the logarithm of the PDF using binomial error estimates.\label{fig:dns_curvature_pdf}}
\end{figure*}

 The geometry of material objects advected and deformed by a turbulent flow can be very complex. While volumes are preserved by incompressible flows, the length of lines and the area of surfaces typically grow exponentially~\cite{batchelor_prsla_1952, girimaji_jfm_1990, drummond_jfm_1990_stretch,ishihara1992jpsj, tabor_csf_1994}, with their geometry appearing fractal~\cite{villermaux_prl_1994,nicolleau_pof_1996,iyer_prf_2020}. 
 Since any curve in space is uniquely described by its curvature and torsion~\cite{baer_2010}, there have been numerous works attempting to characterize the curvature of material lines but also of material surfaces \cite{pope_ijes_1988, pope_pofa_1989, drummond_jfm_1991_curv, girimaji_pofa_1991, drummond_jfm_1993, liu_pof_1996, hobbs1997fractals, hobbs_pof_1998, cerbelli_2000_ces, kivotides_pla_2003, thiffeault_pd_2004, leonard_fdr_2005, leonard_jfm_2009, thiffeault_2009_ghost_rods, ma_siam_2014} and Lagrangian trajectories~\cite{braun_jot_2006,xu_prl_2007,scagliarini_jot_2011}.
 Although material lines seem to become unfathomably complicated over time, the above works  suggest that curvature distributions do in fact settle down to a well defined stationary state which features robust power-law tails (see Fig.~\ref{fig:dns_curvature_pdf}), sparking hope that certain features can be predicted by theory.
 
Here we present a line of arguments based on the dynamical mechanism of sling or fold (i.e.~curvature peak) formation and its relation to finite-time Lyapunov exponents that leads to a quantitative prediction of the power law of the curvature distribution observed in Fig.~\ref{fig:dns_curvature_pdf}, panels b and c. We show that the high-curvature regime of the material line can be understood as an ensemble of persistent parabolic folds, which are formed by random stretching of the line. In this way, we illustrate how understanding dynamical mechanisms can be used to make deductions about statistical geometry. For example, our predicted curvature PDF power-law exponent $-2.54 \pm 0.11$ ($3\%$ relative error to the measured exponent) implies that, in the long-time limit, the average curvature along advected loops is finite but all higher moments diverge.  
The only input of our theory is the distribution of Lyapunov exponents of the underlying flow field and, as such, our results apply to a wide range of chaotic dynamics. Our predictions are confirmed by direct numerical simulations of fully developed homogeneous, isotropic Navier-Stokes turbulence as well as by exact results in the analytically solvable Kraichnan model.

\section*{\pdfbookmark[1]{Results}{Results}Results}

To investigate the evolution of material loops $\mathbf{L}(\phi, t)$ in fully developed turbulence, we consider initially circular loops and parameterize them by the initial angle $\phi\in[0,2\pi)$. Each point of the loop follows the velocity field $\mathbf{u}(\mathbf{x}, t)$ according to the tracer equation
\begin{equation} \label{eq:tracer_eq}
  \partial_t \mathbf{L}(\phi, t) = \mathbf{u}(\mathbf{L}(\phi, t), t).
\end{equation}
The evolution of such a loop is shown in Fig.~\ref{fig:loop_visualization}, which illustrates that the loop rapidly grows in length and diameter, while attaining a complex geometry due to the stretching and folding by the underlying turbulent flow.

As a key metric to characterize the geometry of the loop, we here focus on the curvature
\begin{equation} \label{eq:curv_def}
  \tilde{\kappa}(\phi, t) = \frac{\left|(\partial_{\phi}^2 \mathbf{L}) \times (\partial_{\phi} \mathbf{L})\right|}{\left|\partial_{\phi} \mathbf{L}\right|^3}.
\end{equation}
Material lines grow non-uniformly in length over time. Hence for an evolving ensemble of loops, the distribution of curvature can be defined in different ways, depending on the probability measure we associate with the points along the loop. A simple way of defining the probability density function (PDF) of curvature $f(\kappa; t)$, that does not depend on the initial parameterization, is to take curvature samples uniformly along the arc length of the loops. Specifically,
\begin{equation}\label{eq:curvPDF}
f(\kappa; t) = \frac{1}{\langle L(t)\rangle} \left \langle \int_0^{L(t)}\dif s \ \delta(\kappa -\tilde{\kappa}(s,t)) \right\rangle 
\end{equation}
where $\delta$ is the Dirac delta function, $L(t)$ is the length of the loop at time $t$ and $\tilde{\kappa}(s,t)$ is the curvature of the loop as a function of arc length $s$ at time $t$.  The average $\langle \cdot \rangle$ is taken to be uniform over loops, and we have here used $\tilde{\kappa}$ to distinguish the loop (realization) dependent curvature from its sample-space variable $\kappa$.

We use fully resolved turbulence simulations to investigate this measure of the statistical geometry of material lines (see Methods). Here, we focus on a data set at the Taylor-scale Reynolds number $R_{\lambda} \approx 216$, in which we track 1000 randomly placed loops with an initial diameter of $10\eta$ ($\eta$ is the Kolmogorov length scale). We test the robustness of our results with additional simulations at various Reynolds numbers in Supplementary Note~1.

The resulting curvature PDF at different times is shown in Fig.~\ref{fig:dns_curvature_pdf}b. Remarkably, persistent power-law tails form within a few Kolmogorov time scales $\tau_\eta$, which eventually range over several decades of curvature after the loops have been deformed for $29\tau_\eta$ ($\sim 1.5$ integral times). Within this observation window, the shape of the distribution appears to become stationary, whereas the support, i.e.~the range from minimum to maximum curvature, grows indefinitely in extent. Hence the largest curvatures correspond to structures significantly smaller than the Kolmogorov length scale $\eta$. As we show in Supplementary Note~1, the distributions are almost indistinguishable for different Reynolds numbers when nondimensionalized by $\eta$, but they shift to larger $\kappa$ when displayed in units of the integral length. This is a first indication that the curvature distribution is generated by the smallest scales of the flow, in particular by velocity gradients.
Given the markedly complex shape of the deformed material loop, the universal shape of the distribution calls for a theoretical explanation, which we develop in the following.

\subsection*{Ensemble of material folds}

The high-curvature regime of the curvature distribution is heavy-tailed and characterized by rare events. Over time, the material line will form isolated sites of extremely high curvature~\cite{thiffeault_pd_2004, leonard_fdr_2005, thiffeault_2009_ghost_rods, leonard_jfm_2009, ma_siam_2014}, as can be seen in Fig.~\ref{fig:dns_curvature_pdf}a. Such curvature peaks mark sharp folds in the material line geometry. In the following, we reveal how such folds form stochastically and how this is related to the power-law exponent of the curvature distribution.

This picture in view, we estimate the high-curvature tail of the PDF~\eqref{eq:curvPDF} in the statistically steady state by replacing the ensemble average over entire loops in \eqref{eq:curvPDF} by an ensemble of folds,
\begin{align}
    f(\kappa) \sim \int_0^\infty \dif \kappa_p~f(\kappa_p) \int_{-\infty}^\infty \dif s~
    \delta\left(\kappa - \kappa^{\text{pb}}(s; \kappa_p)\right).
    \label{eq:slingensemble}
\end{align}
Here, $\kappa_p$ is the peak curvature of a fold and $f(\kappa_p)$ its distribution. The second integral is the contribution of curvature around each curvature peak. As we will elaborate in more detail below, high-curvature folds develop a universal, locally parabolic shape. The curvature function around a peak with maximum $\kappa_p$, therefore, can be estimated as~\cite{leonard_jfm_2009}
\begin{align} \label{eq:curvature_parabola}
    \kappa^{\text{pb}}(s; \kappa_p) &= \frac{\kappa_p}{\left(1+F^{-1}(|\kappa_p s|)^2\right)^{3/2}},
\end{align}
where $F^{-1}(x)$ denotes the inverse of the primitive of $\sqrt{1+x^2}$ on the positive real line, originating from parameterizing the parabola by arc length. Remarkably, the curvature profile is characterized by the peak curvature as the only parameter.
To further evaluate \eqref{eq:slingensemble}, we substitute the inner integration variable by $\kappa' = \kappa^{\text{pb}}(s; \kappa_p)$ with the Jacobian
\begin{align} \label{eq:single_peak_contribution}
    \left|\dod{s^{\text{pb}}(\kappa'; \kappa_p)}{\kappa'}\right| = \frac{1}{3\kappa'^2 \sqrt{(\kappa_p/\kappa')^{2/3} - 1}},
\end{align}
which yields
\begin{align}
    f(\kappa) &\sim 
    \int_{\kappa}^\infty \dif \kappa_p~f(\kappa_p) \left|\dod{s^\text{pb}(\kappa; \kappa_p)}{\kappa}\right|.
    \label{eq:peak_ensemble_integral}
\end{align}
This equation expresses the curvature PDF as a composition of the curvature peak PDF with the contribution from the locally parabolic folds.

\subsection*{Statistical evolution of curvature peaks}
In what follows, we determine the curvature peak distribution $f(\kappa_p)$, which can be achieved by capturing the essence of the curvature peak dynamics. Since peaks are generally generated at medium curvature and then grow stochastically, we may define the generation time $t_0$ of a large peak as the time where it has first surpassed an (arbitrary) threshold $\kappa_0$ and its age as $\tau=t-t_0$. At time $t$, the ensemble of peaks larger than $\kappa_0$ can thus be attributed a distribution of ages $f(\tau; t)$. By the law of total probability, the peak distribution above $\kappa_0$ can be estimated as
\begin{align} \label{eq:curvature_peak_total_prob}
    f(\kappa_p; t) &\sim \int_0^t \dif \tau~f(\kappa_p | \tau) f(\tau; t),
\end{align}
where $f(\kappa_p | \tau)$ is the probability of a peak with curvature $\kappa_0$ at time $t_0$ to have curvature $\kappa_p$ at time $t_0+\tau$. This decomposes the curvature peak distribution into a distribution of peaks with a given age and the distribution of ages.
In \eqref{eq:peak_ensemble_integral}, we are interested in the stationary regime $f(\kappa_p):=\lim_{t\to\infty}   f(\kappa_p; t)$, which we expect to be well captured by the estimate~\eqref{eq:curvature_peak_total_prob} and to be independent of the arbitrary threshold $\kappa_0$.  

\begin{figure}
    \centering
    \begin{minipage}{.5\textwidth}
      \includegraphics[width=1\textwidth]{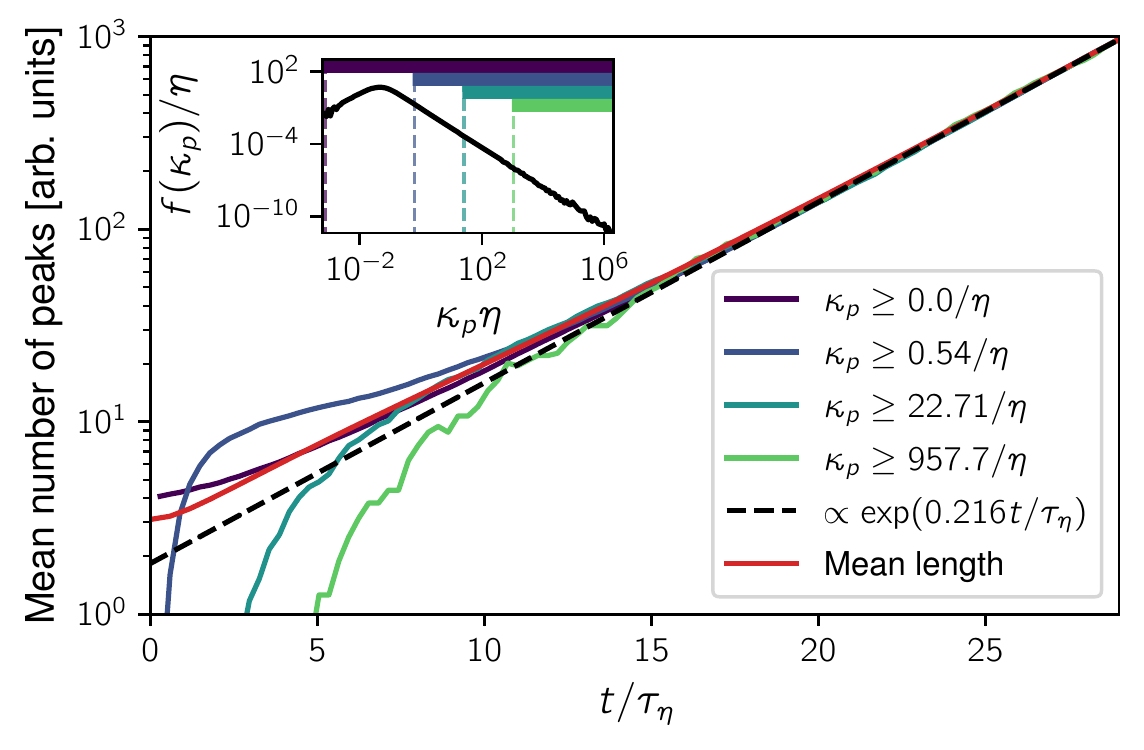}
    \end{minipage}
    \caption{\textbf{Mean number of curvature peaks above different thresholds over time.} The lines are vertically shifted for comparison, showing that the peaks are generated at a clearly defined exponential rate. Moreover, the curves appear to be asymptotically proportional to the mean arc length of loops (red). The dashed line indicates an exponential fit to the last third of the total peak number curve (violet), yielding the rate $\beta =  (0.21619 \pm 0.00014)/\tau_\eta$. The standard error of this rate is so small that we neglect it in the following.  Note that without vertically shifting the curves in the plot, they would remain ordered as a function of the threshold condition. Inset: Curvature peak distribution at $t=29.15\tau_\eta$ indicating the different thresholds. \label{fig:peak_number}}
\end{figure}

The peak age distribution can be estimated from the mean number of curvature peaks. Figure~\ref{fig:peak_number} shows that the mean numbers of curvature maxima above different thresholds grow at the same exponential rate $\beta\approx 0.216/\tau_\eta$, which coincides with the growth rate of the mean length of the loops.
Intuitively, this can be explained by the fact that the generation of folds is a random process along the loop. Since the loop length grows on average exponentially over time, so does the number of folds.
Neglecting the disappearance of peaks, we, therefore, estimate the probability of a high-curvature fold at time $t$ to be generated before some time $t'$ (with $0\leq t' \leq t$) by the fraction of peaks that existed at $t'$, given by
$e^{\beta t'}/e^{\beta t}$.
This cumulative distribution function of peak birth times implies the probability density function of peak age
\begin{align} \label{eq:generation_time_estimate}
    f(\tau; t) &\approx \beta e^{-\beta\tau}, \qquad 0\leq \tau \leq t.
\end{align}
This shows that, since curvature peaks are generated at an exponential rate, their age distribution also decays exponentially, implying that the bulk of the peaks are young even after a long evolution of the loop.

In the following, we investigate the dynamics and statistics of peak curvature in an effort to estimate the remaining conditional probability $f(\kappa_p|\tau)$ and form our theory.

\subsection*{Amplification of folds by turbulent stretching}
\begin{figure}
    \centering
        \begin{minipage}{0.5\textwidth}
    \includegraphics[width=1\textwidth]{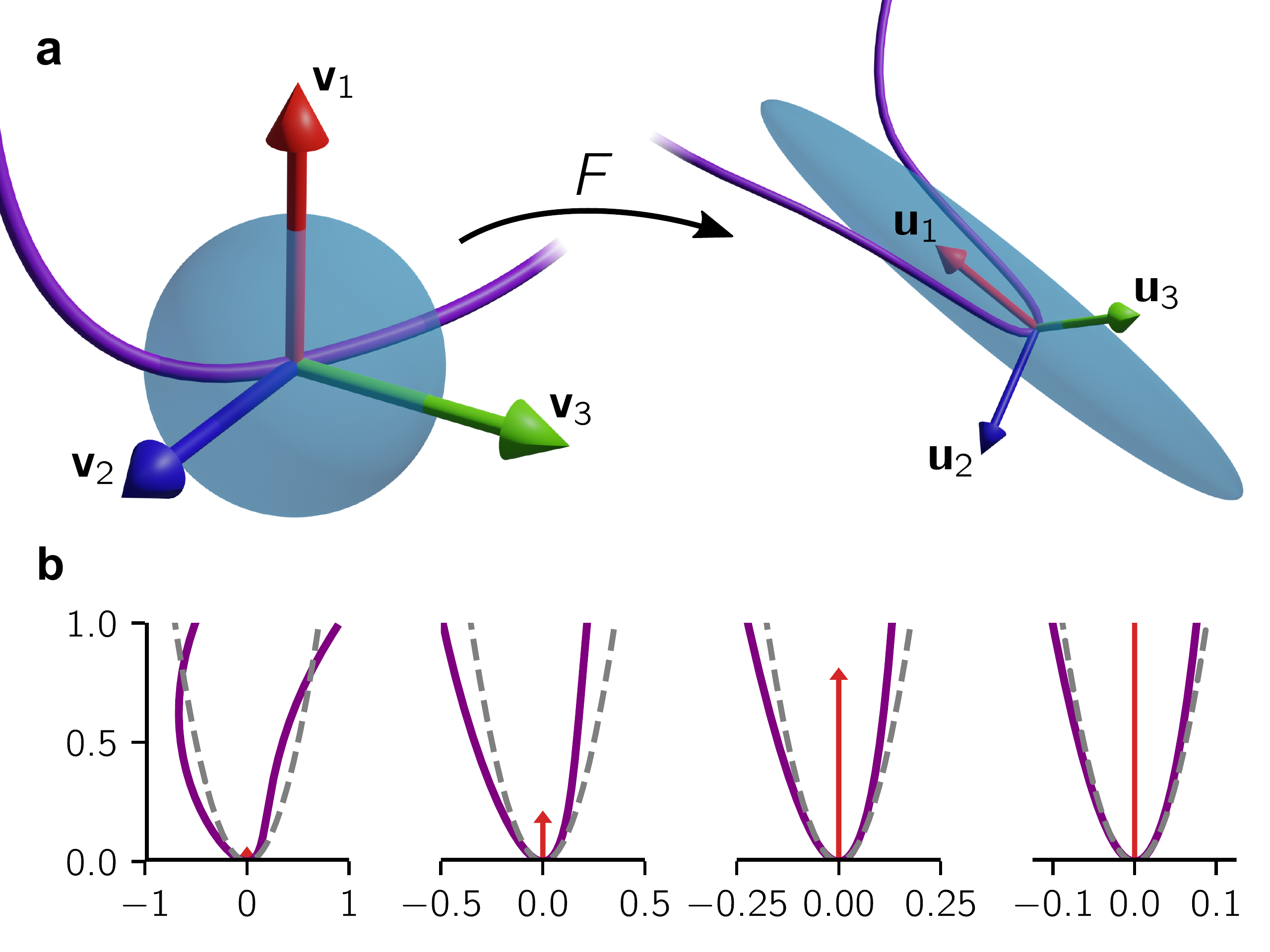}
    \end{minipage}
    \caption{\textbf{Formation of a parabolic fold.} \textbf{a} Illustration of the deformation tensor $F$. $\mathbf{v}_i$ denote the principal axes of stretching before deformation and $\mathbf{u}_i$ the corresponding axes after deformation. A fluid element (blue) will be predominantly stretched along the direction of most stretching $\mathbf{v}_1$ and compressed in the direction of most compression $\mathbf{v}_3$ over time. If a material line element (violet) is initially orthogonal to the direction of most stretching, a fold will form.
    Such a fold is then compressed onto the $\mathbf{u}_1$-$\mathbf{u}_2$ plane and tends to  align with the  $\mathbf{u}_1$ direction along which it is amplified. 
    \textbf{b} Stretching creates a locally parabolic curve. An initially non-parabolic curve is stretched vertically as indicated by the red arrows. Viewed on the appropriate horizontal scale, the line becomes increasingly parabolic. For comparison, the dashed line indicates a parabola with the same peak curvature. 
    \label{fig:align} }
\end{figure}

We observe that those rare peaks that have existed for a long time can exhibit extremely high curvature. This is caused by fluid element stretching, a process quantitatively captured by the deformation tensor
\begin{align} \label{eq:deformation_tensor}
    F_{ij}(\mathbf{x}, t) &= \dpd{X_i(\mathbf{x}, t)}{x_j},
\end{align}
where $\mathbf{X}(\mathbf{x},t)$ is the Lagrangian map, mapping the initial condition $\mathbf{x}$ of a tracer particle to its position $\mathbf{X}$ at time $t$.
The singular value decomposition of the deformation tensor associates two coordinate systems $\mathbf{v}_i$ and $\mathbf{u}_i$ with the deformation (see Methods), as illustrated in Fig.~\ref{fig:align}a. 
The associated exponential stretching rates are given by the finite-time Lyapunov exponents (FTLE) $\rho_i(t)$. 

As discussed in ref.~\cite{leonard_jfm_2009}, 
generically a line element will align with the $\mathbf{u}_1$-direction and become stretched exponentially with $e^{\rho_1(t) t}$ (whose mean asymptotically scales like $e^{\beta t}$). The surrounding curve will be forced into the $\mathbf{u}_1$-$\mathbf{u}_2$ plane by compression in the $\mathbf{u}_3$-direction. The dominant stretching in the $\mathbf{u}_1$-direction locally decreases curvature. However, an exception to this generic setting occurs at a finite number of points along the loop when the initial material line lies perpendicular to $\mathbf{v}_1$ (see Fig.~\ref{fig:align}).  In this case, the line element cannot align with $\mathbf{u}_1$ and will align with $\mathbf{u}_2$ instead. The surrounding curve, however, still experiences the stretching in the $\mathbf{u}_1$-direction. This essentially magnifies the local structure of the curve, which will generically result in a parabolic shape, as illustrated in Fig.~\ref{fig:align}b. Therefore parabolas become increasingly good local approximations of the folds.

To reveal the role of the finite-time Lyapunov exponents, let us consider a parabola $ y=\kappa_0 x^2/2$ which is already initially lying in the $\mathbf v_1$-$\mathbf v_2$ plane. Over time, it is subject to stretching $y' = e^{\rho_1(t) t}y$ and $x' = e^{\rho_2(t) t}x$, which preserves the parabolic shape, i.e.~$y' = e^{[\rho_1(t)-2\rho_2(t)]t} \kappa_0 x'^2/2$.
In this process, the peak curvature increases as long as $\rho_1(t) > 2\rho_2(t)$~\cite{leonard_jfm_2009}, i.e.\ the first FTLE must be more than twice as large as the second one. We illustrate this 
at the example of a parabola in a linearized flow in Methods, showing that its peak curvature grows as
\begin{align}\label{eq:curv_ftle_relation}
    \kappa_p(t) \stackrel{t \gg 0}{\approx} \widetilde{\kappa}_0 e^{[\rho_1(t) - 2\rho_2(t)]t}
\end{align}
for some effective initial peak curvature $\widetilde{\kappa}_0$. This equation can already be found in ref.~\cite{leonard_jfm_2009}, where it is derived for a generic material line.
Let us call the growth rate of peaks $\rho_p(t) = \rho_1(t) - 2\rho_2(t)$. In turbulence, this growth rate is typically asymptotically positive. In our simulation used for obtaining the FTLEs (see Methods), we can estimate the infinite-time Lyapunov exponents, $\lambda_i = \lim_{t \to\infty} \rho_i(t)$, by taking the mean of the FTLEs at the final time of the simulation, which yields $\lambda_1 \approx 0.12/\tau_\eta$, $\lambda_2 \approx 0.03/\tau_\eta$, $\lambda_3 \approx -0.15/\tau_\eta$, in good agreement with previous literature \cite{bec_pof_2006, johnson_pof_2015}, and thus $\lambda_p = \lim_{t\to\infty} \rho_p(t) \approx 0.06/\tau_\eta > 0$.

\subsection*{Connecting the power-law exponent to fluid stretching}

To relate the dynamical formation of folds to the power-law tails of the curvature PDF, we estimate the distribution of $\kappa_p(t)$ by making statements about the distribution of FTLEs. By ergodicity, FTLEs behave like sums of independent and identically distributed random variables at large times~\cite{balkovsky_pre_1999, johnson_pof_2015}. The same is true for the growth rate of peaks $\rho_p(t)$. Using its Cram\'er function $S(\rho_p)$, we make a large-deviations estimate of the PDF,
\begin{align} \label{eq:ftle_pdf_ld_approx}
    f(\rho_p; t) \approx N(t) e^{-tS(\rho_p)},
\end{align}
where $N(t)$ is a normalization. Transforming by \eqref{eq:curv_ftle_relation}, the peak curvature PDF for peaks of age $\tau$ can thus be written as
\begin{align}
    f(\kappa_p|\tau) &\approx \frac{N(\tau)}{\kappa_p \tau} e^{-\tau S\left(\log\left(\frac{\kappa_p}{\kappa_0}\right)/\tau\right)}.
\end{align}
Note that we here identified the peak age $\tau$ with the time $t$ and the curvature threshold $\kappa_0$ with the effective initial peak curvature $\widetilde{\kappa}_0$. For the asymptotics that we are interested in, the distinction does not matter. Inserting this result into \eqref{eq:curvature_peak_total_prob}, combined with \eqref{eq:generation_time_estimate} and letting $t\to\infty$, gives the asymptotic distribution of curvature peaks in the high-curvature regime
\begin{align}
    f(\kappa_p) \sim \int_0^\infty \dif\tau~e^{-\beta\tau} \frac{N(\tau)}{\kappa_p\tau} e^{-\tau S\left(\log\left(\frac{\kappa_p}{\kappa_0}\right)/\tau\right)}.  \label{eq:ld_integral}
\end{align}

\begin{figure}
    \begin{minipage}{.5\textwidth}
      \includegraphics[width=1\textwidth]{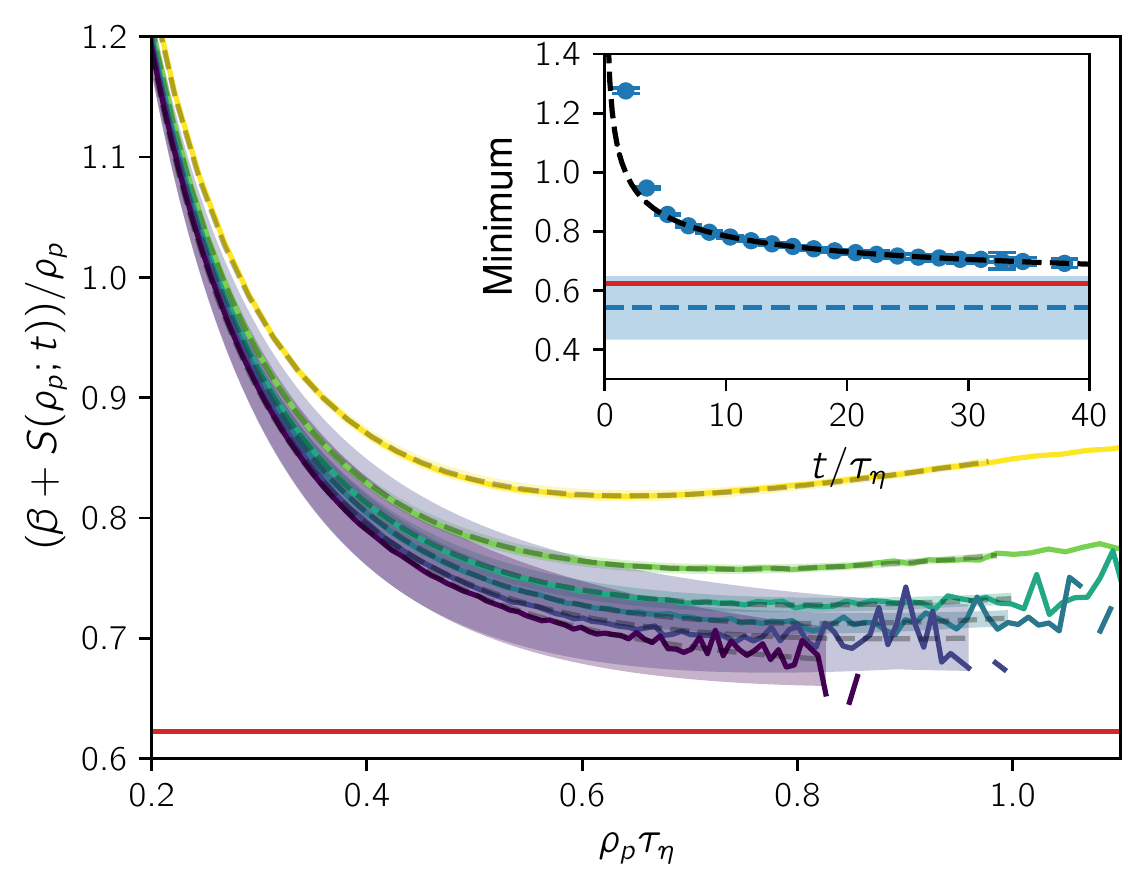}
    \end{minipage}
    \caption{\textbf{Determination of the steepest-descent minimum.} The Cram\'er function is estimated from FTLE histograms by~\eqref{eq:ftle_pdf_ld_approx}. We call these finite-time estimates $S(\rho_p;t)$. Here we show the function minimized in \eqref{alphaeqn} for estimates of the Cram\'er function ranging from $t = 6.90\tau_\eta$ (yellow) up to $t = 39.68\tau_\eta$ (violet). Best fits are indicated by dashed lines with shaded areas showing their error (see Methods for details). Inset: Minima of these functions over time. A simple fit of the decay of minima (black dashed line) yields an estimate of their limiting value $\alpha=0.54 \pm 0.11$ (horizontal blue dashed line and shaded area). For comparison, the red lines show the value of $\alpha$ estimated by subtracting 2 from the observed curvature PDF power-law exponent in Fig.~\ref{fig:dns_curvature_pdf}b, showing a good agreement within uncertainties. For more details, see Methods.\label{fig:cramer_minimum}}
\end{figure}

We now use the method of steepest descent~\cite{olver_1997} in order to extract the large-$\kappa_p$ asymptotics of the peak curvature distribution from our estimate~\eqref{eq:ld_integral}. The result (see Methods) is that the distribution scales as a power law, $f(\kappa_p) \sim \kappa_p^{-1-\alpha}$, with exponent
\begin{align}\label{alphaeqn}
    \alpha = \min_{\rho_p} \left[\tfrac{1}{\rho_p}( \beta + S(\rho_p))\right].
\end{align}
This minimum is estimated for our data in Fig.~\ref{fig:cramer_minimum}, 
where the Cram\'er functions have been estimated via~\eqref{eq:ftle_pdf_ld_approx} using FTLE histograms from an additional simulation (see Methods). While we are interested in finding the minimum for the fully converged Cram\'er function, the amount of samples needed to resolve large-deviations statistics increases exponentially with time, limiting our observation window of the minimum to a maximum time of about 30 to $40\tau_\eta$. In this regime, the minima still lie above the value of $\alpha$ inferred from the loops simulation (red line). However, an analysis of the time evolution of minima (Fig.~\ref{fig:cramer_minimum}, inset) reveals that they are well described by a slow, algebraic decay. Extrapolating the desired minimum towards $t\to\infty$, we get the estimate $\alpha = 0.54\pm 0.11$, slightly below but within error bars of the curvature peak power-law exponent in Fig.~\ref{fig:dns_curvature_pdf}c. For more details on the extrapolation, see Methods.

Given the power-law scaling of the peak distribution, $f(\kappa_p) \sim {\kappa_p}^{-1-\alpha}$, we can perform the integral~\eqref{eq:peak_ensemble_integral} to obtain the prediction for the curvature PDF
\begin{align}
    f(\kappa) \sim \kappa^{-2-\alpha}.
\end{align}
Hence the difference between the curvature power-law exponent and the curvature peak power-law exponent is 1. This difference originates from the curvature contributions of parabolic fold profiles around the peak curvature~\eqref{eq:curvature_parabola}. Comparing Fig.~\ref{fig:dns_curvature_pdf}b and c shows that this result is consistent with the fully resolved loops simulations. Likewise, our prediction based on Lyapunov exponents estimated by extrapolating the minimum in Fig.~\ref{fig:cramer_minimum} captures the observed power-law exponents of both the curvature and curvature peak PDFs very well. In Supplementary Note~1, we explore our result at various Reynolds numbers, with comparable or even better agreement depending on how far the minima can be resolved in time. Therefore, as a central result, we can quantitatively relate the statistical geometry as characterized by the curvature PDF to the formation of folds and the statistics of FTLEs that determine their dynamical evolution.

Interestingly, an alternative formulation of our result can be obtained by using the Legendre transform of the Cram\'er function, which is known as the generalized Lyapunov exponent~\cite{johnson_pof_2015}. It can be shown (see Methods) that $\alpha$ is given implicitly by
\begin{align} \label{eq:stretch_fold_balance_main}
    \left\langle e^{\alpha\rho_p(t)t} \right\rangle &\sim \left\langle e^{\rho_1(t)t} \right\rangle
\end{align}
in the large-deviations approximation, where $\sim$ indicates the same exponential scaling for large~$t$. This can be understood as the statement that the power-law exponent is chosen so that curvature peak generation (represented by the line growth rate $\rho_1(t)$) and peak amplification (represented by the peak curvature growth rate $\rho_p(t) = \rho_1(t) - 2\rho_2(t)$) are on average balanced. For example, in a flow with the same peak amplification (same statistics of $\rho_p(t)$) but stronger line growth (larger $\langle e^{\rho_1(t) t}\rangle$) and thus stronger peak generation, a larger fraction of small-curvature peaks will accumulate until the stationary state is reached. This means that the curvature PDF in the stationary state has to decay faster, corresponding to a larger $\alpha$, as encoded in \eqref{eq:stretch_fold_balance_main}. We explore this result numerically in Supplementary Note~2, showing that this complementary way of computing $\alpha$ comes equally close to the value observed in the loops simulations. 

\subsection*{Exact results in the Kraichnan model}

To demonstrate the robustness of our results beyond Navier-Stokes turbulence, we consider the exactly solvable Kraichnan model~\cite{kraichnan_pof_1968}. 
The Kraichnan model of turbulence replaces the advecting velocity with a spatially correlated Gaussian random field, white in time, which mimics turbulence. While we do not expect the predictions of the curvature PDF power law from the Kraichnan model to be in quantitative agreement with our DNS results, it serves as a test case in which our approach can be compared rigorously against exact independent Fokker-Planck calculations.

In this setting, all of our argumentation about fold formation and its statistical implications can be made exact.  First,
the Cram\'er function takes the parabolic form~\cite{balkovsky_pre_1999}
\begin{align}
    S(\rho_p) &= \frac{(\rho_p - \lambda_p)^2}{2D_p},
\end{align}
with $\lambda_p = 3Q$, $D_p = 14Q$ and $Q$ a constant related to the energy spectrum quantifying fluctuations of the velocity gradient (see Methods). $\lambda_p$ and $D_p/t$ are the mean and variance of the Gaussian distribution of $\rho_p$ that can be computed from the known multivariate Gaussian distribution of the $\rho_i$~\cite{balkovsky_pre_1999}. Now, the integral~\eqref{eq:ld_integral} can be performed exactly, yielding a power law ${\kappa_p}^{-1-\alpha}$ with
\begin{align}
    \alpha = - \frac{\lambda_p}{D_p} + \sqrt{\frac{\lambda_p^2}{D_p^2} + \frac{2\beta}{D_p}}.
\end{align}
The growth rate of the mean length of line elements in the Kraichnan model is $\beta = 4Q$, determined by $e^{\beta t} \sim \langle e^{\rho_1(t) t}\rangle$. This evaluates to $\alpha = 4/7$, a curvature peak PDF power law $-11/7$ and a curvature PDF power law $-18/7 \approx -2.571$. Although this is very close to the exponent $-2.622 \pm 0.002$ that we find in Navier-Stokes turbulence, we believe that our measurements are precise enough to conclude that the exponents are in fact different and that their closeness is coincidental. 

Importantly, this result based on our picture of curvature growth due to fold formation is consistent with an independent, complementary approach facilitated by the rapidly fluctuating  velocity field. Using It\^{o} calculus, one can obtain an exact Fokker-Planck equation for the curvature distribution (see Methods) and study its steady state.  The equation takes the form
\begin{equation} \label{eq:fokker_planck}
  \partial_t f = -\partial_{\kappa} \left(-18Q\kappa f - 7Q\kappa^2 \partial_{\kappa}f + \frac{9P}{\kappa} f - 9P \partial_{\kappa} f\right),
\end{equation}
and features the stationary solution
\begin{equation} \label{eq:fpe_stationary_solution}
  f(\kappa) = \frac{1}{\mathcal{Z}} \kappa\left(9P + 7Q\kappa^2\right)^{-25/14},
\end{equation}
where $P$ is a constant quantifying fluctuations of second-order derivatives of velocity (see Methods) and $\mathcal{Z}$ is the normalization constant. This exact solution transitions between a $\kappa^1$ power law in the small-curvature regime and a $\kappa^{-18/7}$ power law in the large-curvature regime. Hence our framework based on the dynamical evolution of curvature peak statistics and It\^{o} calculus yield exactly the same large-curvature exponent. The shape of the PDF is also in qualitative agreement with our numerical observations in Navier-Stokes turbulence, see Fig.~\ref{fig:dns_curvature_pdf}b. A numerical analysis of the Kraichnan case can be found in Supplementary Note~7. Analogous computations~\cite{schekochihin_pre_2001} have been done for the curvature PDF of magnetic field lines in the context of the turbulent dynamo problem without compensating for arc length.

We remark in passing that it would be interesting to study material line curvature statistics in the compressible $d$-dimensional Kraichnan model \cite{schekochihin_pre_2001} also from the complementary perspective of fold formation. There, the compressibility can be parameterized by an index $\wp$ and Lyapunov exponents can be explicitly computed (see \S 2.4 of ref.~\cite{falkovich_2008}). The chaotic phase characterized by positive leading Lyapunov exponent $\lambda_1>0$ occurs when $\wp< d/4$. In this regime, one can vary $\lambda_p = \lambda_1 -2\lambda_2$ and analytically study its effect on curvature statistics. As such, the compressibility can be used to precisely control the curvature statistics.

\section*{\pdfbookmark[1]{Discussion}{Discussion}Discussion}
We investigated the curvature statistics of material loops in fully developed turbulence to characterize their statistical geometry. We find that the curvature PDF rapidly converges to a stationary distribution and establish a theory of curvature peaks forming along the loop to explain the power law in its high-curvature regime. Using the connection between curvature peak dynamics and finite-time Lyapunov exponents, we are able to theoretically link the power-law exponent to FTLE large-deviations statistics. In Navier-Stokes turbulence, we find our theory to be in very good agreement with direct numerical simulations. In the Kraichnan model, our theoretical prediction agrees precisely with exact analytical calculations.

An important issue concerns how the results presented here depend on the Reynolds number. In Supplementary Note~1, we provide numerical evidence that moderate variations of the Reynolds number lead qualitatively to the same picture with only very slight quantitative changes in the power-law exponents. When nondimensionalized by the Kolmogorov length scale, the curvature PDFs for different Reynolds numbers collapse in very good approximation, consistent with the notion that turbulent stretching and folding is driven by the tentatively universal small-scale velocity gradients in turbulence.  In light of this, it seems plausible to us that the shape of the curvature distribution we observe is universal and will persist in the limit of large Reynolds number.

Our methods and theoretical predictions can be applied to a large class of chaotic flows and can thereby provide a new statistical-geometry perspective on the intricacies of their evolution. Since a host of processes are closely related to the transport of material lines, our results may help to shed light on such problems from biophysics, geophysics and astrophysics. For example, in polymer turbulence, the conformation tensor describing polymeric stresses is a materially transported quantity modified by (internal) restoration forces.  As such, our computational and theoretical techniques used to study ideal material transport in the form of material lines, suitably adapted to accommodate internal degrees of freedom, provide a framework to study fluid-polymer interaction.

Our work may also shed new light on classical questions in magnetohydrodynamic (MHD) turbulence and, in particular, the dynamo problem. For example, curvature PDFs of magnetic field lines in MHD have been observed to form power-law tails in the kinematic stage~\cite{schekochihin_pre_2001}. It would be very interesting to study how this is related to the formation of folds in the magnetic field and how these folds behave in the non-linear stage of the turbulent dynamo. Furthermore, it is well known that flux cancellations in turbulent magnetic dynamos occur in part due to the folding/bundling of magnetic field lines \cite{childress1995stretch,ott_pop_1998}. In fact, our simulations indicate that tightly wound bundles along the loop are in close correspondence with curvature peaks (see Supplementary Note~3).  Thus the statistical attributes (generation and growth rates) of the peaks predicted here may be indicative of the genericity and intensity of configurations that can stifle dynamo growth.  It is also known that magnetic helicity -- a measurement of the linkage,
twist and writhe of magnetic loops -- has a profound effect on the growth rates for the dynamo \cite{boozer1993magnetic}.  The tools developed here can be used to study field lines in MHD in the highly conductive regime.  Conditioning on the level of magnetic helicity, they could thus offer a new geometric perspective on the role that magnetic helicity plays in  dynamo action.

Finally, we remark that it would be of great interest to generalize our framework to accommodate higher-dimensional structures, such as material surfaces. A material surface can be understood as a continuous family of material lines. We therefore expect it to form folds extending as one-dimensional structures across the surface. This could then be applied to study interfacial problems such as the dispersion of algae blooms or oil spills in the ocean, where the description of the boundary's geometry is of crucial importance for prediction.

\section*{\pdfbookmark[1]{Methods}{Methods}Methods}
\subsection*{Navier-Stokes simulations for loop tracking}
\label{sec:loopsim}

For the direct numerical simulations (DNS), we use our code TurTLE~\cite{turtle}. It implements a pseudo-spectral solver for the Navier-Stokes equation in the vorticity formulation with a third-order Runge-Kutta method for time stepping and a high-order Fourier smoothing~\cite{hou_jcp_2007} to reduce aliasing errors. The flow is forced on the large scales by maintaining a fixed energy injection rate in a discrete band of Fourier modes at small wavenumbers $k \in [1.0, 2.0]$ (DNS units). The simulations presented here were computed on $1024^3$ grid points with a small-scale resolution $k_{\max} \eta \approx 2.9$, where $k_{\max}$ is the maximum resolved wavenumber. Using the same initial background flow, we conducted two separate simulations with different sets of Lagrangian tracers.

The first simulation contains $10^3$ initially circular loops of diameter ${\sim} 10\eta$ with random position and orientation. Each sample point of the loops is treated as a Lagrangian tracer particle. Over time, the strongly heterogeneous line stretching necessitates an adaptive refinement of the loops~\cite{kida_pof_2002, goto_jfm_2007}. Using fifth-order B-spline interpolation~\cite{scipy_2020}, we determine the arc length between adjacent sample points in time intervals of $0.16\tau_\eta$. Whenever their distance surpasses $0.1\eta$, we insert new sample points along the smooth spline curves, which ensures that derivatives of the curves up to fourth order and hence their curvature are well-defined. In order to better resolve high-curvature regions, we additionally require that the distance between sample points does not surpass $1/(6\kappa)$. This significantly improves the resolution of the large-curvature tail of the curvature PDF. Due to the refinement, the initial total number of sample points across all loops -- about $3\times 10^5$ -- increases to about $1.5 \times 10^8$ sample points at $29\tau_\eta$. The adaptive insertion of particles prohibits the direct use of multi-step methods for particle time stepping. For this simulation, we therefore resort to first-order Euler time stepping of particle trajectories. They are coupled with spline interpolation of the field with continuous derivatives up to and including third order computed over a kernel of $12^3$ grid points (as detailed in ref.~\cite{lalescu_jcp_2010}). We verify our determination of the curvature distribution for different temporal and spatial resolutions of the loops in Supplementary Note~4.

While the statistical geometry of any type of material line could be equally well studied, we focus here on material loops due to their important role in fluid dynamics. For example, the velocity circulation along any material loop is invariant in inviscid incompressible fluid motion -- a fact known as the Kelvin theorem. While this invariance breaks down in the presence of any non-ideal effect such as viscosity, properties of material loops at high Reynolds number -- a regime in which the flow is nearly inviscid -- may shed light on a variety of features of fully developed turbulence such as anomalous dissipation and spatio-temporal intermittency \cite{eyink2006turbulent}. Material loops also arise naturally in the context of astrophysics where they approximately describe the motion of closed field lines of a magnetic field at high magnetic Reynolds numbers in a stellar or planetary system.

\subsection*{Computation of finite-time Lyapunov exponents and the Cram\'er function}
\label{sec:ftlesim}

The second simulation contains $10^8$ uniformly distributed Lagrangian tracers. Along with their trajectories, we integrate the deformation tensor~\eqref{eq:deformation_tensor}. Time stepping is performed using the Heun method coupled with spline interpolation of the field with continuous derivatives up to and including second order computed over a kernel of $8^3$ grid points. In order to ensure numerical stability, we perform a QR-decomposition of the deformation tensor~\cite{pikovsky_politi_2016} after each time step and store principal axes and logarithmically scaled stretching factors separately. While in theory the FTLEs are defined by the singular value decomposition, we here use the logarithmic stretching factors obtained from the QR-decomposition as proxies (as done in Refs.~\cite{bec_pof_2006,bagheri2012statistics,johnson_pof_2015}). In certain regimes, their large-deviations statistics may differ~\cite{johnson_pof_2015}. However, in Supplementary Note~8, we show that our theoretical argument can also be made for the proxies. We therefore expect no differing results in the two cases. We then determine finite-time Cram\'er functions $S(\rho_p; t)$ from the FTLE histograms $f(\rho_p; t)$ as~\cite{johnson_pof_2015}
\begin{align}
    S(\rho_p; t)= -\log(f(\rho_p; t))/t,
\end{align}
which converge to the actual Cram\'er function over time. Given that the Cram\'er function is known to take its minimum at $S(\lambda_p) = 0$, where $\lambda_p = \lim_{t\to\infty} \rho_p(t)$, we may accelerate convergence by vertically shifting the finite-time Cram\'er functions such that their minimum is zero, as done similarly in ref.~\cite{johnson_pof_2015}. The resulting functions are used as input for Fig.~\ref{fig:cramer_minimum}.

We determine least-square fits of the finite-time Cram\'er functions using a Batchelor interpolation between two power laws (corresponding to stretched exponentials for the FTLE PDF),
\begin{align}
    S(\lambda_p(t)+x/\tau_\eta; t) &= \frac{a x^2}{(b + x^2)^{c}},
\end{align}
where $\lambda_p(t)$ is the position of the minimum of $S(\rho_p; t)$, and $a$, $b$ and $c$ are fitting parameters. In order to obtain fits with reasonable accuracy, we restrict the fitting range to the interval of interest $[\lambda_p(t), 1/\tau_\eta]$. If the finite-time Cram\'er functions take infinite values in this range, then we further restrict the fitting range to their finite values. In order to obtain the error bars in Fig.~\ref{fig:cramer_minimum}, we vary the fitting parameters within their standard error interval and take the minimum and maximum of the resulting functions. Taking the minimum of the best fits and of their error envelopes, we obtain the time series of minima in the inset of Fig.~\ref{fig:cramer_minimum}. If a fit takes its minimum at the last value of the fitting range, then this value is omitted. 

In order to extrapolate the minimum towards $t\to\infty$, we determine the best fit of the minima time series $m(t)$ weighted by the errors using an algebraic decay,
\begin{align} \label{eq:decay_func}
    m(t) &= A + \left(\frac{B}{t}\right)^{C},
\end{align}
where $A$, $B$, and $C$ are fitting parameters. In order to robustly capture the asymptotic decay using this simple fit function, we leave out an initial transient regime of data points for the fit. We choose $t \geq t_\mathrm{min} \approx 6.9\tau_\eta$, where the weighted mean squared error of the fit reaches a plateau, i.e.\ the point at which the fit improvement from removing more data points diminishes (for more details, see Supplementary Note~1). The parameters are estimated as $A=0.54 \pm 0.11$, $B=(0.19 \pm 0.15)\tau_\eta$ and $C=0.36 \pm 0.18$.

Note that the overall fitting procedure is very delicate and different choices may lead to different results. The present analysis is our best effort to systematically compute the limiting value of the minima.

\subsection*{Peak curvature dynamics of a parabola}
\label{sec:parabola}
Here, we determine the evolution of the peak curvature of a fold modeled by a parabola,
\begin{align} \label{eq:parabolic_sling}
    \mathbf{L}(\phi, t) &= \mathbf{L}(\phi_0, t) + (\phi-\phi_0) \mathbf{l}(t) \Delta s \\
    &\quad + \kappa_p(0) \frac{(\phi - \phi_0)^2}{2} \mathbf{k}(t) \Delta s^2 , \nonumber
\end{align}
where $\phi_0$ is the initial peak position, $\kappa_p(0)$ is its initial peak curvature, $\mathbf{l}$ and $\mathbf{k}$ are two initially orthonormal vectors, and $\Delta s$ is the arc length per angle of the initial parameterization at $\phi_0$. In a sufficiently small range of $\phi$ around $\phi_0$, the velocity field can be linearized. Then the parabolic shape is preserved and the dynamics of $\mathbf{l}$ and $\mathbf{k}$ in the Lagrangian frame is determined by the velocity gradient,
\begin{align}\nonumber
    \dod{\mathbf{l}}{t} &= \mathbf{l}\cdot\nabla\mathbf{u}(\mathbf{L}(\phi_0, t), t)\quad\text{and}\\
    \dod{\mathbf{k}}{t} &= \mathbf{k}\cdot\nabla\mathbf{u}(\mathbf{L}(\phi_0, t), t).
\end{align}
By \eqref{eq:curv_def}, the curvature of the fold is given by
\begin{align}
    \kappa(\phi, t) &= \kappa_p(0) \frac{\left|\mathbf{k}(t)\times\mathbf{l}(t)\right|}{\left|\mathbf{l}(t) + \Delta s(\phi-\phi_0) \kappa_p(0)\mathbf{k}(t)\right|^3} \\
    &= \kappa_p(0) \frac{\left(\left|\mathbf{k}(t)\right|^2\left|\mathbf{l}(t)\right|^2 - (\mathbf{k}(t)\cdot\mathbf{l}(t))^2\right)^{1/2}}{\left|\mathbf{l}(t) + \Delta s(\phi-\phi_0) \kappa_p(0)\mathbf{k}(t)\right|^3} .
\end{align}
Over time, $\mathbf{l}(t)$ and $\mathbf{k}(t)$ cease to be orthogonal and the curvature peak position is shifted. Minimizing the denominator yields the new peak position
\begin{align}
    \phi_p(t) &= \phi_0-\frac{\mathbf{k}(t)\cdot\mathbf{l}(t)}{\Delta s \left|\mathbf{k}(t)\right|^2\kappa_p(0)}.
\end{align}
The new peak curvature is therefore given by
\begin{align}
    \kappa_p(t) := \kappa(\phi_p(t), t) = \frac{\left|\mathbf{k}(t)\right|^3}{\left|\mathbf{k}(t)\right|^2\left|\mathbf{l}(t)\right|^2 - (\mathbf{k}(t)\cdot\mathbf{l}(t))^2} \kappa_p(0). \label{eq:parabola_peak_curv}
\end{align}
Since $\mathbf{l}$ and $\mathbf{k}$ behave like passive vectors, their dynamics can be described by the deformation tensor
\begin{align}
    F_{ij}(t) &= \dpd{X_i (\mathbf{L}(\phi_0,0), t)}{x_j},
\end{align}
where $\mathbf{X}(\mathbf{x}, t)$ is the Lagrangian map. The singular value decomposition of $F$,
\begin{align} \label{eq:deformation_tensor_svd}
    F(t) &= U(t) \Lambda(t) V^T(t),
\end{align}
defines the orthonormal bases $(\mathbf{u}_j(t))_i = U_{ij}(t)$ and $(\mathbf{v}_j(t))_i  = V_{ij}(t)$ and the finite-time Lyapunov exponents $\rho_i(t)$ by $\Lambda_{ii}=e^{\rho_i(t)t}$ where $\Lambda$ is diagonal. Expanding $\mathbf{l}(0)$ and $\mathbf{k}(0)$ in the $\mathbf{v}_j$-coordinate system yields
\begin{align} \nonumber
    \mathbf{l}(0) &= \sum_i a_i(t) \mathbf{v}_i(t)\quad\text{and} \\
    \mathbf{k}(0) &= \sum_i b_i(t) \mathbf{v}_i(t).
\end{align}
Observing that $F(t)\mathbf{l}(0) = \mathbf{l}(t)$ and $F(t)\mathbf{k}(0) = \mathbf{k}(t)$, and applying the deformation tensor to the previous equations, we get
\begin{align} \nonumber
    \mathbf{l}(t) &= \sum_i a_i(t) e^{\rho_i(t)t} \mathbf{u}_i(t)\quad\text{and} \\
    \mathbf{k}(t) &= \sum_i b_i(t) e^{\rho_i(t)t} \mathbf{u}_i(t).
\end{align}
Inserting these expansions into \eqref{eq:parabola_peak_curv} yields
\begin{align} \label{eq:peak_curv_exact}
  \kappa_p(t) = \frac{\left(\sum_i b_i^2 e^{2\rho_i t}\right)^{3/2}}{\sum_{i\neq j} a_j b_i (a_j b_i - a_i b_j)e^{(2\rho_i + 2\rho_j)t}} \kappa_p(0).
\end{align}
As long as the infinite-time Lyapunov exponents (the $t\to\infty$-limits of the FTLEs) are distinct from each other, we will have $e^{\rho_1(t)t} \gg e^{\rho_2(t)t} \gg e^{\rho_3(t)t}$ for large $t$. Assuming furthermore that the random coefficients in \eqref{eq:peak_curv_exact} are non-zero, we can drop those terms with slower exponential growth:
\begin{align}
  \kappa_p(t) \stackrel{t \gg 0}{\approx} \frac{|b_1(t)|^3}{(a_2(t) b_1(t) -a_1(t) b_2(t))^2} \kappa_p(0) e^{[\rho_1(t) - 2\rho_2(t)]t}.
\end{align}
While the FTLEs are known to converge slowly, $V(t)$ and thus $a_i(t)$ and $b_i(t)$ converge exponentially fast~\cite{thiffeault_pd_2002, goldhirsch_physd_1987}. We therefore have (cf.\ ref.~\cite{leonard_jfm_2009})
\begin{align} \label{eq:peak_curv_growth}
    \kappa_p(t) \stackrel{t \gg 0}{\approx} \widetilde{\kappa}_0 e^{[\rho_1(t) - 2\rho_2(t)]t},
\end{align}
for some effective initial peak curvature
\begin{align}
    \widetilde{\kappa}_0 = \lim_{t\to\infty} \frac{|b_1(t)|^3}{(a_2(t) b_1(t) -a_1(t) b_2(t))^2} \kappa_p(0),
\end{align}
which may differ from the actual initial peak curvature $\kappa_p(0)$ depending on the relative orientation of the initial parabola and the converged basis vectors $\lim_{t\to\infty} \mathbf{v}_j(t)$.

\subsection*{Extracting the power law by the method of steepest descent}
\label{sec:steepestdescent}
In order to extract the asymptotic regime of the integral~\eqref{eq:ld_integral}, we substitute the integration variable
\begin{align}
\rho_p = \frac{1}{\tau}\log\left(\frac{\kappa_p}{\kappa_0}\right),
\end{align}
which yields
\begin{align}
    f(\kappa_p) \sim \frac{1}{\kappa_p} \int_0^\infty \dif \rho_p \frac{N(\log(\kappa_p/\kappa_0)/\rho_p)}{\rho_p} e^{-\log\left(\frac{\kappa_p}{\kappa_0}\right) \frac{(\beta + S(\rho_p))}{\rho_p}}.
\end{align}
We now explore the regime where $\log(\kappa_p/\kappa_0)$ becomes large. Assuming that the normalization function $N(\tau)$ is algebraic, the scaling of the integral with $\kappa_p$ is dominated by the exponential, and in particular by the part that has the slowest decay. To first order, we therefore have~\cite[Chapter 9, Theorem 2.1]{olver_1997}
\begin{align}\nonumber
    f(\kappa_p) &\sim \frac{1}{\kappa_p} \exp\left(-\log\left(\frac{\kappa_p}{\kappa_0}\right) \min_{\rho_p}\left[\tfrac{1}{\rho_p}( \beta + S(\rho_p))\right]\right) \\
    &\propto \kappa_p^{-1-\alpha},
\end{align}
with
\begin{align}
    \alpha = \min_{\rho_p} \left[\tfrac{1}{\rho_p}( \beta + S(\rho_p))\right].
\end{align}

\subsection*{Relating our results to generalized Lyapunov exponents}
\label{sec:gle_derivation_main}
Let us define a generalized Lyapunov exponent of curvature peaks by
\begin{align} \label{eq:gle_def}
    L_p(q) &= \lim_{t\to\infty} \frac{1}{t} \log \left\langle \exp(q\rho_p(t)t)\right\rangle.
\end{align}
It differs from the usual definition of generalized Lyapunov exponents only by the fact that we have replaced the standard FTLE by our curvature peak FTLE $\rho_p(t)=\rho_1(t) - 2\rho_2(t)$. It is related to the Cram\'er function by a Legendre transform~\cite{johnson_pof_2015},
\begin{align} \label{eq:gle_legendre}
    L_p(q) &= \sup_{\rho_p} \left[q\rho_p - S(\rho_p)\right].
\end{align}
This strongly resembles our steepest-descent formula established in the main text (cf.~\eqref{alphaeqn}),
\begin{align}\label{alpheqn}
    \alpha &= \min_{\rho_p} \left[\tfrac{1}{\rho_p}( \beta + S(\rho_p))\right]
\end{align}
where, recall, $\beta$ is identified with line growth quantified by the first FTLE (see Fig.~\ref{fig:peak_number} and subsequent discussion)
\begin{align}\label{betaeqn}
    \beta = \lim_{t\to\infty} \frac{1}{t} \log \left\langle \exp(\rho_1(t)t)\right\rangle.
\end{align}
We claim $  L_p(\alpha) =\beta$.  If so, then equating \eqref{eq:gle_def} evaluated at $\alpha$ with $\beta$ given by \eqref{betaeqn}, we find
\begin{align}\label{eq:stretch_fold_balance_explicit}
    \lim_{t\to\infty} \frac{1}{t} \log \left\langle \exp(\alpha\rho_p(t)t)\right\rangle &= \lim_{t\to\infty} \frac{1}{t} \log \left\langle \exp(\rho_1(t)t)\right\rangle,
\end{align}
which we write in short form as~\eqref{eq:stretch_fold_balance_main}.

To verify that $L_p(\alpha) =\beta$, we insert $\alpha$ into \eqref{eq:gle_legendre} to find
\begin{align}\label{Lalph}
    L_p(\alpha) = \sup_{\rho_p} \left[\alpha\rho_p - S(\rho_p)\right].
\end{align}
Assuming that $S(\rho_p)$ is differentiable and strictly convex, the supremum in \eqref{Lalph} occurs at a unique value $ \rho_p^*$.  Moreover, somewhat remarkably, we will show that this value coincides with that at which the minimum of \eqref{alpheqn} occurs. Once established, this gives the claimed result upon substitution of  
$
    \alpha = \tfrac{1}{ \rho_p^*}( \beta + S( \rho_p^*))
$
into $ L_p(\alpha) =\alpha\rho_p^* - S(\rho_p^*)$.

To see that the extrema in  \eqref{Lalph} and \eqref{alpheqn} occur at the same point $ \rho_p^*$, we note that under our assumptions \eqref{Lalph} is minimized at the $\rho= \rho^*$ for which
\begin{align}
    0 &= \dod{}{\rho} \left[\alpha\rho - S(\rho)\right]\Big|_{\rho=\rho^*}= \alpha - S'(\rho^*). \label{e1}
\end{align}
Uniqueness follows from our assumption that $S'(\rho)$ is an invertible function of $\rho$. On the other hand, the minimum in \eqref{alpheqn}  occurs for $\rho=\rho^{**}$ satisfying 
\begin{align}\nonumber
    0 &= \dod{}{\rho} \left[\tfrac{1}{\rho}(\beta + S(\rho))\right] \Big|_{\rho=\rho^{**}}\\ \nonumber
    &= -\frac{1}{\rho^{**}}\left(\tfrac{1}{ \rho^{**}}( \beta + S( \rho^{**}))  - S'(\rho^{**})\right) \\
   &= -\frac{1}{\rho^{**}}\left(\alpha - S'(\rho^{**})\right) \label{e2}
\end{align}
where we have inserted the expression for $\alpha$ in terms of the minimizing argument $\rho^{**}$ given by  \eqref{alpheqn}.  It is clear from comparing \eqref{e1} and \eqref{e2} that the extrema are realized at the same value $\rho^{*}=\rho^{**}=:\rho^{*}_p$. This concludes the proof.

\subsection*{Fokker-Planck equation of curvature in the Kraichnan model}
\label{sec:fpe_basic_derivation}
In the Kraichnan model, the velocity field $\mathbf{u}(\mathbf{x}, t)$ is Gaussian with correlation tensor
\begin{align} \label{eq:kraichnan_correlation}
    \left\langle u_i(\mathbf{x}, t) u_j(\mathbf{x}', t') \right\rangle
    &= \delta(t-t') R_{ij}(\mathbf{x} - \mathbf{x}'),
\end{align}
where $R_{ij}(\mathbf{r})$ denotes the spatial part of the correlation tensor.

Equivalent to \eqref{eq:curvPDF}, the curvature PDF weighted by arc length can be defined by
\begin{equation} \label{eq:curvature_pdf}
  f(\kappa; t) = \frac{\left\langle |\partial_{\phi} \mathbf{L}| \delta(\kappa - \tilde{\kappa}(\phi, t)) \right\rangle}{\left\langle |\partial_{\phi} \mathbf{L}| \right\rangle},
\end{equation}
where we distinguish between the realization $\tilde{\kappa}$ and the sample-space variable $\kappa$. Angular brackets $\langle \cdot \rangle$ denote an average along $\phi$ and over realizations of the velocity field.

In order to derive the Fokker-Planck equation of curvature, we take the time derivative of \eqref{eq:curvature_pdf}, which yields
\begin{align} \label{eq:fpe_arclength_terms}
    \partial_t f(\kappa; t) &= \frac{\left\langle \delta(\kappa - \tilde{\kappa}) \partial_t |\partial_{\phi} \mathbf{L}| \right\rangle}{\left\langle |\partial_{\phi} \mathbf{L}| \right\rangle} - f(\kappa; t) \frac{\partial_t \left\langle |\partial_{\phi} \mathbf{L}| \right\rangle}{\left\langle |\partial_{\phi} \mathbf{L}| \right\rangle} \\
    &\quad- \frac{1}{\left\langle |\partial_{\phi} \mathbf{L}| \right\rangle}
    \partial_{\kappa}\left\langle \delta(\kappa - \tilde{\kappa}) |\partial_{\phi} \mathbf{L}| \partial_t \tilde{\kappa} \right\rangle.\nonumber
\end{align}
The averages can be evaluated using the Gaussian integration by parts formula~\cite{furutsu_1964, donsker_mate_1967, novikov_sjp_1965} and the evolution equations~\cite{drummond_jfm_1991_curv}
\begin{align}
    \partial_t \partial_{\phi}\mathbf{L} &= ((\partial_{\phi}\mathbf{L})\cdot \nabla)\mathbf{u}, \label{eq:gradient_evolution}\\  \label{eq:tangent_evolution}
    \partial_t \hat{\mathbf{t}} &= (\hat{\mathbf{t}}\cdot\nabla)\mathbf{u} - \hat{\mathbf{t}}(\hat{\mathbf{t}}\cdot(\hat{\mathbf{t}}\cdot\nabla)\mathbf{u}),\\ \nonumber
    \partial_t \hat{\mathbf{n}} &= \hat{\mathbf{b}}(\hat{\mathbf{b}}\cdot(\hat{\mathbf{n}}\cdot\nabla)\mathbf{u}) - \hat{\mathbf{t}}(\hat{\mathbf{n}}\cdot(\hat{\mathbf{t}}\cdot\nabla)\mathbf{u})  \label{eq:normal_evolution}\\
    &\quad\ \ +\frac{1}{\tilde{\kappa}} \hat{\mathbf{b}}(\hat{\mathbf{b}}\cdot(\hat{\mathbf{t}}\cdot\nabla)^2\mathbf{u}),\\ \nonumber
      \partial_t \tilde{\kappa} &= \tilde{\kappa}\left(\hat{\mathbf{n}}\cdot(\hat{\mathbf{n}}\cdot\nabla)\mathbf{u} - 2\hat{\mathbf{t}}\cdot(\hat{\mathbf{t}}\cdot\nabla)\mathbf{u}\right)\\
&\quad \ \ + \hat{\mathbf{n}}\cdot(\hat{\mathbf{t}}\cdot\nabla)^2 \mathbf{u}.
\label{eq:curvature_evolution}
\end{align}
Here, $\hat{\mathbf{t}}$, $\hat{\mathbf{n}}$ and $\hat{\mathbf{b}}$ denote the tangent, normal and binormal vector of the Frenet-Serret frame, respectively. As shown in Supplementary Note~5, the evolution equations derive from the definitions of the various quantities combined with the tracer equation~\eqref{eq:tracer_eq}. All quantities are evaluated along the same Lagrangian trajectory.

In order to simplify the resulting expressions, we need to further restrict the spatial correlation structure of the model. Isotropy and incompressibility determine the form of the even derivatives of the spatial correlation tensor $R_{ij}(\mathbf{r})$ at $\mathbf{0}$ (odd numbers of derivatives vanish) to be~\cite{pumir_prf_2017}
\begin{align} \label{eq:Q_def}
    -\partial_k \partial_l R_{ij}(\mathbf{0}) &= Q(4\delta_{ij}\delta_{kl} - \delta_{ik}\delta_{jl} - \delta_{il}\delta_{jk})
\end{align}
and~\cite{kearsley_jrnbs_1975}
\begin{align} \label{eq:P_def}
    \partial_k \partial_l \partial_m \partial_n R_{ij}(\mathbf{0}) &=
    P(6\delta_{ij}\delta_{kl}\delta_{mn}+6\delta_{ij}\delta_{km}\delta_{ln}\\
    &\quad+6\delta_{ij}\delta_{kn}\delta_{lm} - (\text{all others})), \nonumber
\end{align}
with $Q$ and $P$ scalar constants that depend on the exact form of $R_{ij}(\mathbf{r})$.
The last pair of brackets contains all twelve other permutations of Kronecker deltas. All terms arising from the Gaussian integration by parts formula can be evaluated using this result and the orthonormality of the Frenet-Serret frame. The resulting Fokker-Planck equation is~\eqref{eq:fokker_planck}. In Supplementary Note~6, we list results for all terms and exemplify computing one of them.

\section*{\pdfbookmark[1]{Data availability}{Data availability}Data availability}
The data that support the findings of this study are available from the corresponding author upon reasonable request.

\section*{\pdfbookmark[1]{Code availability}{Code availability}Code availability}
The simulation results have been generated with our code TurTLE~\cite{turtle}, which is available on \url{https://gitlab.mpcdf.mpg.de/TurTLE/turtle}. The loop refinement and post-processing codes are available from the corresponding author upon reasonable request.

\bibliography{../refs.bib}

\pdfbookmark[1]{Acknowledgments}{Acknowledgments}
\begin{acknowledgments}
We would like to acknowledge interesting and useful discussions with Maurizio Carbone. We thank Itzhak Fouxon, Perry Johnson, and Jean-Luc Thiffeault for comments on the manuscript. We thank B\'erenger Bramas for his implementation of the particle tracking framework used in our simulations. Computational resources from the Max Planck Computing and Data Facility and support by the Max Planck Society are gratefully acknowledged. The authors gratefully acknowledge the Gauss Centre for Supercomputing e.V. (www.gauss-centre.eu) for funding this project by providing computing time on the GCS Supercomputer SuperMUC-NG at Leibniz Supercomputing Centre (www.lrz.de). TD was partially supported by NSF grant DMS-2106233 and the Charles Simonyi Endowment at the Institute for Advanced Study. 3D visualizations have been created with Blender~\cite{blender_2020}.
\end{acknowledgments}

\section*{\pdfbookmark[1]{Author contributions}{Author contributions}Author contributions}
LB, TD and MW designed the study. LB carried out the numerical simulations and analysis. CL helped with code development. All authors analyzed the data and wrote the manuscript.

\section*{\pdfbookmark[1]{Competing interests}{Competing interests}Competing interests}
The authors declare no competing interests.

\newcommand*{\wc}{R}
\newcommand*\vdif[1][3]{\mathrm{d^#1}}
\newcommand*{\bv}{\hat{\mathbf{b}}}
\newcommand*{\bc}{\hat{b}}
\newcommand*{\tv}{\hat{\mathbf{t}}}
\newcommand*{\tc}{\hat{t}}
\newcommand*{\nv}{\hat{\mathbf{n}}}
\newcommand*{\nc}{\hat{n}}
\newcommand*{\xv}{\mathbf{x}}
\newcommand*{\zv}{\mathbf{z}}
\newcommand*{\uv}{\mathbf{u}}
\newcommand*{\lv}{\mathbf{l}}
\newcommand*{\kv}{\mathbf{k}}
\newcommand*{\grrate}{\beta}
\newcommand*{\ftle}{\rho}
\newcommand*{\ftlesamp}{\rho}
\newcommand*{\ksamp}{\kappa}
\newcommand*{\dfd}[2]{\frac{\delta #1}{\delta #2}}
\newcommand*{\loo}{\mathbf{L}}
\newcommand*{\angl}{\phi}
\renewcommand{\norm}[1]{\left|#1\right|}
\newcommand{\norms}[1]{|#1|}
\onecolumngrid
\newpage

\setcounter{page}{1}
{
\centering
\date{\today}
    {\pdfbookmark[-1]{Supplementary Material}{Supplementary Material}\large \textbf{Supplementary Material for \\``The statistical geometry of material loops in turbulence''}}

    \vspace{.5cm}
    Lukas Bentkamp, Theodore D. Drivas, Cristian C. Lalescu, and Michael Wilczek
    
    \vspace{.5cm}
    
}

\renewcommand{\figurename}{Supplementary Figure}
\renewcommand{\tablename}{Supplementary Table}
\renewcommand{\thesection}{\arabic{section}}
\setcounter{table}{0}
\renewcommand{\thetable}{S\arabic{table}}%
\setcounter{figure}{0}
\renewcommand{\thefigure}{S\arabic{figure}}%
\setcounter{equation}{0}
\def\theequation{S\arabic{equation}}
\setcounter{section}{0}
\titleformat{\section}{\centering\small\bfseries\uppercase}{Supplementary Note \thesection.}{1em}{}

\section{Curvature statistics and determination of power-law exponent at various Reynolds numbers}
\label{sec:set_of_sims}
\begin{figure*}[p]
    \centering
    \begin{minipage}{0.49\textwidth}
    \includegraphics[width=1.\textwidth]{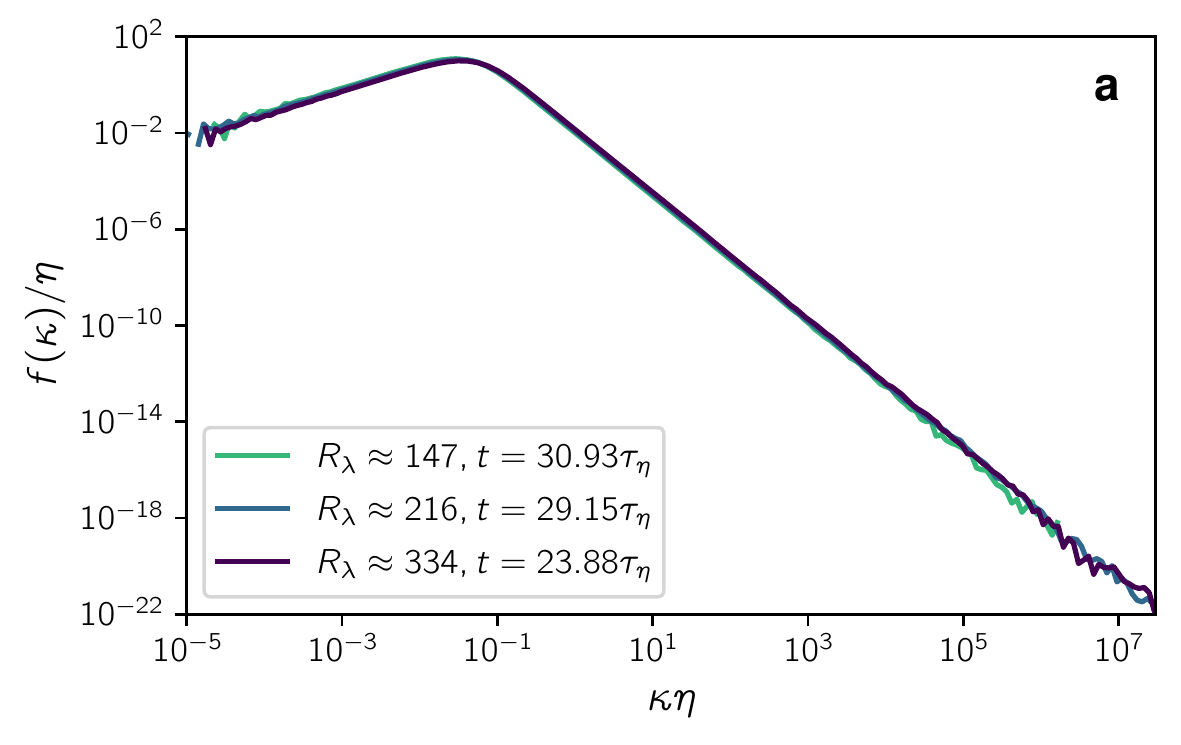}
    \end{minipage}
    \begin{minipage}{0.49\textwidth}
    \includegraphics[width=1.\textwidth]{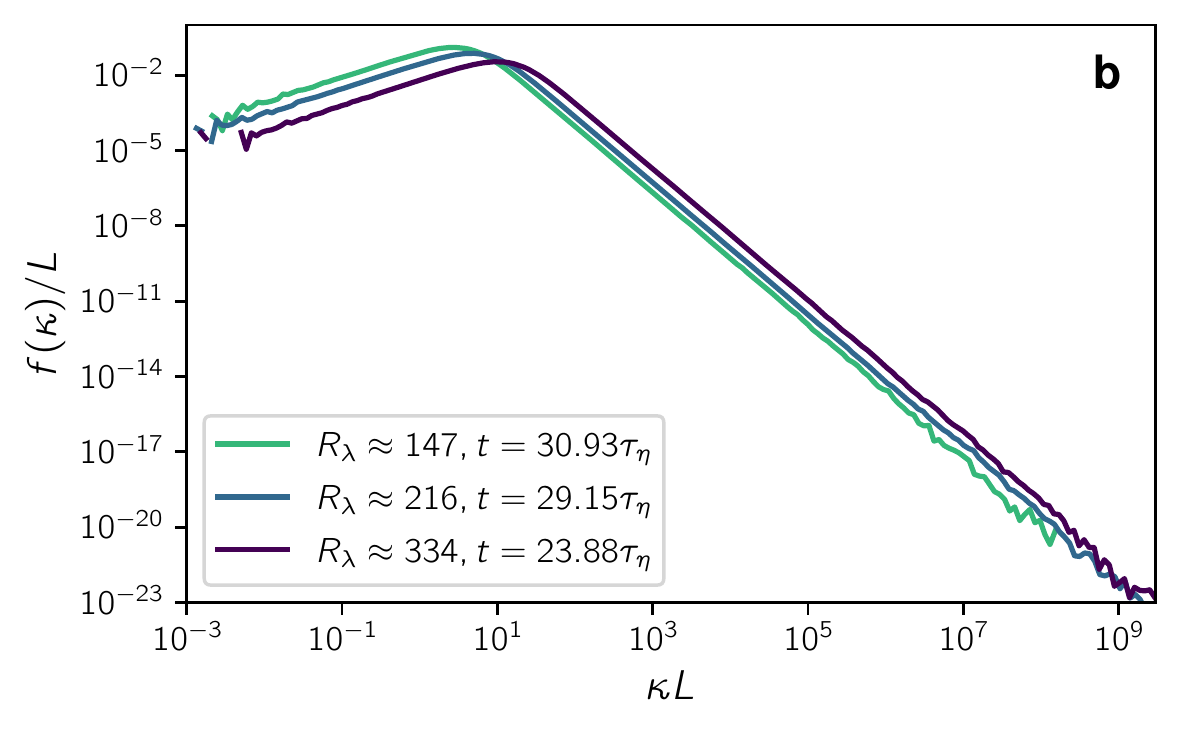}
    \end{minipage} \vspace{-0.2cm}
    \caption{\textbf{Comparison of curvature PDFs at the latest simulation times}, scaled by the Kolmogorov length (\textbf{a}) and the integral length (\textbf{b}). In Kolmogorov units, hardly any trend is visible whereas the peak of the PDF shifts toward larger curvature values with increasing Reynolds number when nondimensionalized in integral units.\label{fig:curv_pdf_by_Re}}
\end{figure*}

\begin{figure*}[p]
    \centering
    \begin{minipage}{0.49\textwidth}
    \includegraphics[width=1.\textwidth]{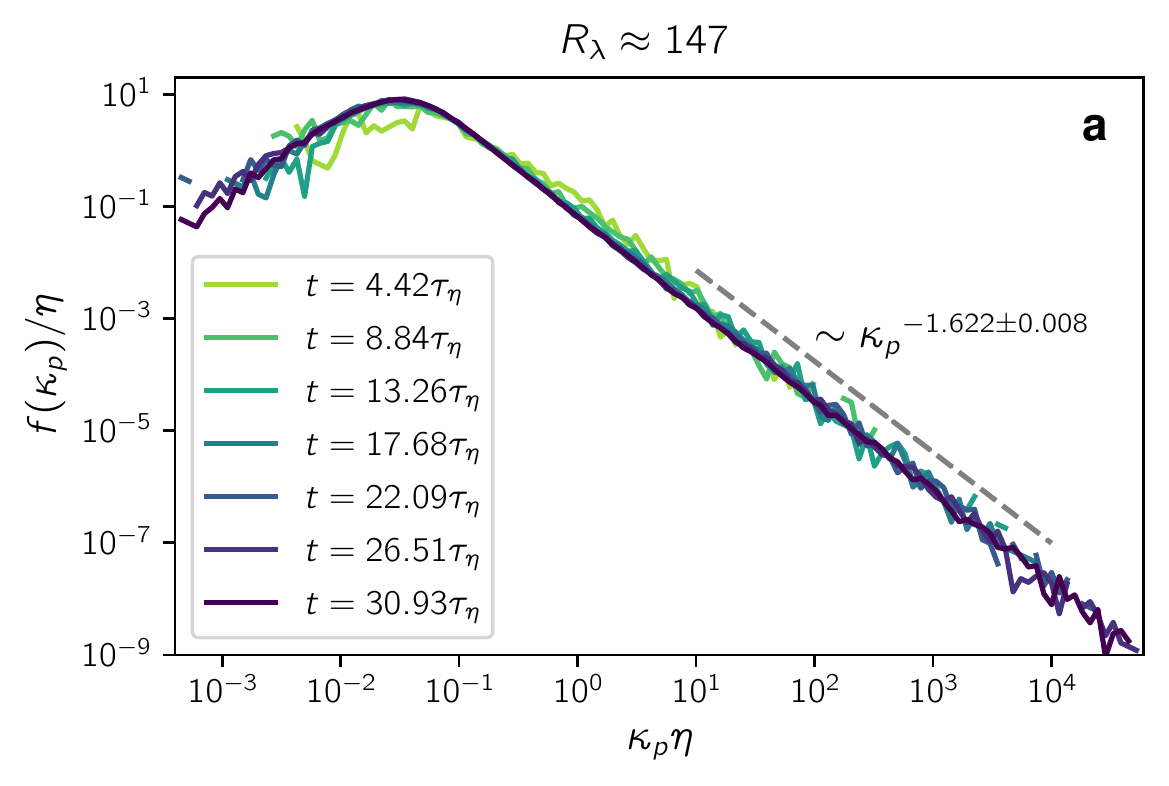}
    \end{minipage}
    \begin{minipage}{0.49\textwidth}
    \includegraphics[width=1.\textwidth]{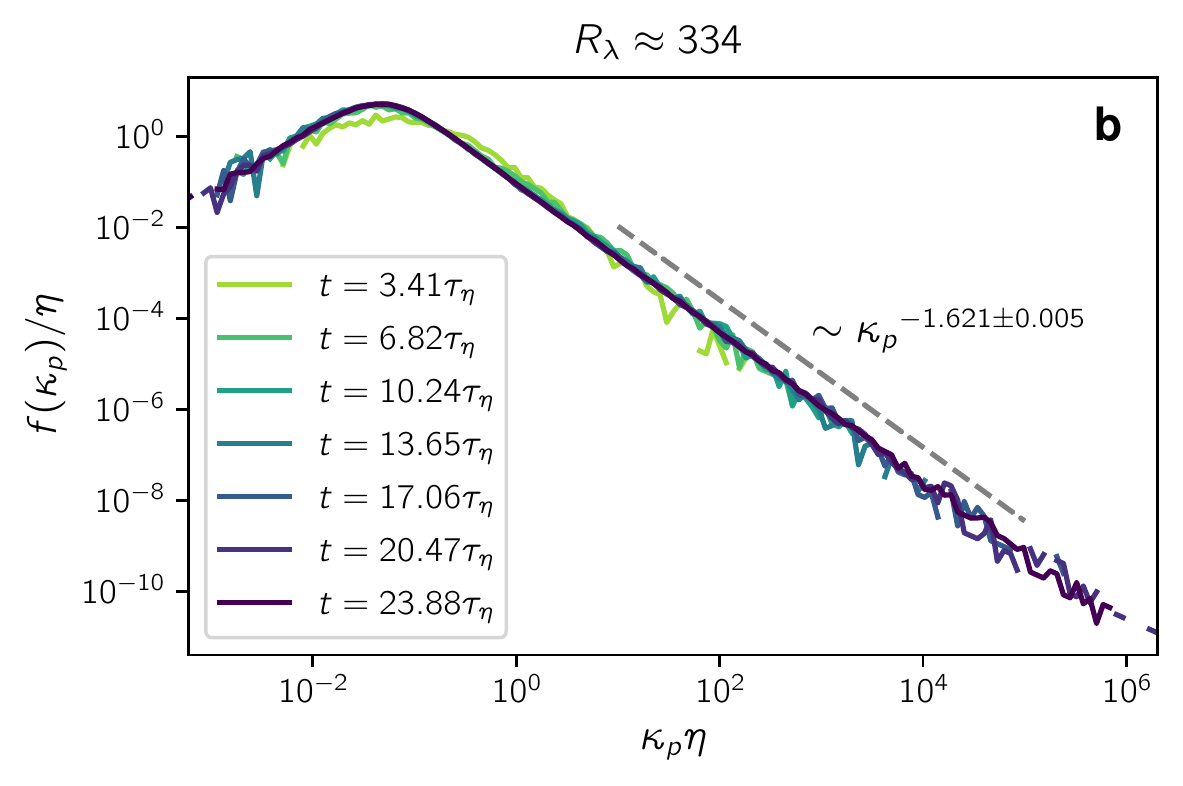}
    \end{minipage}\vspace{-0.2cm}
    \caption{\textbf{Curvature peak statistics} in the low-Reynolds-number (\textbf{a}) and the high-Reynolds-number (\textbf{b}) supplementary simulation look the same as in the main simulation. Like for the curvature PDF, the large-curvature power-law exponents, determined by best fits, are almost identical across all simulations.~\label{fig:supp_curv_peak_pdf}}
\end{figure*}

We carried out our analysis for two additional direct numerical simulations (DNS) of the Navier-Stokes equation, one at Taylor-scale Reynolds number $R_\lambda \approx 147$ and one at $R_\lambda \approx 334$. The simulation details are summarized in Table~\ref{tab:DNS_parameters}. The curvature statistics of an ensemble of material loops were determined in each simulation as described in Methods. The curvature distributions of the supplementary simulations are shown in Fig.~\ref{fig:supp_curv_pdf}. Remarkably, the large-curvature power-law exponents, determined by best fits, are almost the same across the simulations, indicating that there may be no significant Reynolds-number dependence in this range. This becomes even more apparent in Fig.~\ref{fig:curv_pdf_by_Re}a, where the curvature distributions are shown to almost perfectly collapse for the latest point in time in the three simulations when nondimensionalized by $\eta$. As a function of the integral length, the distributions shift toward larger $\kappa$ with increasing Reynolds number (Fig.~\ref{fig:curv_pdf_by_Re}b). Also the curvature peak PDFs in Figure~\ref{fig:supp_curv_peak_pdf} show no measurable Reynolds-number dependence. Consistent with our theory, their high-curvature exponent differs from the curvature PDF exponent by 1, mostly within error bars.
    \begin{table}[b!]
        \centering
\begin{tabular}{c|cccccccc|ccccccc}
    \multicolumn{1}{c}{} & \multicolumn{8}{c}{Loops simulations} & \multicolumn{7}{c}{FTLE simulations} \\
    $N$ & $k_{\max}\eta$ & $R_{\lambda}$ & $\langle u^2 \rangle^{1/2}$ & $L$ & $L / \eta$ & $T / \tau_{\eta}$ & $n_L$ & $\tau_\mathrm{ref}/\tau_\eta$
    & $k_{\max}\eta$ & $R_{\lambda}$ & $\langle u^2 \rangle^{1/2}$ & $L$ & $L / \eta$ & $T / \tau_{\eta}$ & $n$\\
    \hline
    $512$ &  2.0 & 147 & 0.96 & 0.93 & 94  & 15.2 & 250 & 0.15
    & 2.0 & 142 & 0.96 & 0.91 & 93 & 15.4 & $2.5\times 10^7$\\
    $1024$ & 2.9 & 216 & 1.09 & 1.06 & 148 & 19.8 & 1000 & 0.16
    & 2.9 & 215 & 1.09 & 1.06 & 148 & 19.8 & $1\times 10^8$\\
    $2048$ & 2.9 & 334 & 1.07 & 1.04 & 289 & 31.2 & 1000 & 0.17
    & 3.0 & 335 & 1.07 & 1.04 & 289 & 31.1 & $1\times 10^9$\\
    \hline
\end{tabular}
        \caption{
            \textbf{Main DNS Parameters.} Our simulations are run on three-dimensional periodic domains of side length $2\pi$ discretized on a real-space grid with $N^3$ points. 
            The Kolmogorov length and time scales, $\eta$ and $\tau_\eta$, respectively, are computed from the mean kinetic energy dissipation $\varepsilon$ and the kinematic viscosity $\nu$. Based on the largest wavenumber $k_{\max}$ resolved by our code, we compute the resolution criterion $k_{\max}\eta$. Using the root-mean-squared velocity component $\langle u^2 \rangle^{1/2}$ and the energy spectrum $E(k)$, we define the integral length $L=\frac{\pi}{2\langle u^2 \rangle} \int \frac{\mathrm{d}k}{k} E(k)$. The integral time scale is computed as $T=L \langle u^2\rangle^{-1/2}$. Although loops and FTLE simulations are initialized with identical fields and parameters, the flows eventually diverge due to numerical rounding errors and chaos. The loops simulations contain $n_L$ material loops, initially sampled by 300 tracer particles per loop whose number increases roughly exponentially due to refinement in intervals of $\tau_\mathrm{ref}$. For the FTLE simulations, $n$ tracer trajectories are integrated along with the flow field.
        }
        \label{tab:DNS_parameters}
    \end{table}
\begin{figure*}[b!]
    \centering
    \begin{minipage}{0.49\textwidth}
    \includegraphics[width=1.\textwidth]{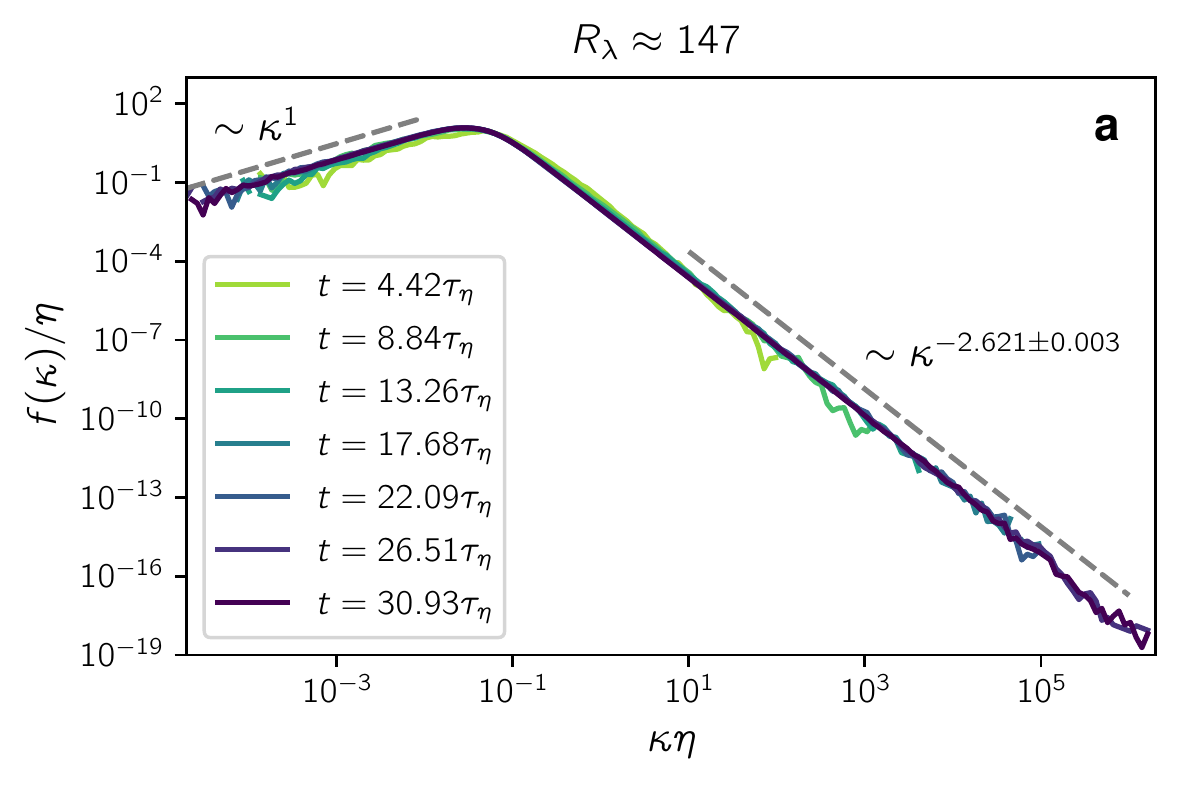}
    \end{minipage}
    \begin{minipage}{0.49\textwidth}
    \includegraphics[width=1.\textwidth]{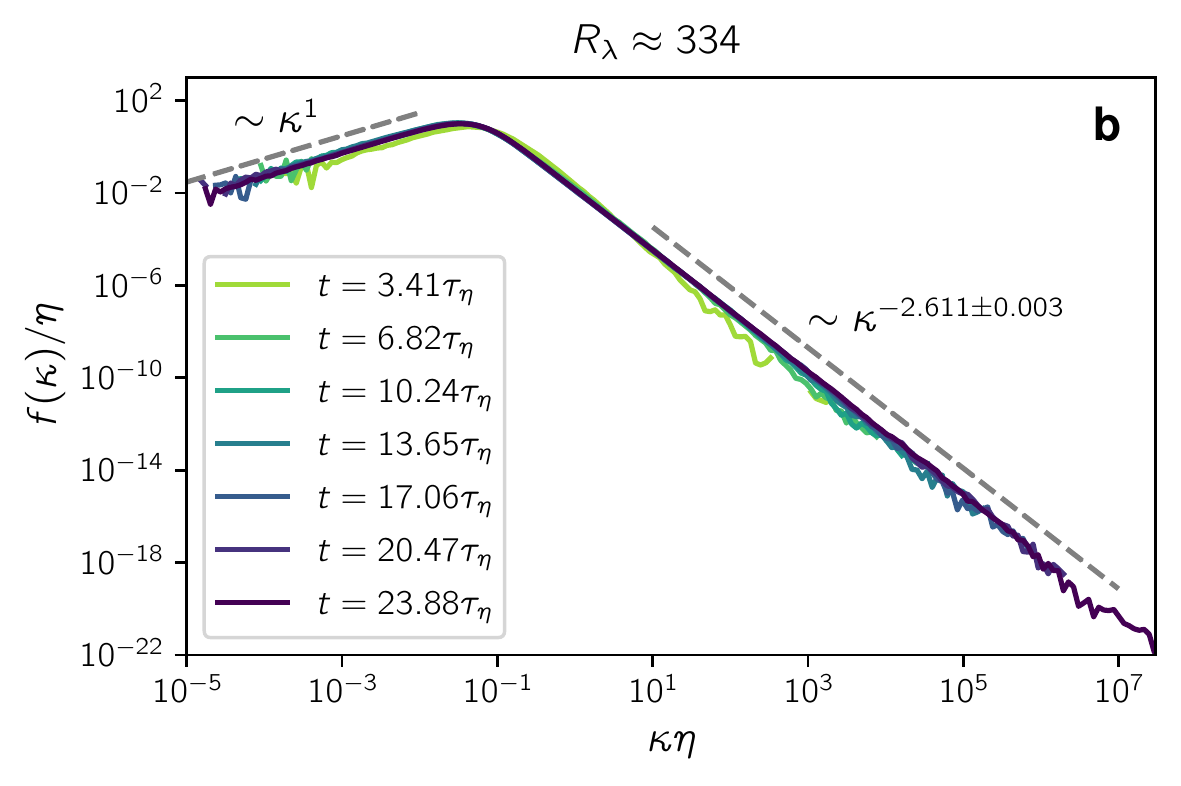}
    \end{minipage}
    \caption{\textbf{Curvature PDFs in supplementary simulations.} Both in the low-Reynolds-number (\textbf{a}) and the high-Reynolds-number (\textbf{b}) supplementary simulation, they strongly resemble the distribution of the main simulation. Their power-law exponents, determined by best fits, are almost identical across all simulations. \label{fig:supp_curv_pdf}}
\end{figure*}

\begin{figure*}[p]
    \centering
    \begin{minipage}{0.49\textwidth}
    \includegraphics[width=1.\textwidth]{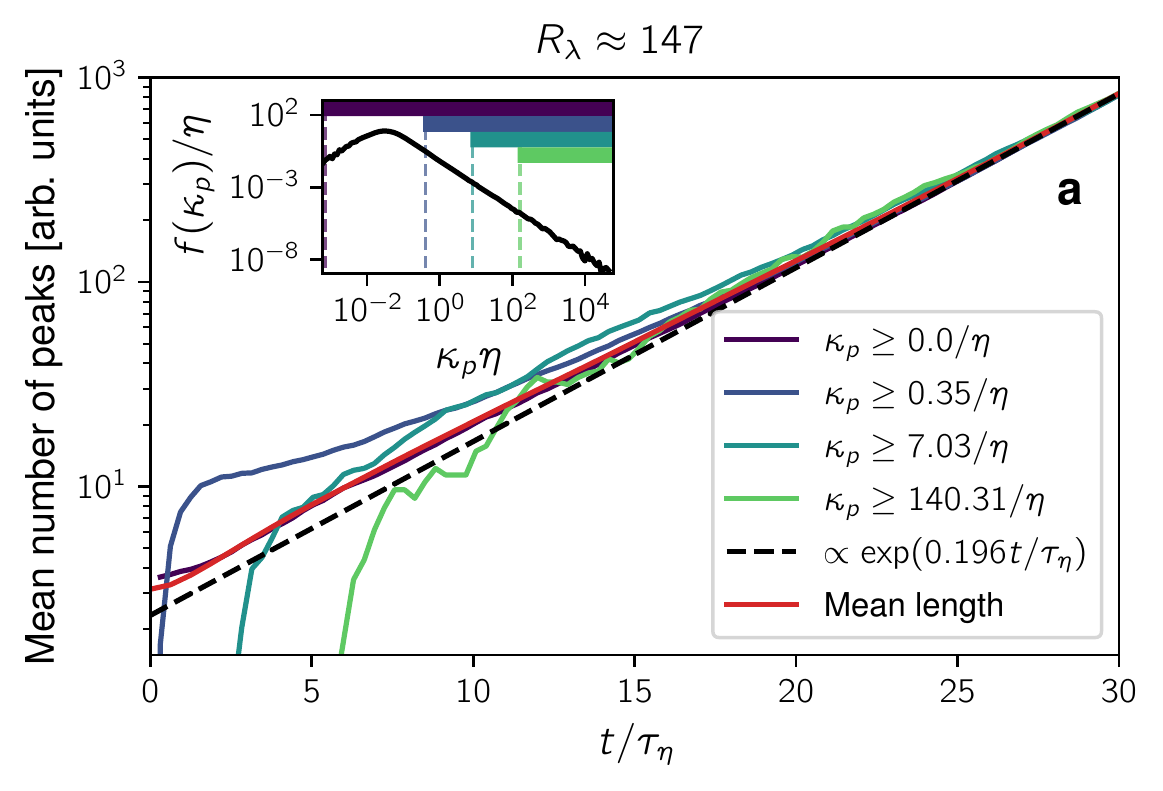}
    \end{minipage}
    \begin{minipage}{0.49\textwidth}
    \includegraphics[width=1.\textwidth]{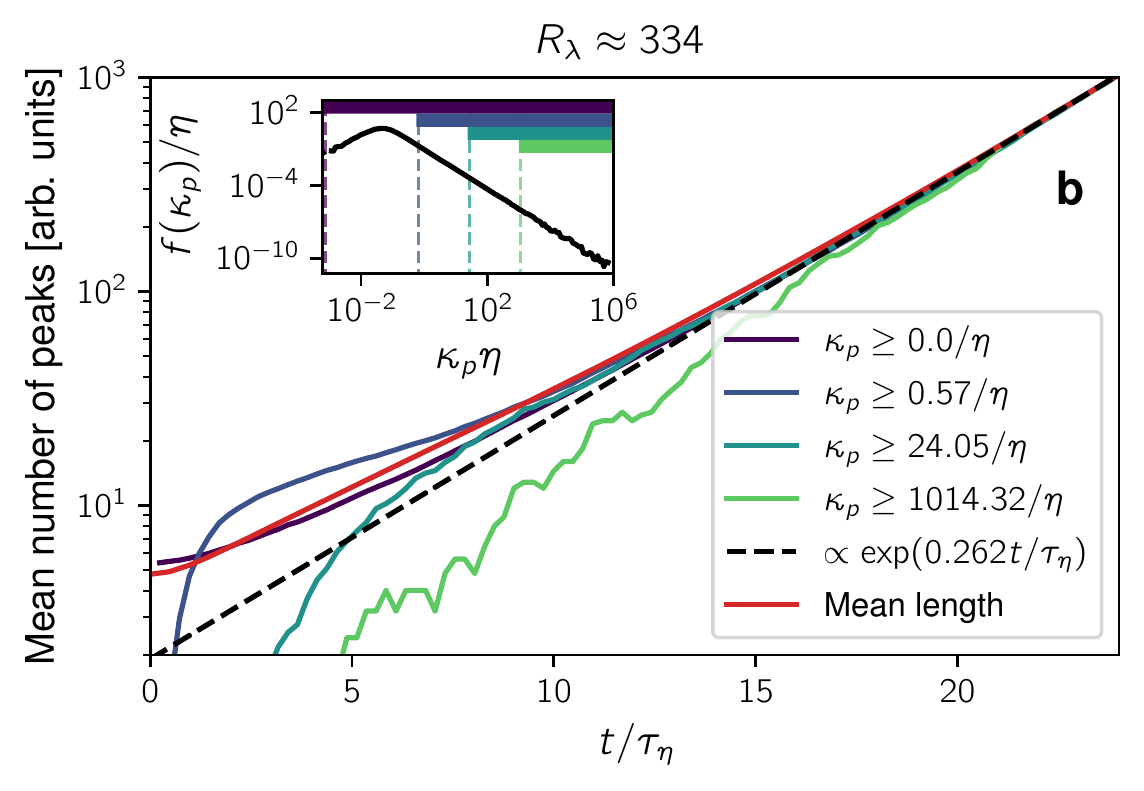}
    \end{minipage}\vspace{-0.2cm}
    \caption{\textbf{Mean curvature peak number as a function of time} in the low-Reynolds-number (\textbf{a}) and the high-Reynolds-number (\textbf{b}) supplementary simulation. The mean number of peaks above different thresholds grows exponentially, proportional to the mean length of the loops. Lines are vertically shifted to compare their growth rate, which is why they are not necessarily ordered as a function of the threshold condition. Best fits to the last third of the total peak number curve yield $\grrate = (0.19580 \pm 0.00028)/\tau_\eta$ (\textbf{a}) and $\grrate = (0.26207 \pm 0.00014)/\tau_\eta$ (\textbf{b}). As in the main simulation, the standard error from the fit is so small that we neglect it in the following. Insets: Curvature peak distribution at the latest simulation time indicating the different thresholds. \label{fig:supp_peak_number}}
\end{figure*}

The curvature peak number above different thresholds as a function of time is shown in Figure~\ref{fig:supp_peak_number} for the two supplementary simulations. Notably, our observation that the curvature peak number grows proportionally to the mean length of the loops carries over to these Reynolds numbers. The corresponding growth rate $\grrate$ in units of the Kolmogorov time appears to increase as a function of Reynolds number.

\begin{figure*}[h]
    \centering
    \begin{minipage}{0.49\textwidth}
    \includegraphics[width=1.\textwidth]{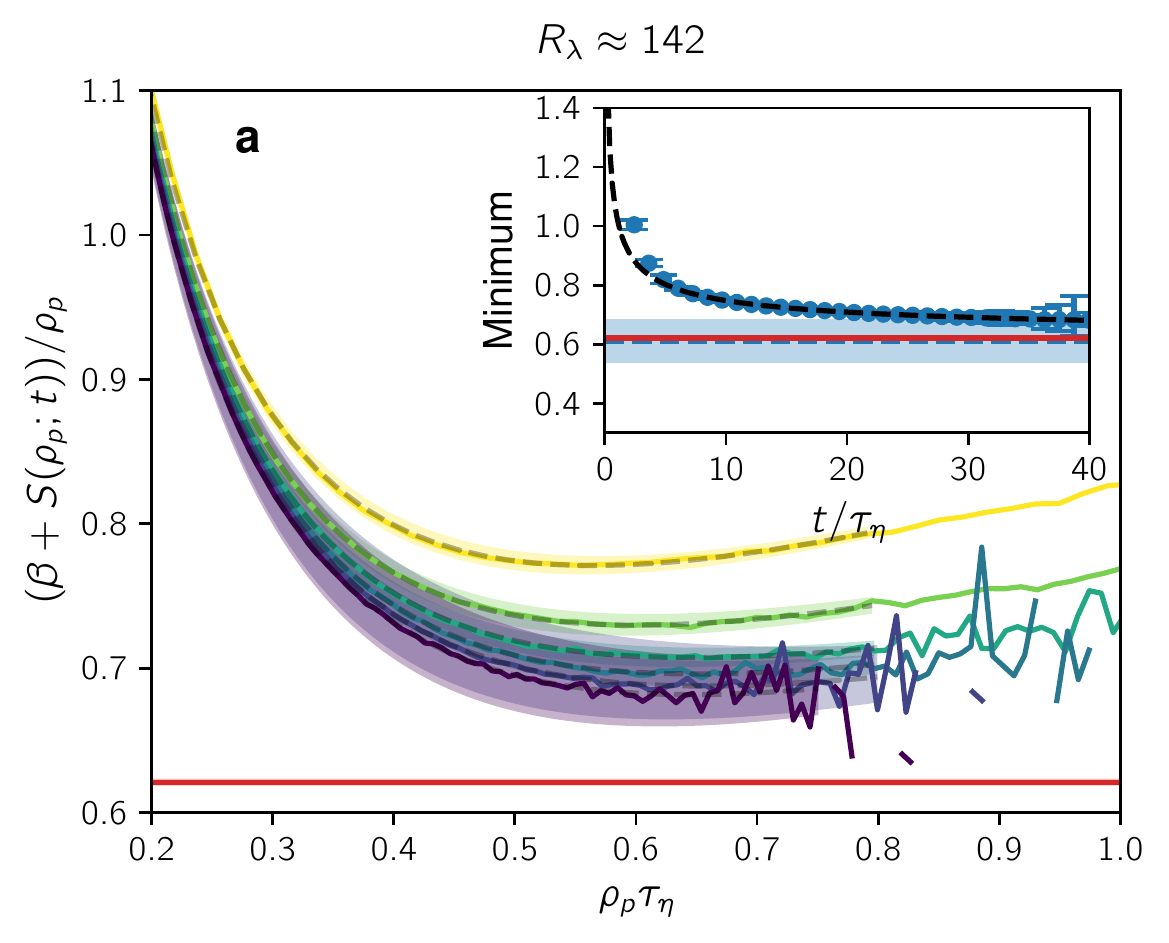}
    \end{minipage}
    \begin{minipage}{0.49\textwidth}
    \includegraphics[width=1.\textwidth]{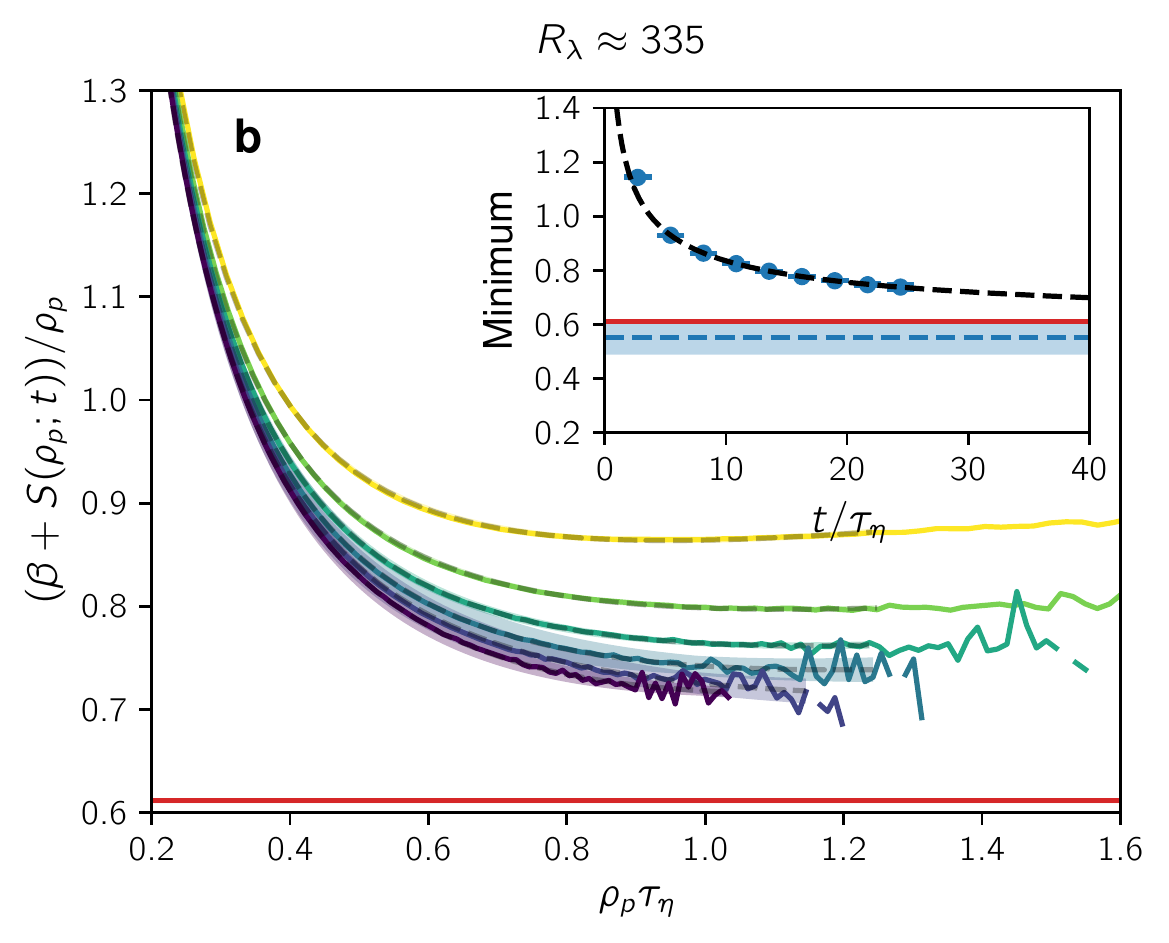}
    \end{minipage}
    \caption{\textbf{Determination of the steepest-descent minimum for the supplementary simulations.} \textbf{a} Low-Reynolds-number simulation with Cram\'er functions ranging from $7.26\tau_\eta$ (yellow) up to $39.94\tau_\eta$ (violet). Their fits are restricted to the range $[\lambda_p(t), 0.8/\tau_\eta]$, where $\lambda_p(t)$ is the position of the minimum of $S(\rho_p;t)$. Inset: Extrapolation of the minimum yields $\alpha = 0.61 \pm 0.08$ (dashed blue line). \textbf{b} High-Reynolds-number simulation with Cram\'er functions at times ranging from $8.14\tau_\eta$ (yellow) up to $35.26\tau_\eta$ (violet). Their fits are restricted to the range $[\lambda_p(t), 1.25/\tau_\eta]$. Inset: Extrapolation of the minimum yields $\alpha = 0.55 \pm 0.07$ (dashed blue line). For comparison, in each plot the red line indicates the value of $\alpha$ estimated from the curvature PDF. For details on the procedure, see Methods. \label{fig:supp_cramer_minimum}}
\end{figure*}

\begin{figure*}[ht]
    \centering
    \begin{minipage}{0.334\textwidth}
    \includegraphics[width=1\textwidth]{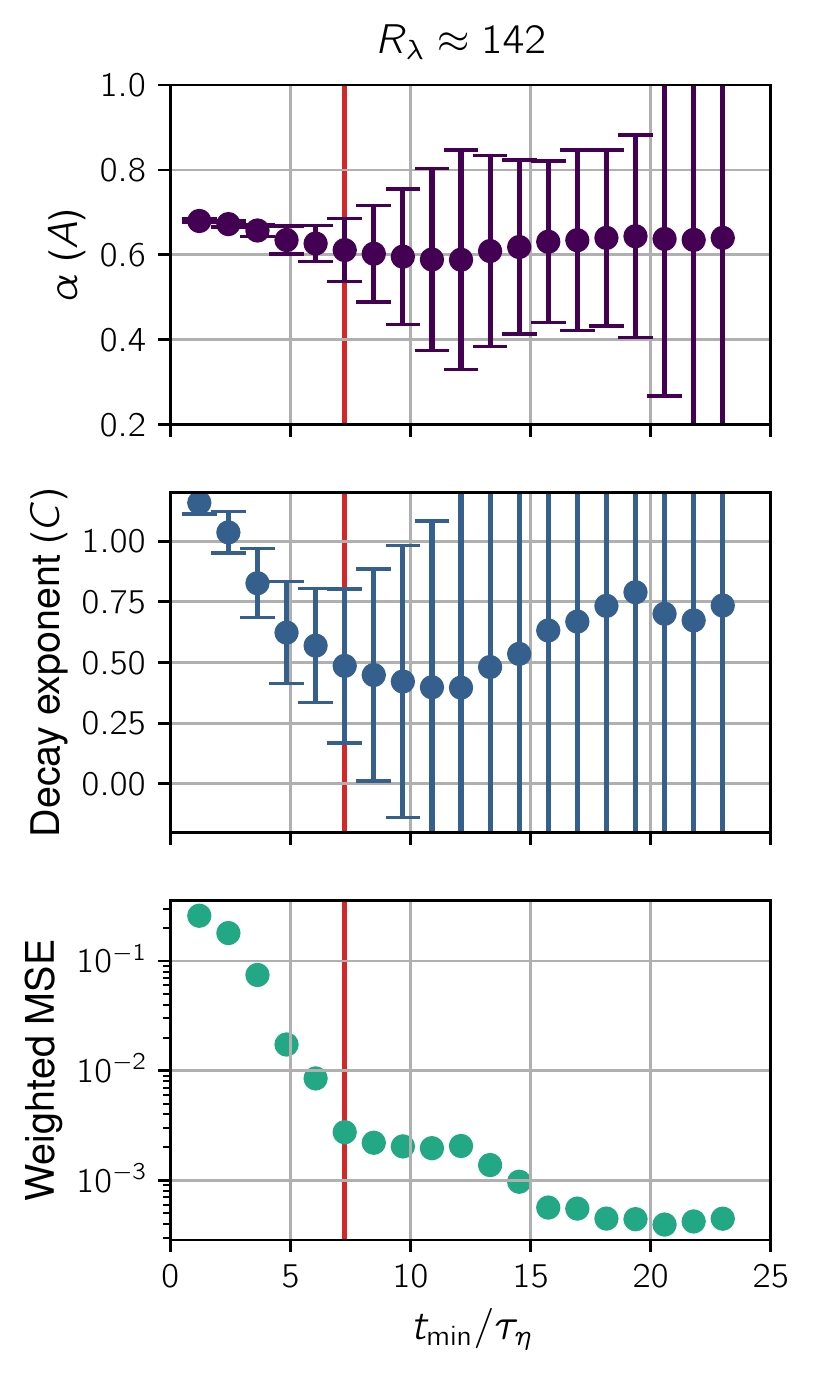}
    \end{minipage}
    \begin{minipage}{0.313\textwidth}
    \includegraphics[width=1\textwidth]{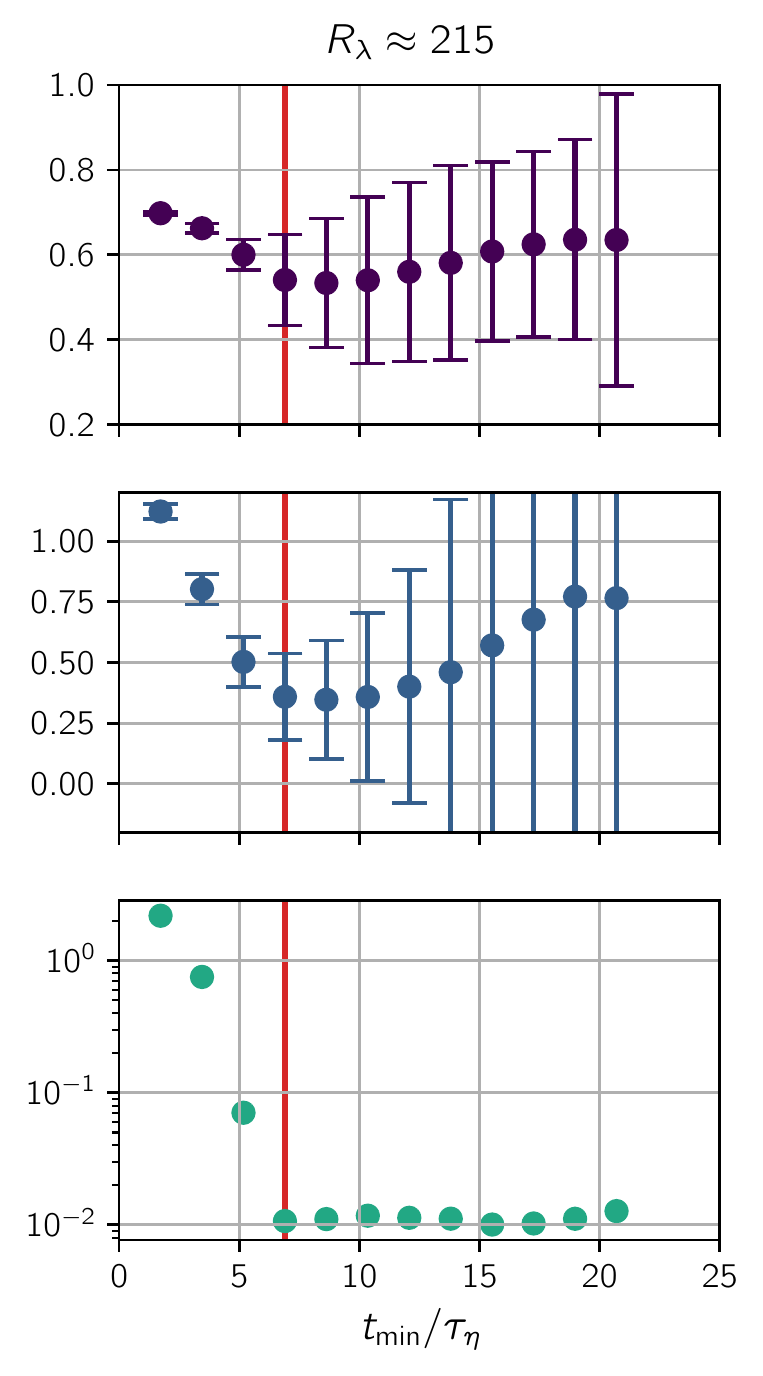}
    \end{minipage}
    \begin{minipage}{0.313\textwidth}
    \includegraphics[width=1\textwidth]{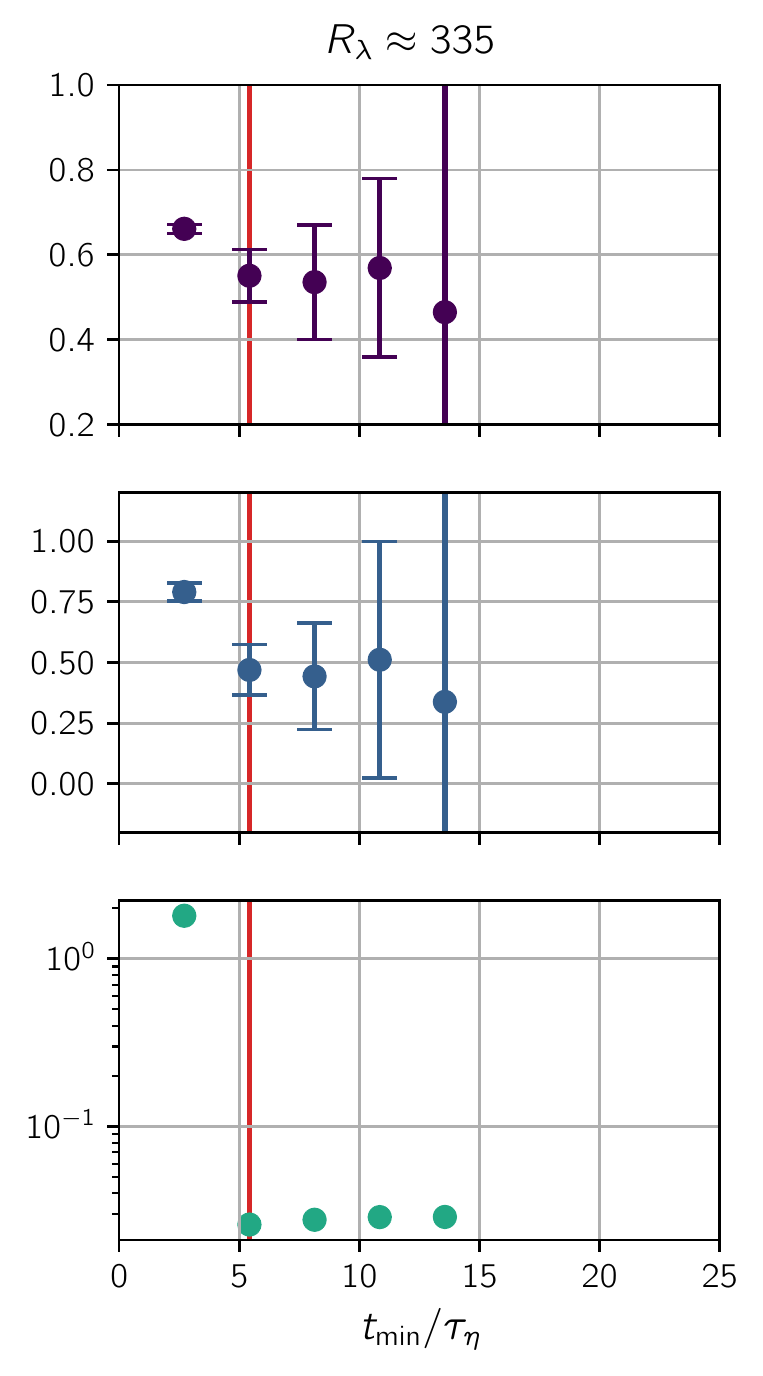}
    \end{minipage}
    \caption{\textbf{Different fit choices for extrapolating the steepest-descent minimum.} In order to compute the steepest-descent minimum~\eqref{alphaeqn}, 
    in principle we need a fully resolved Cram\'er function. In practice, however, we only have finite-time estimates of the Cram\'er function, which yield finite-time estimates of the minimum (our ``data points'' in the insets of Figs.~\ref{fig:cramer_minimum} 
    and \ref{fig:supp_cramer_minimum}). We use the simple decay function~\eqref{eq:decay_func} 
    to capture and extrapolate the evolution of these data points as a function of the time at which the finite-time Cram\'er function is computed. In order to obtain a satisfactory fit, it is helpful to leave out a transient regime of data points $t < t_\mathrm{min}$ such that only the asymptotic behavior is captured by the fit. Here we show the resulting values of $\alpha=A$ (top) and the decay exponent $C$ (middle) for different choices of $t_\mathrm{min}$ in all of the simulations (left to right). The time scale $B$ is not shown. We justify the choice of $t_\mathrm{min}$ by computing the weighted mean squared error (MSE, bottom) given by $\sum_{i=1}^N (\delta_i/\sigma_i)^2/(N-3)$. Here, $N$ is the number of data points included, $\delta_i$ is the deviation of the fit from the $i$-th data point, and $\sigma_i$ is the error of the $i$-th data point, given by the maximum of the two-sided error computed from the error envelopes in Figs.~\ref{fig:cramer_minimum} 
    and \ref{fig:supp_cramer_minimum}. The quantity $\sum_{i=1}^N (\delta_i/\sigma_i)^2$ is minimized by the fit, which we divide by the number of degrees of freedom $N-3$ (number of data points minus number of fit parameters) for comparability. We choose $t_\mathrm{min}$ to be the start of the first plateau of the weighted MSE (red lines). \label{fig:cramer_leave_out}}
\end{figure*}

Finally, we also determine FTLE statistics by integrating the deformation tensor along trajectories of randomly distributed particles. This is done for $25$ million tracer particles in an additional simulation at the smaller Reynolds number and for one billion tracer particles in an additional simulation at the larger Reynolds number that use the same initial condition as the corresponding loops simulations. In Figure~\ref{fig:supp_cramer_minimum}, we determine the steepest-descent minima needed for our theoretical prediction. Notice that the minimum is taken at values of $\rho_p$ that increase with Reynolds number and therefore reach further into the tail of the FTLE distribution. Hence in the large-Reynolds-number simulation, despite the enormous number of tracer particles tracked, the minimum can only be resolved up to $t\approx 25\tau_\eta$. As in the main simulation, we determine the asymptotic value of the minimum by fitting the algebraic decay function~\eqref{eq:decay_func} 
to those data points with $t\geq t_\mathrm{min}$ (Figure~\ref{fig:supp_cramer_minimum}, insets). The time $t_\mathrm{min}$ is chosen based on the weighted mean squared error, as explained in Figure~\ref{fig:cramer_leave_out}, in order to filter out a transient regime of the decay. The resulting exponents are $\alpha = A = 0.61 \pm 0.08$ for the low-Reynolds-number simulation, $\alpha = A = 0.54 \pm 0.11$ for the main simulation and $\alpha = A = 0.55 \pm 0.07$ for the high-Reynolds-number simulation, all of them consistent with the measured curvature PDF power-law exponents.


\section{Numerical analysis of generalized Lyapunov exponents} \label{sec:gle_numeric}

\begin{figure*}[ht]
    \centering
    \begin{minipage}{0.49\textwidth}
    \includegraphics[width=1.\textwidth]{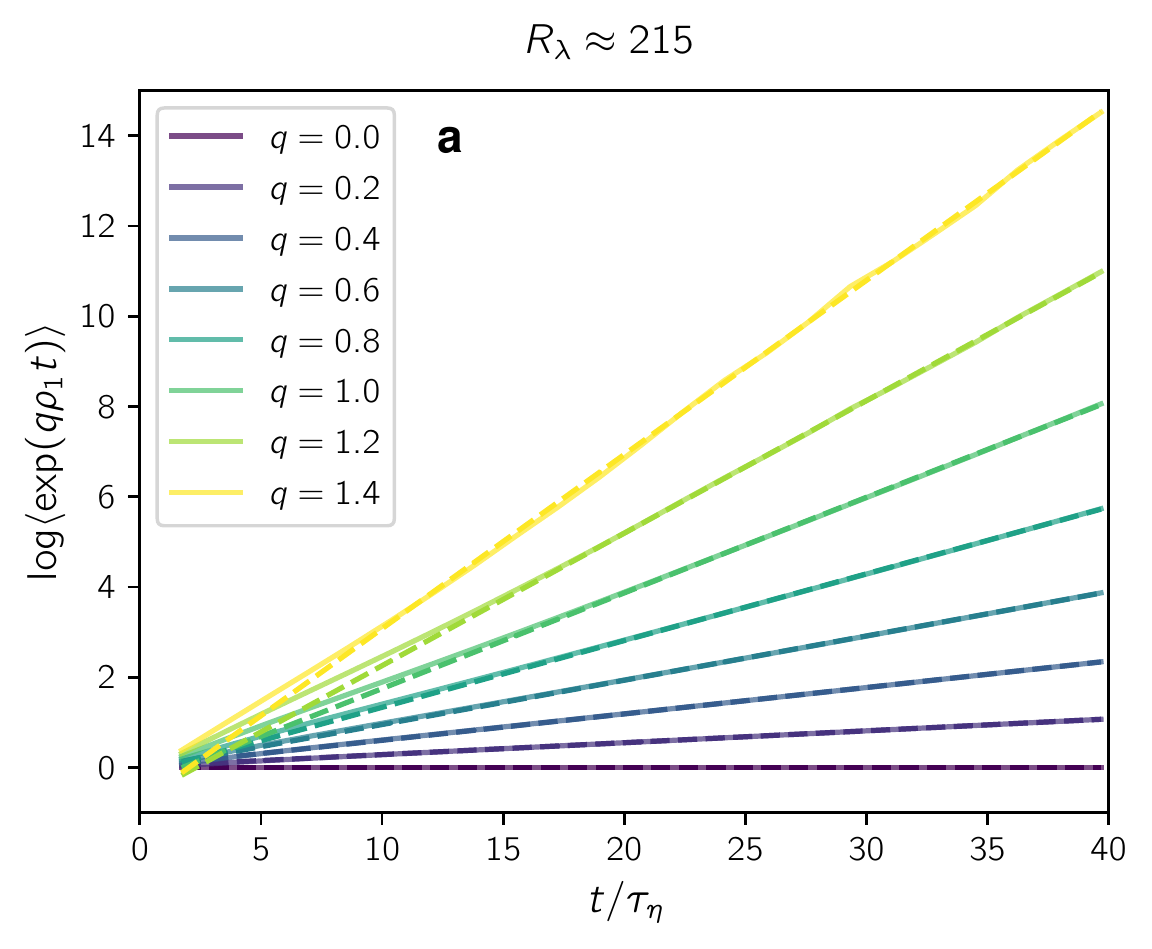}
    \end{minipage}
    \begin{minipage}{0.49\textwidth}
    \includegraphics[width=1.\textwidth]{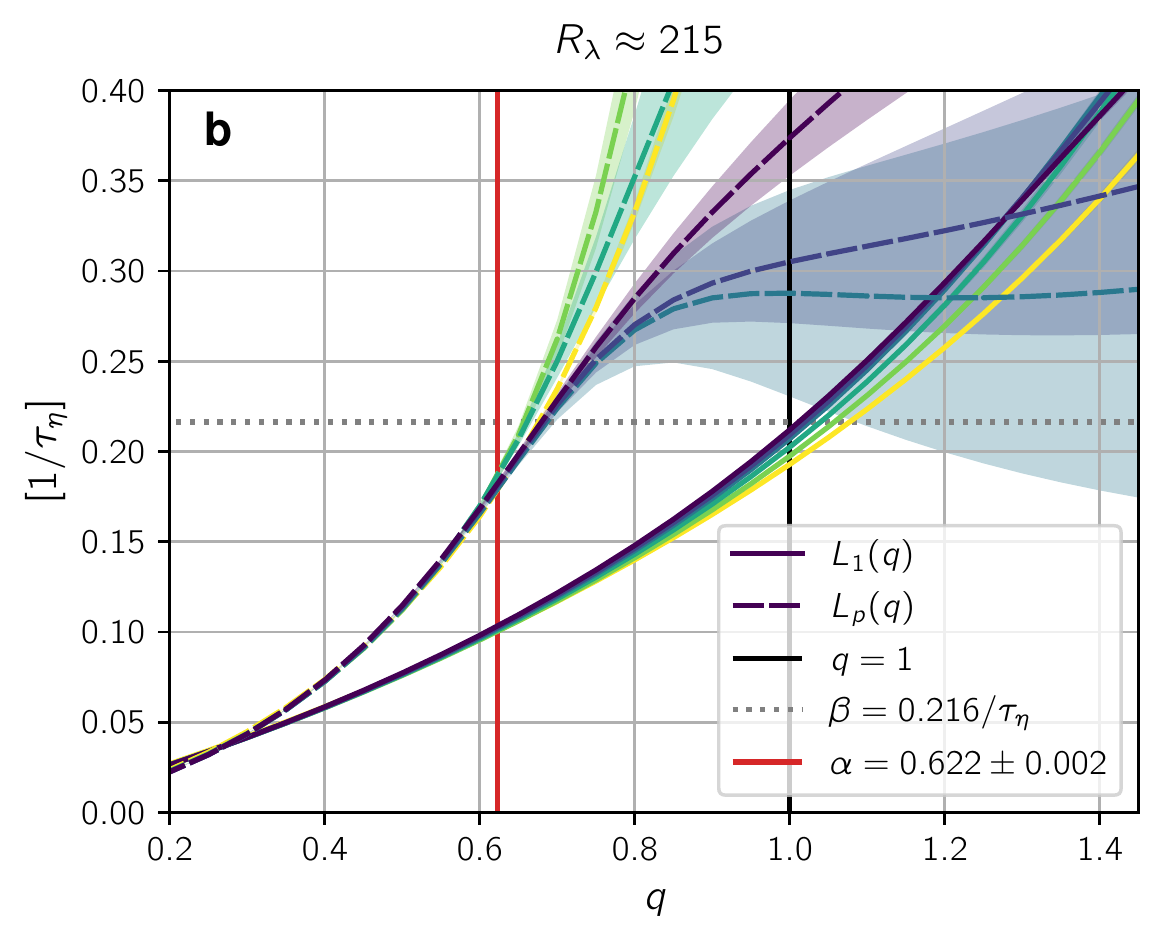}
    \end{minipage}
    \caption{\textbf{Generalized Lyapunov exponents in the main simulation.} \textbf{a} Plotting the cumulant-generating function of $\rho_1(t)t$ for fixed argument $q$ as a function of time (solid lines), the GLE $L_1(q)$ can be estimated as the asymptotic slope of the curve by an affine fit (dashed lines) on the interval $t \in [t_\text{max}/2, t_\text{max}]$ where in this case $t_\text{max} = 40\tau_\eta$. \textbf{b} Generalized Lyapunov exponents can be used to estimate $\grrate$ and $\alpha$. The first standard GLE $L_1(q)$ (solid lines) is shown for $t_\text{max}$ ranging from $12.08\tau_\eta$ (yellow) to $39.68\tau_\eta$ (violet). The curvature-peak GLE $L_p(q)$ is shown for the same times (dashed lines). For comparison, we also show the line $q=1$ (solid, black), the value of $\grrate$ estimated from curvature peak number (dotted, grey) and the value of $\alpha$ estimated from the curvature PDF (solid, red). \label{fig:gle_1024}}
\end{figure*}

As a complementary approach to computing Cram\'er functions, we may also use generalized Lyapunov exponents (GLE) in order to determine the exponent $\alpha$ based on the implicit equation \eqref{eq:stretch_fold_balance_explicit}, 
in short: $L_p(\alpha) = \grrate = L_1(1)$, where
\begin{align}
    L_1(q) = \lim_{t\to\infty} \frac{1}{t} \log \left\langle \exp(q\ftle_1(t)t)\right\rangle
\end{align}
is the first standard GLE as opposed to 
\begin{align}
    L_p(q) = \lim_{t\to\infty} \frac{1}{t} \log \left\langle \exp(q\ftle_p(t)t)\right\rangle,
\end{align}
the curvature-peak GLE. For their numerical computation, we adopt the method from ref.~\cite{johnson_pof_2015}. We first compute the cumulant-generating function of $\rho_1(t) t$, given by $\log\langle\exp(q\rho_1(t) t)\rangle$, as a function of $q$ and $t$. In order to estimate $L_1(q)$, we perform an affine fit of the cumulant-generating function in the range $t \in [t_\text{max}/2, t_\text{max}]$, as exemplified in Fig.~\ref{fig:gle_1024}a for $t_\text{max} = 40\tau_\eta$. The slope of each fit including its standard error becomes our estimate of $L_1(q)$. The same procedure is applied to $L_p(q)$.

\begin{figure*}[ht]
    \centering
    \begin{minipage}{0.49\textwidth}
    \includegraphics[width=1.\textwidth]{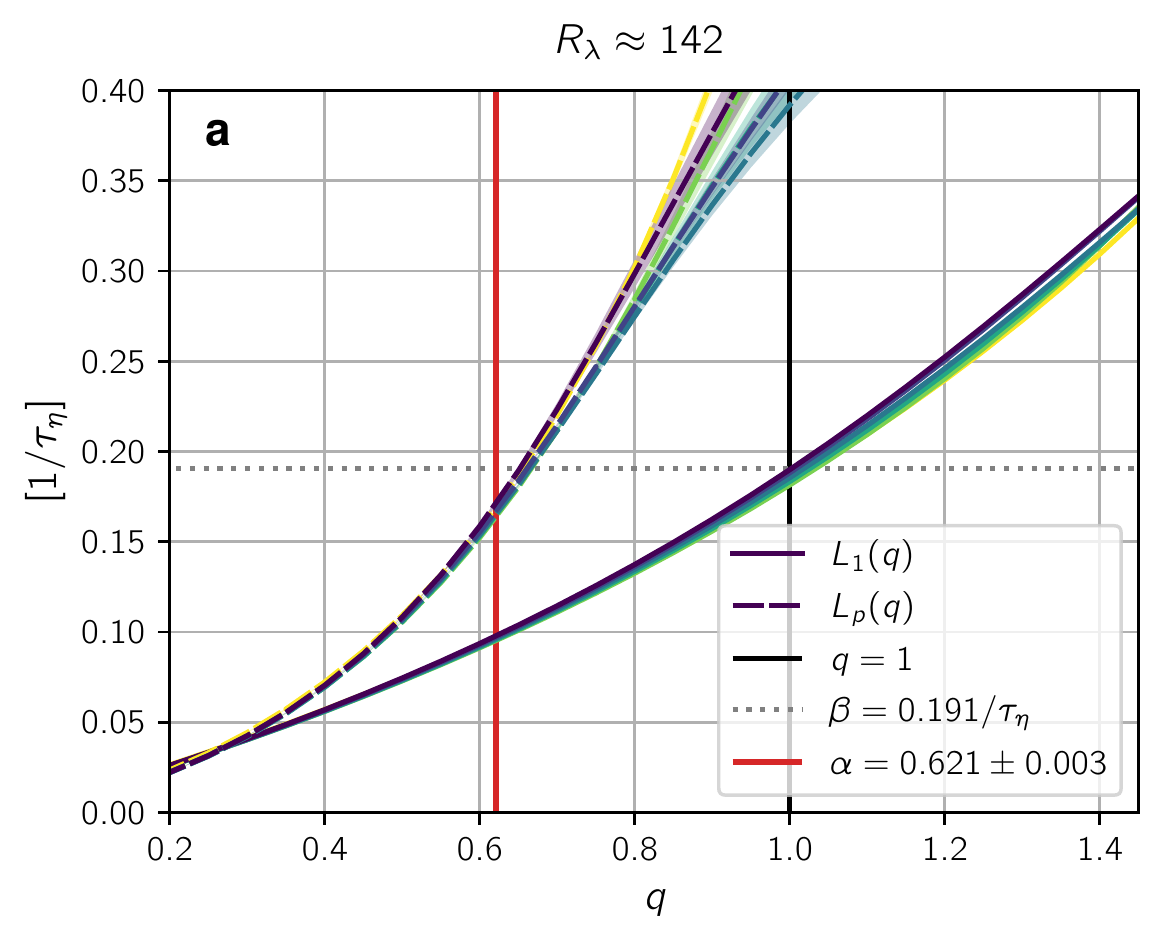}
    \end{minipage}
    \begin{minipage}{0.49\textwidth}
    \includegraphics[width=1.\textwidth]{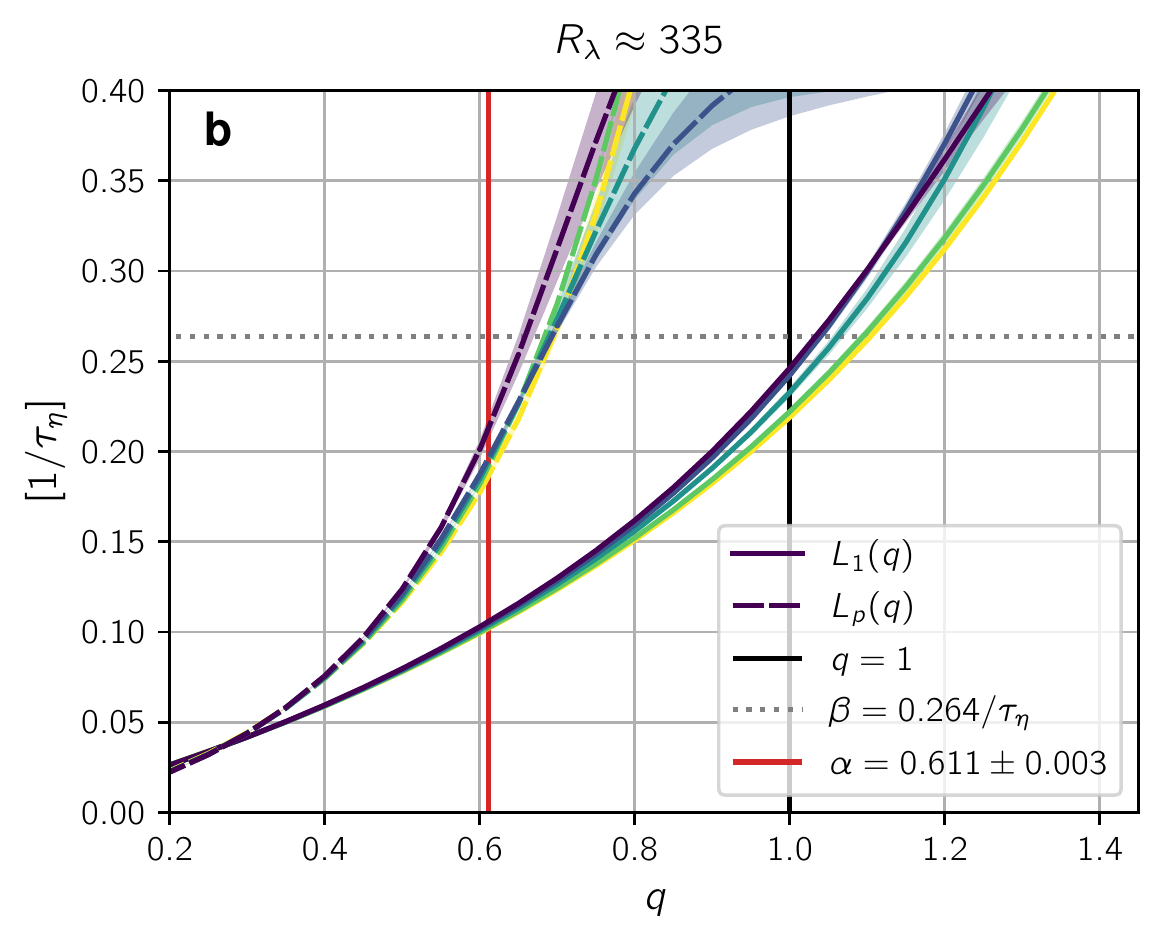}
    \end{minipage}
    \caption{\textbf{Generalized Lyapunov exponents in the supplementary simulations}, at  low Reynolds number (\textbf{a}) and at high Reynolds number (\textbf{b}). As in the main simulation, the first standard GLE $L_1(1)$ converges toward the previously estimated value of $\grrate$, but the convergence is slower in the high-Reynolds-number simulation. Similarly, the value of $\alpha$ is slightly overestimated in both simulations. In \textbf{a}, $t_\text{max}$ ranges from $12.10\tau_\eta$ (yellow) to $39.94\tau_\eta$ (violet). In \textbf{b}, $t_\text{max}$ ranges from $16.27\tau_\eta$ (yellow) to $35.26\tau_\eta$ (violet). Note that the value of $\grrate$ as a function of the Kolmogorov time scale $\tau_\eta$ differs from the one in Fig.~\ref{fig:supp_peak_number} because $\tau_\eta$ is slightly different in the FTLE simulation. \label{fig:gle_supp}}
\end{figure*}

For the main simulation, the results are shown in Fig.~\ref{fig:gle_1024}b. We can first read off the value of $\grrate$ by evaluating $L_1(q)$ at $q=1$. Indeed, the different estimates for $L_1(1)$ appear to converge toward the value of $\grrate$ previously estimated by other means. In order to estimate $\alpha$, we need to read off the intersection of the $\grrate$-line with $L_p(q)$. For this curvature-peak GLE, we observe stronger fluctuations as a function of $t_\text{max}$. For small $t_\text{max}$, we expect the estimates of the cumulant-generating function to be accurate. However, if $t_\text{max}$ is too small, we have not yet reached convergence of the $t\to\infty$ limit in the GLE. For larger $t_\text{max}$, we improve on the convergence of the GLE, but we also rely more heavily on extreme values of $\rho_p(t)$ (especially for large $q$), which are limited by our sample size. Hence we expect the best estimate to be found at intermediate $t_\text{max}$. For the main simulation, we indeed find those intermediate curves to come closest to the value of $\alpha$ estimated from the curvature PDF. For the supplementary simulations (Fig.~\ref{fig:gle_supp}), the same argumentation holds and the estimates fluctuate as a function of $t_\text{max}$ to a certain extent. If we simply read off $\alpha$ from the intersection of lines, the GLE method slightly overestimates $\alpha$ for all simulations, possibly due to both sampling and time-convergence limitations.\newpage

\newpage
\section{Material line bundles and flux cancellations} \label{sec:bundles}
\begin{figure*}[ht]
    \centering
    \begin{minipage}{0.6\textwidth}
    \includegraphics[width=1.\textwidth]{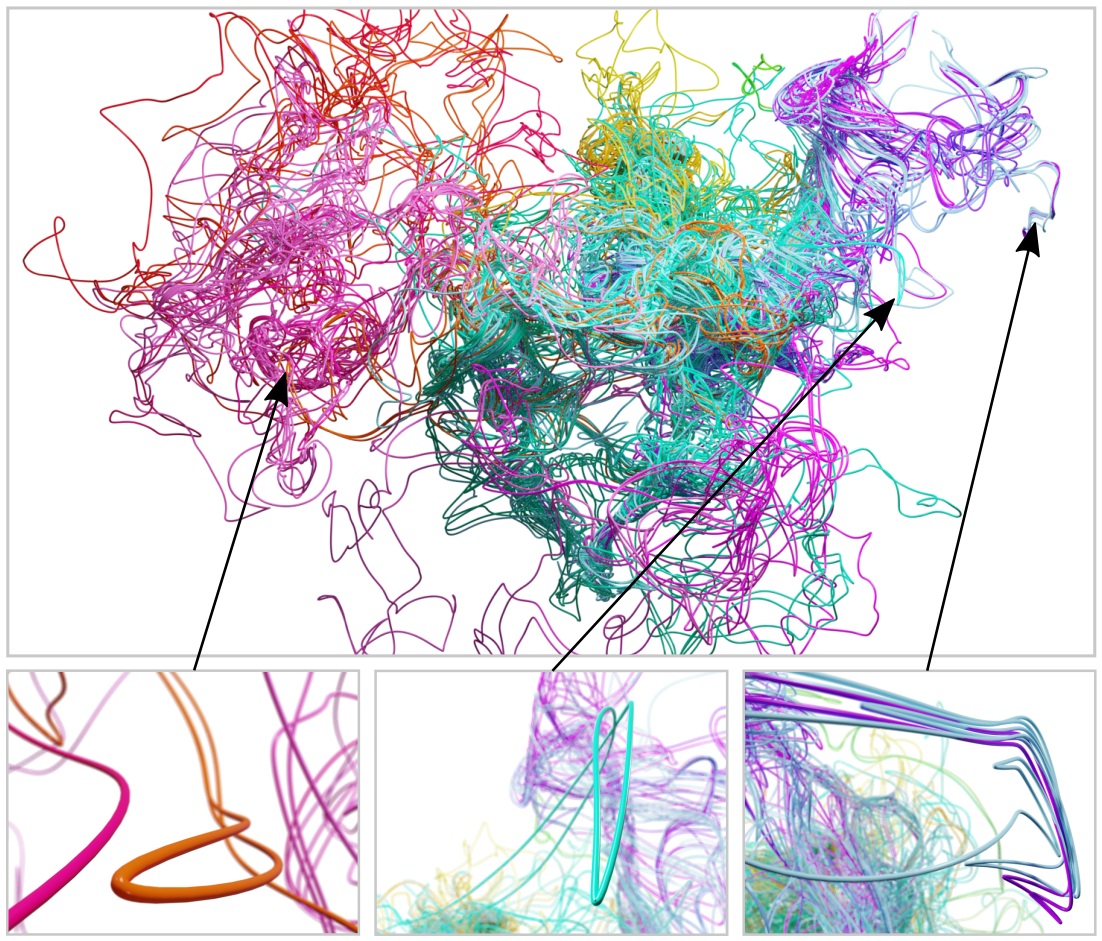}
    \end{minipage}
    \caption{\textbf{Loops form pairs and bundles of almost parallel material lines}, which are closely associated with sharp folds. The loop snapshot is identical to Fig.~\ref{fig:loop_visualization} 
    of the manuscript, taken at $t=27\tau_\eta$ in the main simulation at $R_\lambda \approx 216$. \label{fig:bundles}}
\end{figure*}

In the discussion, we explain how our results may help to shed light on the problem of flux cancellations in magnetohydrodynamics (MHD). Flux cancellations occur when magnetic field lines with opposite orientation are brought closely together by the flow and thus cancel each other in the integration of magnetic flux. In our simulations, we observe that the material lines are brought into such a configuration quite frequently (see Fig.~\ref{fig:bundles}). They even tend to form bundles of lines where about half of the lines has opposite orientation. Remarkably, these bundles appear to be strongly related to the sharp folds that lead to curvature peaks since the folds are typically observed in the middle or at the end of such bundles. Therefore, it may be worthwhile studying folds of the magnetic field in MHD (possibly detected by curvature peaks) and their relation to flux cancellations.

\newpage
\section{Spatial and temporal resolution of loop tracking}
\begin{figure*}[h]
    \centering
    \begin{minipage}{0.6\textwidth}
    \includegraphics[width=1.\textwidth]{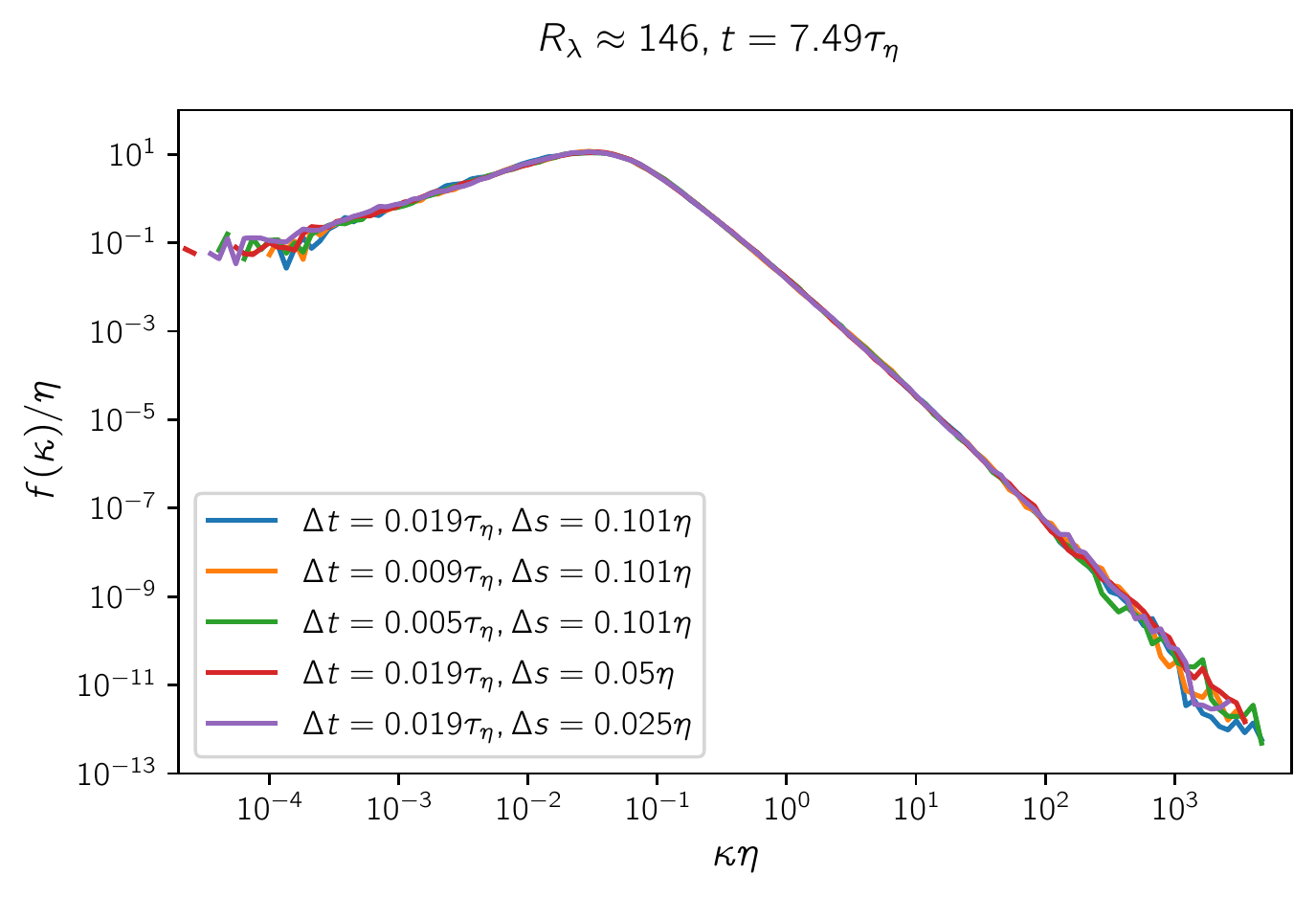}
    \end{minipage}
\caption{\textbf{Curvature distributions for different temporal and spatial resolutions of the loops}, taken from 1000 loops after $7.49\tau_\eta$ in simulations at $R_\lambda\approx 146$. $\Delta t$ is the time-step size of the simulation, including field and particle time stepping. $\Delta s$ denotes the maximum distance of sampling points of the loops enforced by the refinement. We adjusted the additional refinement condition based on curvature accordingly, effectively doubling the density of sample points uniformly along the loops when moving from $\Delta s = 0.1\eta$ to $\Delta s = 0.05\eta$ and further to $\Delta s = 0.025\eta$. Other parameters are identical with the low-Reynolds-number simulation shown in Supplementary Note~\ref{sec:set_of_sims}, which uses $\Delta t = 0.018\tau_\eta$ and $\Delta s = 0.1\eta$. Note that the curved shape of the distribution can be attributed to the early time ($t=7.49\tau_\eta$) in the loops' evolution. \label{fig:convergence_curv_pdf}}
\end{figure*}

For tracking material loops in our simulations, we treated the sample points of the loops as Lagrangian tracers. Over time, new particles are inserted to sustain the necessary loop resolution. Since using higher-order time-stepping methods for the particles would require histories that are not available for newly inserted particles, we resorted to first-order Euler time stepping. In order to ensure the quality of our results, we tested the code with different temporal resolutions of flow and particle time stepping and different spatial resolution conditions of the loops. The results for simulations at $R_\lambda\approx 146$ are shown in Fig.~\ref{fig:convergence_curv_pdf} where we compare curvature statistics for the different resolutions. We observe that neither an improved temporal resolution nor an improved spatial resolution of the loops leads to noticeable changes in the curvature distribution. 

\newpage
\section{Geometric evolution equations}
\label{ch:geom_evol_eq}
Here, we derive the evolution equations \eqref{eq:gradient_evolution}--\eqref{eq:curvature_evolution}. 
Implicitly, all quantities considered in these equations are evaluated along a material line element $\mathbf{L}=\mathbf{L}(\angl, t)$, whose evolution is given by the tracer equation~\eqref{eq:tracer_eq}. 
Therefore 
\begin{align}
    \partial_t \partial_{\angl} \mathbf{L}(\angl, t) 
    &= ((\partial_{\angl} \mathbf{L}) \cdot \nabla)\mathbf{u}(\mathbf{L}(\angl, t), t)
\end{align}
which is \eqref{eq:gradient_evolution}. 
The evolution equation~\eqref{eq:tangent_evolution} 
for the tangent vector of the Frenet-Serret frame~\cite{baer_2010},
\begin{align}
 \tv(\angl, t) = \frac{\partial_\angl \mathbf{L}}{\norm{\partial_\angl \mathbf{L}}},
\end{align}
can then be directly computed
\begin{align}
  \partial_t \tv(\angl, t) &= \partial_t \frac{\partial_\angl \mathbf{L}}{\norm{\partial_\angl \mathbf{L}}}= \frac{1}{\norm{\partial_\angl \mathbf{L}}}(\partial_\angl \mathbf{L} \cdot \nabla) \uv
  - \frac{\partial_\angl \mathbf{L}}{\norm{\partial_\angl \mathbf{L}}^3} \partial_\angl \mathbf{L} \cdot (\partial_\angl \mathbf{L} \cdot \nabla) \uv = (\tv\cdot\nabla)\uv - (\tv\cdot(\tv\cdot\nabla)\uv) \tv.\label{tevol}
\end{align}
The normal vector of the Frenet-Serret frame is defined as
\begin{align} \label{eq:n_def}
    \nv(\angl, t) = \frac{\partial_s \tv}{\norms{\partial_s\tv}},
\end{align}
where $s$ denotes an arc-length parameterization of the line, i.e.\ $|\partial_s \mathbf{L}| = 1$. Since the transform from $s$ to $\angl$ is time-dependent, evaluating $\partial_t$ at constant $\angl$ and at constant $s$ are different operations. Here, we always want to take $\partial_t$ at constant $\angl$ (i.e.~at the same tracer particle). Then, $\partial_s$ and $\partial_t$ do not commute. Having this in mind, we compute (summation over repeated indices implied)
\begin{align}\nonumber
  \partial_t \nc_i &= \partial_t \frac{\partial_s \tc_i}{\norms{\partial_s \tv}} = \frac{(\partial_t \partial_s \tc_i)\norms{\partial_s \tv} - (\partial_s \tc_i) \partial_t \norms{\partial_s \tv}}{\norms{\partial_s \tv}^2} = \frac{\partial_t \partial_s \tc_i}{\norms{\partial_s \tv}} - \nc_i \frac{\partial_t \norms{\partial_s \tv}}{\norms{\partial_s \tv}} = \frac{\partial_t \partial_s \tc_i}{\norms{\partial_s \tv}} - \nc_i \nc_j \frac{1}{|\partial_s \tv|} \partial_t \partial_s \tc_j \\
  &= \frac{1}{|\partial_s \tv|}\left(\delta_{ij} - \nc_i \nc_j\right) \partial_t \partial_s \tc_j.
\end{align}
In order to swap the $t$- and $s$-derivatives, we notice that $\partial_s = \tod{\angl}{s} \partial_\angl 
    = \frac{1}{|\partial_\angl \mathbf{L}|} \partial_\angl$.
Therefore
\begin{align}
    \partial_t \partial_s \tc_j &= \left(\partial_t \frac{1}{|\partial_\angl \mathbf{L}|}\right)\partial_\angl \tc_j +  \partial_s \partial_t \tc_j= - \tc_k \tc_l (\partial_k u_l) (\partial_s \tc_j) + \partial_s \partial_t \tc_j.
\end{align}
We then insert the evolution equation \eqref{tevol} for the tangent vector $\tv$,
\begin{align} \nonumber
  \partial_t \partial_s \tc_j
  &= - \tc_k \tc_l (\partial_k u_l) (\partial_s \tc_j) + \partial_s \left((\delta_{jk} - \tc_j \tc_k) \tc_l \partial_l u_k \right)\\ \nonumber
  &= - \tc_k \tc_l (\partial_k u_l) (\partial_s \tc_j) -\left((\partial_s \tc_j) \tc_k + \tc_j (\partial_s \tc_k)\right) \tc_l \partial_l u_k
  + (\delta_{jk} - \tc_j \tc_k) \left((\partial_s \tc_l) \partial_l u_k + \tc_l \partial_s \partial_l u_k\right) \\
  &= \norms{\partial_s \tv} \Bigl[- \tc_k \tc_l \nc_j (\partial_k u_l) - (\nc_j \tc_k + \tc_j \nc_k) \tc_l \partial_l u_k
  + (\delta_{jk} - \tc_j \tc_k)\Bigl(\nc_l \partial_l u_k + \frac{1}{\norms{\partial_s \tv}}\tc_l \tc_m \partial_l \partial_m u_k\Bigr)\Bigr],
\end{align}
where in the last step we used~\eqref{eq:n_def} and $\partial_s u_k = \tc_m \partial_m u_k$.
Since $\norms{\partial_s \tv} = \tilde{\kappa}$ and $\delta_{ij} = \tc_i\tc_j + \nc_i \nc_j + \bc_i \bc_j$, we have
\begin{align}\nonumber
    \partial_t \nc_i &= (\tc_i \tc_j + \bc_i \bc_j) \Bigl[- \tc_k \tc_l \nc_j (\partial_k u_l) - (\nc_j \tc_k + \tc_j \nc_k) \tc_l \partial_l u_k
  + (\nc_j \nc_k + \bc_j \bc_k)\Bigl(\nc_l \partial_l u_k + \frac{1}{\tilde{\kappa}}\tc_l \tc_m \partial_l \partial_m u_k\Bigr)\Bigr]\\
    &= -\tc_i\nc_k\tc_l \partial_l u_k + \bc_i \bc_k \nc_l \partial_l u_k + \frac{1}{\tilde{\kappa}} \bc_i \bc_k \tc_l \tc_m \partial_l \partial_m u_k,
\end{align}
where we also used orthonormality of the Frenet-Serret frame. This is \eqref{eq:normal_evolution}. 

Finally, in order to derive the curvature evolution equation, we use a simple definition as a function of the tangent vector with arc-length parameterization, 
\begin{equation}\tilde{\kappa}(\phi,t) = |\partial_s \tv|,
\end{equation} which is equivalent to definition~\eqref{eq:curv_def}. 
Using the previous results, we obtain
\begin{align}\nonumber
  \partial_t \tilde{\kappa} &= \partial_t \norms{\partial_s \tv}= \nc_j \partial_t \partial_s \tc_j = \nc_j \tilde{\kappa} \Bigl[- \tc_k \tc_l \nc_j (\partial_k u_l) - (\nc_j \tc_k + \tc_j \nc_k) \tc_l \partial_l u_k  + (\nc_j \nc_k + \bc_j \bc_k)\Bigl(\nc_l \partial_l u_k + \frac{1}{\norms{\partial_s \tv}}\tc_l \tc_m \partial_l \partial_m u_k\Bigr)\Bigr] \\
  &= \tilde{\kappa} \Bigl( -2\tc_k \tc_l + \nc_k \nc_l \Bigr) \partial_k u_l + \nc_k \tc_l \tc_m \partial_l \partial_m u_k. \label{eq:curv_evo_eq}
\end{align}
This is \eqref{eq:curvature_evolution}, 
which can also be found in ref.~\cite{drummond_jfm_1991_curv}.

\section{Fokker-Planck equation in the Kraichnan model}
In Methods, we laid out the terms that need to be calculated in order to arrive at the Fokker-Planck equation in the Kraichnan model. In order to proceed, we combine \eqref{eq:fpe_arclength_terms} 
with the evolution equations derived in the previous section and get
\begin{align}
    \partial_t f(\kappa; t) &= \frac{\left\langle \delta(\kappa - \tilde{\kappa}) \partial_t \norms{\partial_{\angl} \mathbf{L}} \right\rangle}{\left\langle \norms{\partial_{\angl} \mathbf{L}} \right\rangle} - f(\kappa; t) \frac{\partial_t \left\langle \norms{\partial_{\angl} \mathbf{L}} \right\rangle}{\left\langle \norms{\partial_{\angl} \mathbf{L}} \right\rangle}
    - \frac{1}{\left\langle \norms{\partial_{\angl} \mathbf{L}} \right\rangle}
    \partial_{\kappa}\left\langle \delta(\kappa - \tilde{\kappa}) \norms{\partial_{\angl} \mathbf{L}} \partial_t \tilde{\kappa} \right\rangle \\
    &= \frac{1}{\left\langle \norms{\partial_{\angl} \mathbf{L}} \right\rangle} \Biggl( \left\langle  |\partial_\angl \mathbf{L}| \delta(\kappa - \tilde{\kappa}) \tc_i \tc_j \partial_j u_i\right\rangle - f(\kappa; t) \left\langle |\partial_\angl \mathbf{L}| \tc_i \tc_j \partial_j u_i \right\rangle \label{eq:fpe_all_averages} \\
    &\quad - \partial_{\kappa} \Bigl( -2\kappa \Bigl\langle |\partial_\angl \mathbf{L}| \delta(\kappa - \tilde{\kappa}) \tc_i \tc_j \partial_j u_i \Bigr\rangle + \kappa \Bigl\langle |\partial_\angl \mathbf{L}| \delta(\kappa - \tilde{\kappa}) \nc_i \nc_j \partial_j u_i \Bigr\rangle + \Bigl\langle |\partial_\angl \mathbf{L}| \delta(\kappa - \tilde{\kappa}) \nc_i \tc_j \tc_k \partial_j \partial_k u_i \Bigr\rangle \Bigr) \Biggr). \nonumber
\end{align}
We want to evaluate these averages using the Gaussian integration by parts formula~\cite{furutsu_1964, donsker_mate_1967, novikov_sjp_1965} combined with the correlation tensor~\eqref{eq:kraichnan_correlation}. 
This works analogously for all of them. So let us focus on one of the averages. By introducing delta functions, we can consider the velocity field at the Eulerian coordinate $\mathbf x$ and take the derivative out of the average (the other quantities are still evaluated at $\mathbf{L}(\phi, t)$),
\begin{align}
  \langle |\partial_\angl \mathbf{L}| \delta(\kappa - \tilde{\kappa}) \nc_i \nc_j \partial_j u_i \rangle
  &= \int\vdif \mathbf{x}\int\vdif\mathbf{y}~\delta(\mathbf{x}-\mathbf{y})\dpd{}{x_j} \left\langle |\partial_\angl \mathbf{L}| \delta(\mathbf{y} - \mathbf{L}(\phi,t)) \delta(\kappa - \tilde{\kappa}) \nc_i \nc_j u_i(\mathbf{x},t) \right\rangle.
\end{align}
Then Gaussian integration by parts yields
\begin{align} \nonumber
  \langle |\partial_\angl \mathbf{L}| \delta(\kappa - \tilde{\kappa}) \nc_i \nc_j \partial_j u_i \rangle
  &= \int\vdif \mathbf{x}\int\vdif\mathbf{y}~\delta(\mathbf{x}-\mathbf{y})\dpd{}{x_j} \int\vdif\mathbf{z}~\wc_{ik}(\mathbf{x} - \mathbf{z}) \left\langle \dfd{\left[|\partial_\angl \mathbf{L}|\delta(\mathbf{y} - \mathbf{L}(\phi,t))\delta(\kappa -\tilde{\kappa}) \nc_i \nc_j\right]}{u_k(\mathbf {z}, t)} \right\rangle \\
  &= \int\vdif \mathbf{x} \int\vdif\zv~\left(\partial_j\wc_{ik}(\xv - \zv)\right) \left\langle \dfd{\left[|\partial_\angl \mathbf{L}|\delta(\xv - \mathbf{L}(\phi,t))\delta(\kappa -\tilde{\kappa}) \nc_i \nc_j\right]}{u_k(\zv, t)} \right\rangle.
\end{align}
The product rule for the functional derivative yields five different terms, which can all be treated in the same way. Let us again focus on a single one of them, namely
\begin{align}\nonumber
  M &= \int\vdif \mathbf{x} \int\vdif\zv~\left(\partial_j\wc_{ik}(\xv - \zv)\right)
  \left\langle |\partial_\angl \mathbf{L}| \delta(\xv - \mathbf{L}(\phi,t)) \nc_i \nc_j \dfd{\left[\delta(\kappa -\tilde{\kappa})\right]}{u_k(\zv, t)} \right\rangle \\
  &= \int\vdif\zv~\left\langle \left(\partial_j\wc_{ik}(\mathbf{L}(\phi,t) - \zv)\right)
  |\partial_\angl \mathbf{L}| \nc_i \nc_j \delta'(\tilde{\kappa} - \kappa) \dfd{\tilde{\kappa}}{u_k(\zv, t)} \right\rangle.
\end{align}
In order to determine the response function $\dfd{\tilde{\kappa}}{u_k(\zv, t)}$, we formally integrate the curvature evolution equation~\eqref{eq:curv_evo_eq},
\begin{align}
  \tilde{\kappa}(\angl, t) &= \tilde{\kappa}(\angl, 0) + \int_0^t \dif t'~\eval{\left(\tilde{\kappa} \nc_m\nc_n\partial_n u_m - 2\tilde{\kappa}\tc_m\tc_n\partial_n u_m + \nc_m\tc_n\tc_o\partial_n\partial_o u_m\right)}_{(\mathbf{L}(\angl, t'), t')}.
\end{align}
By causality, the initial condition will not depend on $u_k(\zv, t)$ for $t>0$. The integrand will not depend on $u_k(\zv, t)$ for all $t'<t$ either, and the only contribution to the functional derivative can come from the time $t'=t$. Since $\tilde{\kappa}$, $\nv$, and $\tv$ are integrated quantities of the delta-correlated field $\uv$, we expect them to be continuous in time, just like a Wiener process is continuous while its differential is not. Hence their response functions will only be finite, thus not contribute to the integral. Using that
\begin{align}
  \dfd{u_m(\mathbf{L}(\angl, t'), t')}{u_k(\zv, t)} &= \delta(\mathbf{L}(\angl, t') - \zv)\delta(t-t')\delta_{mk},
\end{align}
the response function becomes
\begin{align}
  \dfd{\tilde{\kappa}(\angl, t)}{u_k(\zv, t)} &= \frac{1}{2}\Bigl(\tilde{\kappa} \nc_k\nc_n \partial_n \delta(\mathbf{L}(\angl, t) - \zv) - 2\tilde{\kappa} \tc_k\tc_n \partial_n \delta(\mathbf{L}(\angl, t) - \zv) + \nc_k\tc_n\tc_o \partial_n\partial_o \delta(\mathbf{L}(\angl, t) - \zv)\Bigr),\nonumber
\end{align}
where the factor $\frac{1}{2}$ comes from the fact that only half of the delta function $\delta(t-t')$ is contained in the integration range $[0,t]$. Using integration by parts and the sifting property of the delta function, our term $M$ can thus be simplified to
\begin{align} 
  M &= \frac{1}{2} \partial_{\kappa} \Biggl( \left(\partial_j\partial_n \wc_{ik}(\mathbf{0})\right) \kappa \left\langle |\partial_\angl \mathbf{L}| \nc_i\nc_j \delta(\kappa-\tilde{\kappa}) (\nc_k\nc_n - 2\tc_k\tc_n) \right\rangle - \left(\partial_j\partial_n\partial_o \wc_{ik}(\mathbf{0})\right) \left\langle |\partial_\angl \mathbf{L}| \nc_i\nc_j \delta(\kappa - \tilde{\kappa})\nc_k\tc_n\tc_o \right\rangle \Biggr). \nonumber
\end{align}

Although the spatial correlation tensor $\wc_{ik}(\mathbf{r})$ can be freely chosen, we can restrict its functional form by assuming isotropy and incompressibility of the Kraichnan field. By isotropy the correlation tensor must be even, hence odd derivatives vanish at zero, e.g.\ $\partial_j\partial_n\partial_o \wc_{ik}(\mathbf{0}) = 0$.
For the second derivatives, we know that they must have the general form of an isotropic rank-4 tensor~\cite{kearsley_jrnbs_1975},
\begin{align}
  Q_{jn}^{ik} = -\partial_j \partial_n \wc_{ik}(\mathbf{0}) = A\delta_{ik}\delta_{jn} + B\delta_{ij}\delta_{kn} + C\delta_{in}\delta_{jk}.
\end{align}
By definition, this tensor must be symmetric under exchange of $j$ and $n$, which implies $B=C$. Finally, incompressibility implies $  \delta_{ij} Q_{jn}^{ik}=0$ so that
\begin{align}
 A &= -4B,
\end{align}
leading to the general form~\cite{pumir_prf_2017}
\begin{align}
  Q_{jn}^{ik} = Q(4\delta_{ik}\delta_{jn} - \delta_{ij}\delta_{kn} - \delta_{in}\delta_{jk}).
\end{align}
Analogous arguments can be used to deduce the general form~\eqref{eq:P_def} 
of the fourth-order derivatives of the correlation tensor. Using this expression, our term $M$ can be evaluated,
\begin{align}\nonumber
  M &= -\frac{1}{2}Q\partial_{\kappa}\left((4\delta_{ik}\delta_{jn} - \delta_{ij}\delta_{kn} - \delta_{in}\delta_{jk})\kappa \left\langle |\partial_\angl \mathbf{L}| \nc_i\nc_j \delta(\kappa-\tilde{\kappa}) (\nc_k\nc_n - 2\tc_k\tc_n) \right\rangle \right) \\
  &= -2Q\partial_{\kappa} \left(\kappa f(\kappa; t)\right) \left\langle |\partial_\angl \mathbf{L}| \right\rangle,
\end{align}
by orthonormality of the Frenet-Serret frame.

In the following, we list all the terms that need to be computed along with the results of their evaluation. By Gaussian integration by parts and the product rule for functional derivatives, the averages of \eqref{eq:fpe_all_averages} split into
\begin{align}
\langle |\partial_\angl \mathbf{L}| \delta(\kappa - \tilde{\kappa}) \tc_i \tc_j \partial_j u_i \rangle &= \int\vdif\zv~\left\langle \left(\partial_j\wc_{ik}(\mathbf{L}(\phi,t) - \zv)\right)
  \delta(\kappa -\tilde{\kappa}) \tc_i \tc_j \dfd{|\partial_\angl \mathbf{L}|}{u_k(\zv, t)}\right\rangle \label{eq:curv_tt_arc}\\
  & \qquad + \int\vdif\zv~\left\langle \left(\partial_j\wc_{ik}(\mathbf{L}(\phi,t) - \zv)\right)
  |\partial_\angl \mathbf{L}| \tc_i \tc_j \dfd{\left[\delta(\kappa -\tilde{\kappa})\right]}{u_k(\zv, t)}\right\rangle \label{eq:curv_tt_kappa}\\
  & \qquad + \int\vdif\zv~\left\langle \left(\partial_j\wc_{ik}(\mathbf{L}(\phi,t) - \zv)\right)
  |\partial_\angl \mathbf{L}| \delta(\kappa -\tilde{\kappa})  \tc_j \dfd{\tc_i}{u_k(\zv, t)}\right\rangle  \label{eq:curv_tt_t_vel}\\
  & \qquad + \int\vdif\zv~\left\langle \left(\partial_j\wc_{ik}(\mathbf{L}(\phi,t) - \zv)\right)
  |\partial_\angl \mathbf{L}| \delta(\kappa -\tilde{\kappa})  \tc_i \dfd{\tc_j}{u_k(\zv, t)}\right\rangle  \label{eq:curv_tt_t_diff}\\
  & \qquad + \int\vdif \mathbf{x} \int\vdif\zv~\left(\partial_j\wc_{ik}(\xv - \zv)\right)
  \left\langle |\partial_\angl \mathbf{L}| \delta(\kappa -\tilde{\kappa}) \tc_i \tc_j \dfd{\left[\delta(\xv - \mathbf{L}(\phi,t))\right]}{u_k(\zv, t)} \right\rangle, \label{eq:curv_tt_lagr}
\end{align}
\begin{align}
\langle |\partial_\angl \mathbf{L}| \tc_i \tc_j \partial_j u_i \rangle &= \int\vdif\zv~\left\langle \left(\partial_j\wc_{ik}(\mathbf{L}(\phi,t) - \zv)\right)
  \tc_i \tc_j \dfd{|\partial_\angl \mathbf{L}|}{u_k(\zv, t)}\right\rangle \label{eq:curv_arc_arc}\\
  & \qquad + \int\vdif\zv~\left\langle \left(\partial_j\wc_{ik}(\mathbf{L}(\phi,t) - \zv)\right)
  |\partial_\angl \mathbf{L}|  \tc_j \dfd{\tc_i}{u_k(\zv, t)}\right\rangle  \label{eq:curv_arc_t_vel}\\
  & \qquad + \int\vdif\zv~\left\langle \left(\partial_j\wc_{ik}(\mathbf{L}(\phi,t) - \zv)\right)
  |\partial_\angl \mathbf{L}|  \tc_i \dfd{\tc_j}{u_k(\zv, t)}\right\rangle  \label{eq:curv_arc_t_diff}\\
  & \qquad + \int\vdif \mathbf{x} \int\vdif\zv~\left(\partial_j\wc_{ik}(\xv - \zv)\right)
  \left\langle |\partial_\angl \mathbf{L}| \tc_i \tc_j \dfd{\left[\delta(\xv - \mathbf{L}(\phi,t))\right]}{u_k(\zv, t)} \right\rangle, \label{eq:curv_arc_lagr}
\end{align}
\begin{align}
\langle |\partial_\angl \mathbf{L}| \delta(\kappa - \tilde{\kappa}) \nc_i \nc_j \partial_j u_i \rangle 
  &= \int\vdif\zv~\left\langle \left(\partial_j\wc_{ik}(\mathbf{L}(\phi,t) - \zv)\right)
  \delta(\kappa -\tilde{\kappa}) \nc_i \nc_j \dfd{|\partial_\angl \mathbf{L}|}{u_k(\zv, t)}\right\rangle \label{eq:curv_nn_arc} \\
  & \qquad + \int\vdif\zv~\left\langle \left(\partial_j\wc_{ik}(\mathbf{L}(\phi,t) - \zv)\right)
  |\partial_\angl \mathbf{L}| \nc_i \nc_j \dfd{\left[\delta(\kappa -\tilde{\kappa})\right]}{u_k(\zv, t)}\right\rangle \label{eq:curv_nn_kappa} \\
  & \qquad + \int\vdif\zv~\left\langle \left(\partial_j\wc_{ik}(\mathbf{L}(\phi,t) - \zv)\right) |\partial_\angl \mathbf{L}|
  \delta(\kappa -\tilde{\kappa}) \nc_j \dfd{\nc_i}{u_k(\zv, t)}\right\rangle \label{eq:curv_nn_n_vel}\\
  & \qquad + \int\vdif\zv~\left\langle \left(\partial_j\wc_{ik}(\mathbf{L}(\phi,t) - \zv)\right) |\partial_\angl \mathbf{L}|
  \delta(\kappa -\tilde{\kappa}) \nc_i \dfd{\nc_j}{u_k(\zv, t)}\right\rangle \label{eq:curv_nn_n_diff}\\
  & \qquad + \int\vdif \mathbf{x} \int\vdif\zv~\left(\partial_j\wc_{ik}(\xv - \zv)\right)
  \left\langle |\partial_\angl \mathbf{L}| \delta(\kappa -\tilde{\kappa}) \nc_i \nc_j \dfd{\left[\delta(\xv - \mathbf{L}(\phi,t))\right]}{u_k(\zv, t)} \right\rangle, \label{eq:curv_nn_lagr}
\end{align}
and
\begin{align}
\langle |\partial_\angl \mathbf{L}| \delta(\kappa - \tilde{\kappa}) \nc_i \tc_j \tc_k \partial_j \partial_k u_i \rangle &= \int\vdif\zv~\left\langle \left(\partial_j \partial_k \wc_{il}(\mathbf{L}(\phi,t) - \zv)\right)
  \delta(\kappa -\tilde{\kappa}) \nc_i \tc_j \tc_k \dfd{|\partial_\angl \mathbf{L}|}{u_l(\zv, t)}\right\rangle  \label{eq:curv_ntt_arc}\\
  & \qquad + \int\vdif\zv~\left\langle \left(\partial_j \partial_k \wc_{il}(\mathbf{L}(\phi,t) - \zv)\right)
  |\partial_\angl \mathbf{L}| \nc_i \tc_j \tc_k \dfd{\left[\delta(\kappa -\tilde{\kappa})\right]}{u_l(\zv, t)}\right\rangle  \label{eq:curv_ntt_kappa}\\
  & \qquad + \int\vdif\zv~\left\langle \left(\partial_j \partial_k \wc_{il}(\mathbf{L}(\phi,t) - \zv)\right)
  |\partial_\angl \mathbf{L}| \delta(\kappa -\tilde{\kappa}) \tc_j \tc_k \dfd{\nc_i}{u_l(\zv, t)}\right\rangle \label{eq:curv_ntt_n_vel}\\
  & \qquad + \int\vdif\zv~\left\langle \left(\partial_j \partial_k \wc_{il}(\mathbf{L}(\phi,t) - \zv)\right)
  |\partial_\angl \mathbf{L}| \delta(\kappa -\tilde{\kappa}) \nc_i \tc_k \dfd{\tc_j}{u_l(\zv, t)}\right\rangle \label{eq:curv_ntt_t_diff1} \\
  & \qquad + \int\vdif\zv~\left\langle \left(\partial_j \partial_k \wc_{il}(\mathbf{L}(\phi,t) - \zv)\right)
  |\partial_\angl \mathbf{L}| \delta(\kappa -\tilde{\kappa}) \nc_i \tc_j \dfd{\tc_k}{u_l(\zv, t)}\right\rangle \label{eq:curv_ntt_t_diff2} \\
  & \qquad + \int\vdif \mathbf{x} \int\vdif\zv~\left(\partial_j \partial_k \wc_{il}(\xv - \zv)\right)
  \left\langle |\partial_\angl \mathbf{L}| \delta(\kappa -\tilde{\kappa}) \nc_i \tc_j \tc_k \dfd{\left[\delta(\xv - \mathbf{L}(\phi,t))\right]}{u_l(\zv, t)} \right\rangle.  \label{eq:curv_ntt_lagr}
\end{align}
Evaluating each of these terms as explained previously yields
\begin{alignat*}{3}
 \text{\eqref{eq:curv_tt_arc}} &= Q f(\kappa; t) \left\langle |\partial_\angl \mathbf{L}| \right\rangle
 &&\qquad \text{\eqref{eq:curv_tt_kappa}} = \frac{5}{2} Q\partial_{\kappa} (\kappa f(\kappa; t)) \left\langle |\partial_\angl \mathbf{L}| \right\rangle\\
  \text{\eqref{eq:curv_tt_t_vel}} &= 4Qf(\kappa; t) \left\langle |\partial_\angl \mathbf{L}| \right\rangle
   &&\qquad \text{\eqref{eq:curv_tt_t_diff}} = -Qf(\kappa; t) \left\langle |\partial_\angl \mathbf{L}| \right\rangle\\
  \text{\eqref{eq:curv_tt_lagr}} &= 0
   &&\qquad \text{\eqref{eq:curv_arc_arc}} = Q\left\langle |\partial_\angl \mathbf{L}| \right\rangle\\
  \text{\eqref{eq:curv_arc_t_vel}} &= 4Q\left\langle |\partial_\angl \mathbf{L}| \right\rangle
   &&\qquad \text{\eqref{eq:curv_arc_t_diff}} = -Q\left\langle |\partial_\angl \mathbf{L}| \right\rangle\\
  \text{\eqref{eq:curv_arc_lagr}} &= 0
   &&\qquad \text{\eqref{eq:curv_nn_arc}} = -\frac{1}{2}Q f(\kappa; t) \left\langle |\partial_\angl \mathbf{L}| \right\rangle\\
  \text{\eqref{eq:curv_nn_kappa}} &= -2Q\partial_{\kappa} (\kappa f(\kappa; t))\left\langle |\partial_\angl \mathbf{L}| \right\rangle
   &&\qquad \text{\eqref{eq:curv_nn_n_vel}} = \frac{5}{2} Q f(\kappa; t)\left\langle |\partial_\angl \mathbf{L}| \right\rangle\\
  \text{\eqref{eq:curv_nn_n_diff}} &= -\frac{5}{2}Q f(\kappa; t)\left\langle |\partial_\angl \mathbf{L}| \right\rangle
   &&\qquad \text{\eqref{eq:curv_nn_lagr}} = 0\\
  \text{\eqref{eq:curv_ntt_arc}} &= 0
   &&\qquad \text{\eqref{eq:curv_ntt_kappa}} = -9P \partial_{\kappa} f(\kappa; t)\left\langle |\partial_\angl \mathbf{L}| \right\rangle\\
  \text{\eqref{eq:curv_ntt_n_vel}} &= \frac{9}{\kappa}P f(\kappa; t)\left\langle |\partial_\angl \mathbf{L}| \right\rangle
   &&\qquad \text{\eqref{eq:curv_ntt_t_diff1}} = 0\\
  \text{\eqref{eq:curv_ntt_t_diff2}} &= 0
   && \qquad \text{\eqref{eq:curv_ntt_lagr}} = 0,
\end{alignat*}
where $Q$ and $P$ are defined through~\eqref{eq:Q_def} 
and~\eqref{eq:P_def}, 
respectively. Inserting these results into \eqref{eq:fpe_all_averages} yields the Fokker-Planck equation~\eqref{eq:fokker_planck}. 

\newpage
\section{Numerical results in the Kraichnan model}
Here, we present a numerical analysis of curvature statistics in the Kraichnan model. To this end, we interpret the tracer equation~\eqref{eq:tracer_eq} 
as a Langevin equation. Since It\^o and Stratonovich interpretations coincide for this equation, we may use the Euler-Maruyama scheme~\cite{glassermann_2003} to integrate particle trajectories. In every time step, the Gaussian flow field is computed on $1024^3$ grid points with a model energy spectrum as described by Pope~\cite[p.~232]{pope_2000},
\begin{align} \label{eq:model_spectrum}
  E(k) &\propto k^{-5/3} f_L(kL) f_{\eta}(k\eta),
\end{align}
with $\eta$ a viscous length scale, $L \approx 946 \eta$ an integral length scale and the functions
\begin{align}
  f_L(x) &= \left(\frac{x}{(x^2 + c_L)^{1/2}}\right)^{11/3}
\end{align}
and
\begin{align}
  f_\eta(x) &= \exp\left(-\beta\left(x^4 + c_\eta^4\right)^{1/4} - c_\eta\right),
\end{align}
which determine the large- and small-scale behavior. The spectrum~\eqref{eq:model_spectrum} integrates to the total energy $E = 1.05$ (code units). The temporal resolution is $\Delta t = 7.15\times 10^{-7}$ (code units) and the spatial resolution can be quantified by $k_{\max}\eta \approx 2.0$, where $k_{\max}$ is the maximum resolved wavenumber. Furthermore, we choose $\beta=5.2$, $c_L = 6.03$ and $c_\eta = 0.40$. For particle time stepping, the field is interpolated using spline interpolation with continuous derivatives up to and including third order computed over a kernel of $12^3$ grid points. The loops are adaptively refined as described for the Navier-Stokes simulations in Methods.

\begin{figure}[p]
  \vspace{-0.25cm}
  \centering
  \begin{minipage}{.5\textwidth}
  \includegraphics[width=1\textwidth]{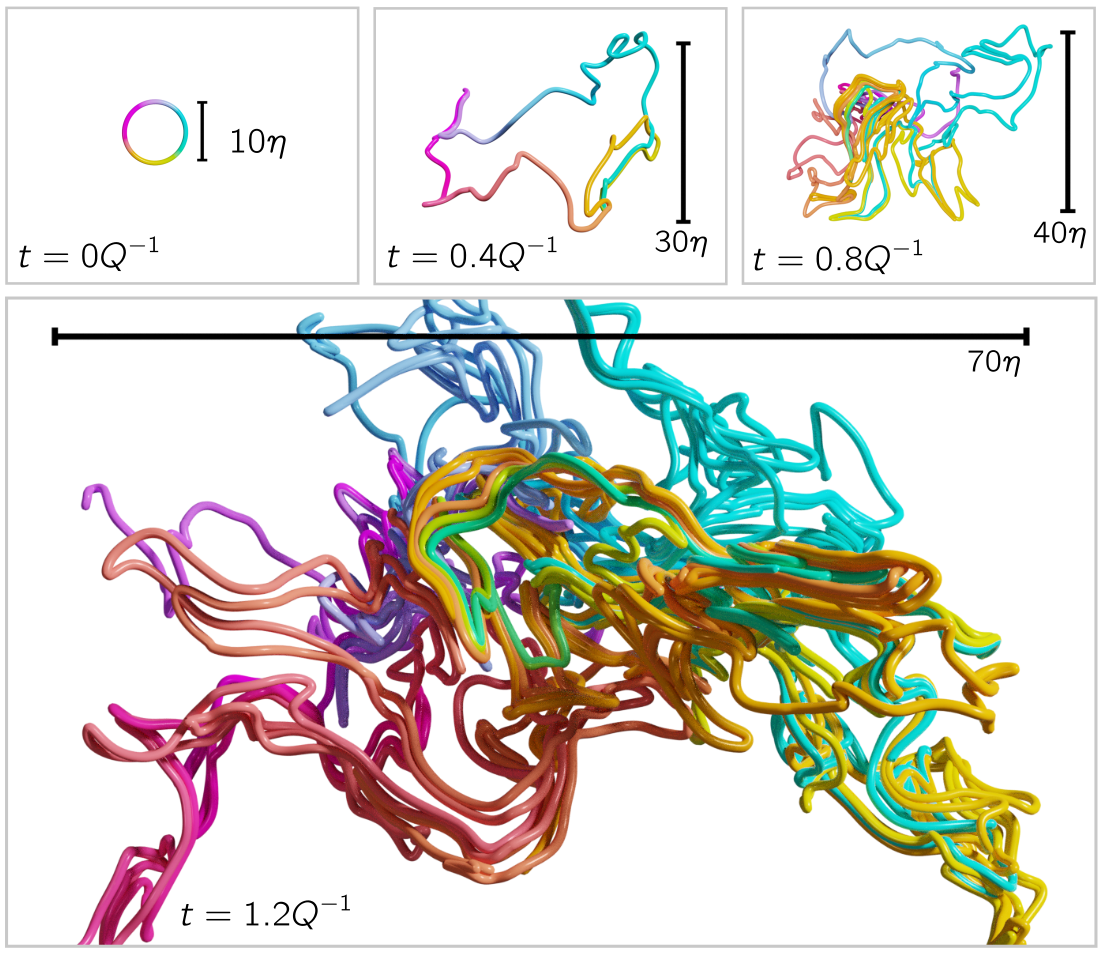}
  \end{minipage}
  \caption{\textbf{Visualization of a material loop advected by a Kraichnan field} for $1.2Q^{-1}$.  \label{fig:kraichnan_loop}}
\end{figure}

\begin{figure*}[p]
    \centering
    \begin{minipage}{0.49\textwidth}
    \includegraphics[width=1.\textwidth]{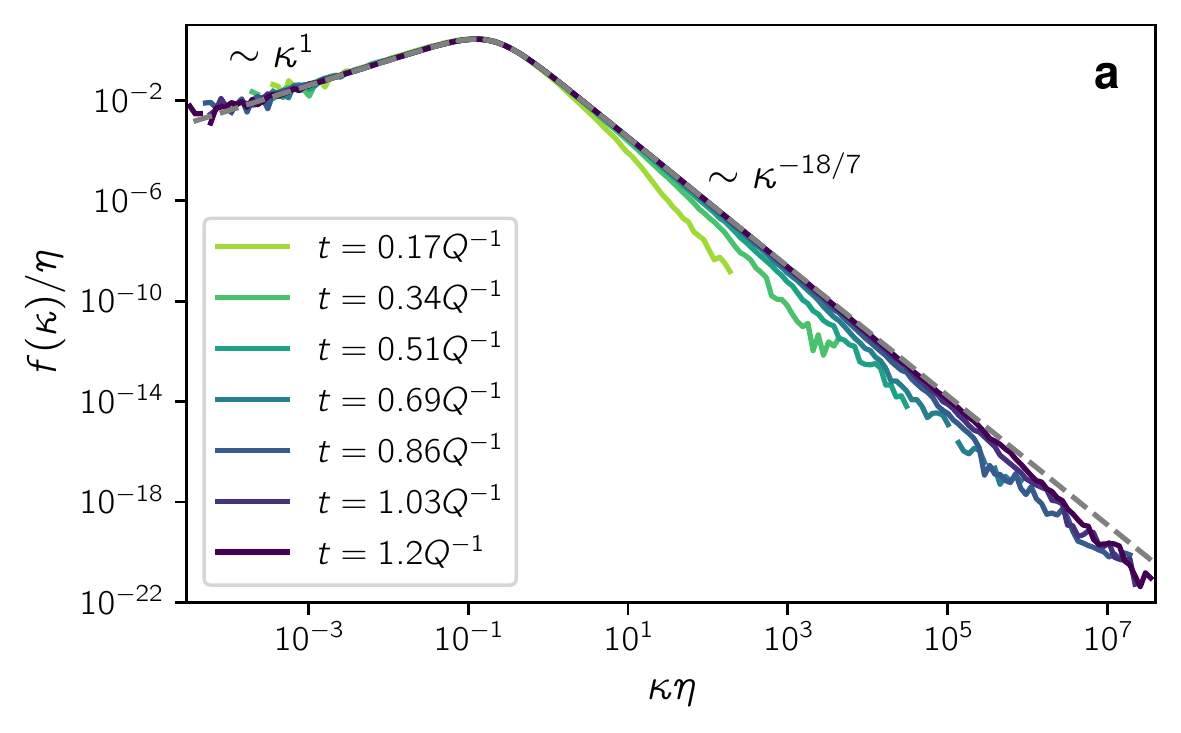}
    \end{minipage}
    \begin{minipage}{0.49\textwidth}
    \includegraphics[width=1.\textwidth]{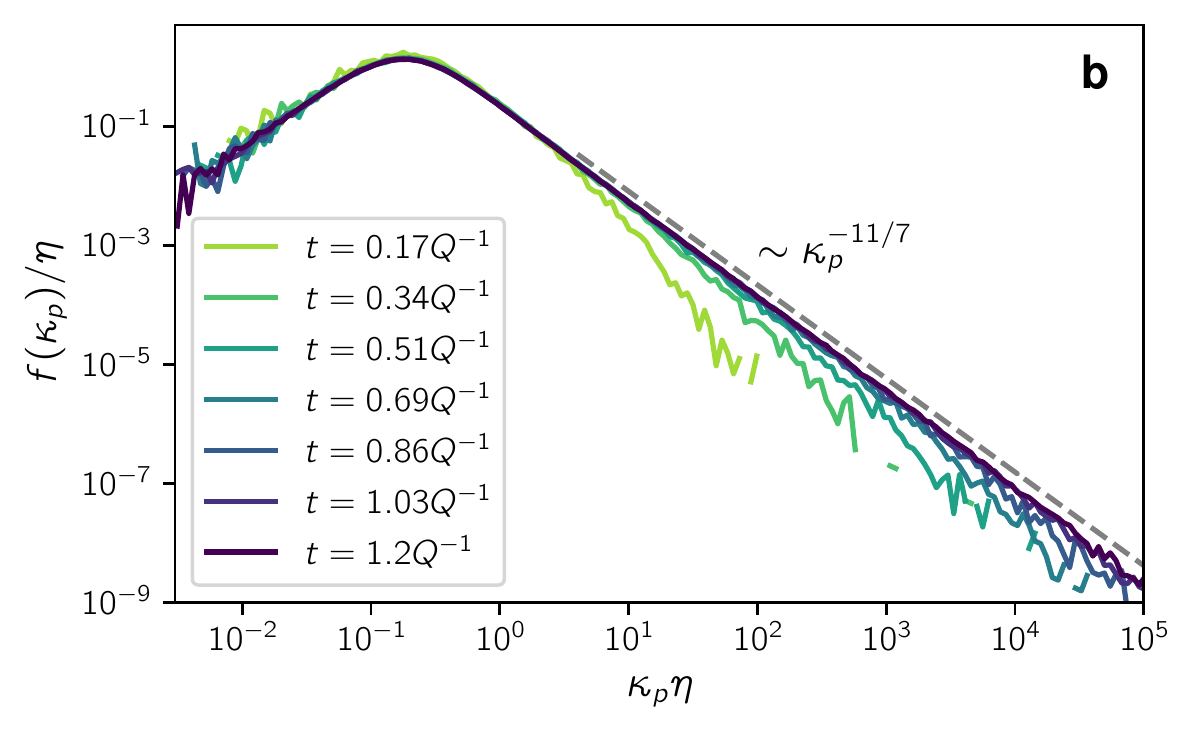}
    \end{minipage}
    \caption{\textbf{Curvature statistics in the Kraichnan model.} \textbf{a} Curvature PDF of material loops at different times. The PDF converges to the stationary solution~\eqref{eq:fpe_stationary_solution} 
    of the Fokker-Planck equation, indicated by the dashed line. \textbf{b} PDF of curvature maxima of material loops at the same times. The high-curvature tail scales as a power law, in agreement with our theoretical prediction $\kappa_p^{-11/7}$. \label{fig:kraichnan_curvature}}
    \vspace{-0.25cm}
\end{figure*}

Figure~\ref{fig:kraichnan_loop} shows a visualization of an initially circular loop deformed by the Kraichnan field for $1.2Q^{-1}$. Visually, it shares many features of material loops in Navier-Stokes turbulence but appears slightly more compact (compare Fig.~\ref{fig:loop_visualization}). 
The geometric similarities also manifest in the curvature statistics (Figure~\ref{fig:kraichnan_curvature}a), which display the same type of unimodal distribution with power-law tails. Over time, the PDF converges to the stationary solution~\eqref{eq:fpe_stationary_solution} 
of the Fokker-Planck equation, featuring the power-law tails $\kappa^1$ and $\kappa^{-18/7}$. 

\begin{figure}[p]
    \centering
    \begin{minipage}{.5\textwidth}
    \includegraphics[width=1\textwidth]{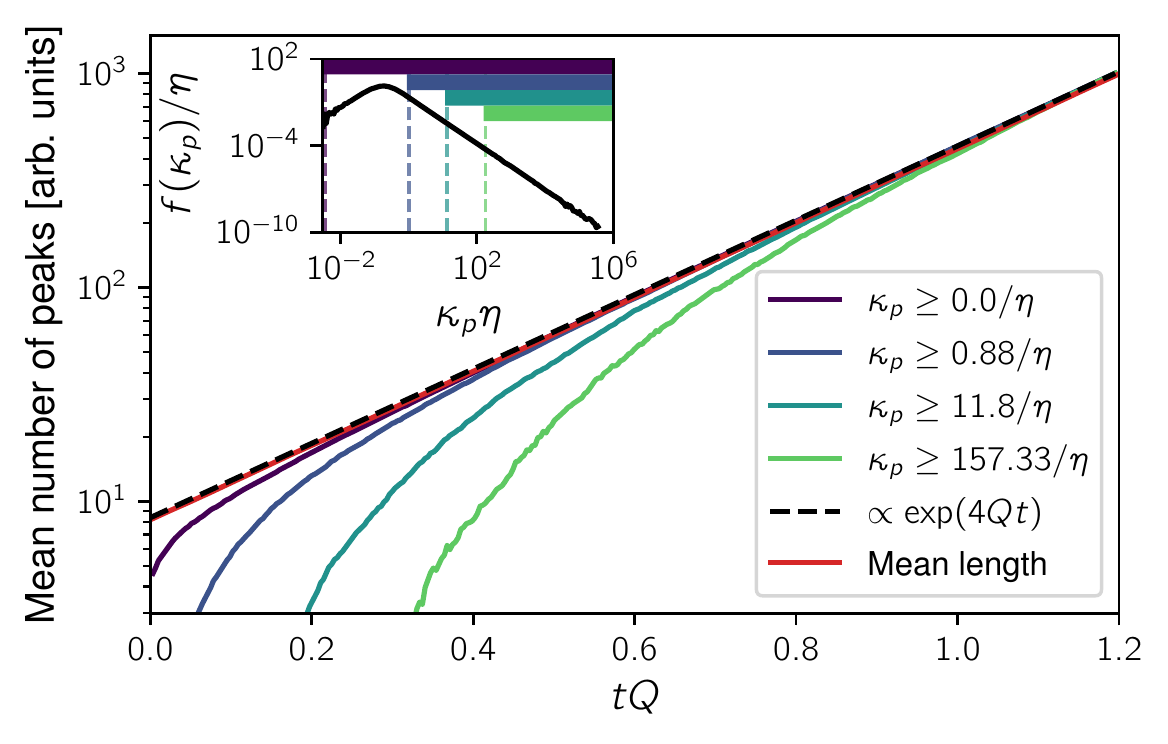}
    \end{minipage}\vspace{-0.25cm}
    \caption{\textbf{Mean number of curvature peaks in the Kraichnan model} above different thresholds over time, vertically shifted for comparison. Consistent with our observation in Navier-Stokes turbulence, the curves become asymptotically proportional to the mean arc length of loops, which grows as $e^{4Qt}$ as predicted by theory. Inset: Curvature peak distribution at $t=1.2Q^{-1}$ indicating the different thresholds. \label{fig:kraichnan_peak_number}}\vspace{-0.25cm}
\end{figure}

The Kraichnan model also forms curvature peaks, whose distribution (Figure~\ref{fig:kraichnan_curvature}b) qualitatively resembles the one in Navier-Stokes turbulence. Our theory predicts the high-curvature tail to scale as a power law with exponent $-11/7$, which is confirmed by the simulation. In order to form our theory, we made the empirical observation that the curvature peak number grows proportional to the mean line length. This is also what we observe in the Kraichnan model (Figure~\ref{fig:kraichnan_peak_number}), where the mean line length can be computed analytically to be proportional to $e^{4Qt}$~\cite{balkovsky_pre_1999}.

\newpage
\section{Lyapunov exponents by QR-decomposition}
\label{ch:peak_curvature_qr}
In Methods, we have defined finite-time Lyapunov exponents as the growth rate of singular values of the deformation tensor \eqref{eq:deformation_tensor_svd}. 
In practice, however, we instead compute the QR-decomposition of the deformation tensor for numerical stability,
\begin{align}
  F(t) = Q(t)R(t),
\end{align}
where $Q(t)$ is orthogonal and $R(t)$ is upper triangular. Note that the matrix $Q(t)$ is unrelated to the constant $Q$ introduced previously, which quantifies velocity gradient fluctuations in the Kraichnan model. The growth rate of the diagonal elements of $R(t)$ can then be interpreted as an alternative definition of FTLEs~\cite{johnson_pof_2015},
\begin{align} \label{eq:qr_ftle}
    \rho_i'(t) = \frac{1}{t} \log R_{ii}(t).
\end{align}
Note that this definition of FTLEs depends on the choice of the coordinate system. Complementing the Methods section, we here show that peak curvature dynamics of a parabola are exactly captured by this alternative definition of FTLEs.

We start from a parabolic material line as defined in \eqref{eq:parabolic_sling}. 
Given that the flow is statistically isotropic, FTLE statistics should not depend on the choice of the coordinate system. We can therefore assume that the line is aligned with the coordinate axes, i.e.~$\kv(0) = \mathbf{e}_1$ and $\lv(0) = \mathbf{e}_2$. In this case we have
\begin{align}
  \norm{\mathbf{k}(t)}^2 &= \norm{Q(t)R(t)\kv(0)}^2 \\
  &= Q_{ij}(t) R_{j1}(t) Q_{ik}(t) R_{k1}(t)\\
  &= R_{11}^2(t),
\end{align}
where we have used the fact that $Q$ is orthogonal and $R$ is upper triangular. Furthermore, we get
\begin{align}
  \norm{\kv(t)}^2\norm{\lv(t)}^2 - (\kv(t)\cdot\lv(t))^2
  &= R_{11}^2 Q_{ij} R_{j2} Q_{ik}R_{k2} - (Q_{ij}R_{j1} Q_{ik}R_{k2})^2 \\
  &= R_{11}^2 (R_{j2} R_{j2}) - (R_{j1} R_{j2})^2 \nonumber\\
  &= R_{11}^2 R_{12}^2 + R_{11}^2 R_{22}^2 - R_{11}^2 R_{12}^2 \\
  &= R_{11}^2(t) R_{22}^2(t).
\end{align}
Hence, the peak curvature given by \eqref{eq:parabola_peak_curv} 
can be written as
\begin{align}
  \kappa_p(t) &= \frac{R_{11}(t)}{R_{22}^2(t)}\kappa_p(0).
\end{align}
Inserting the alternative definition of FTLEs~\eqref{eq:qr_ftle} yields
\begin{align}
  \kappa_p(t) &= e^{[\rho_1'(t) - 2\rho_2'(t)]t} \kappa_p(0),
\end{align}
an exact analog to the approximate equation~\eqref{eq:peak_curv_growth}. 
This means that the FTLEs defined by the QR-decomposition precisely capture the curvature growth of parabolic line elements. In that sense, the QR-definition of FTLEs is very suitable for our purposes.
\end{document}